\pdfoutput=1
\RequirePackage{xcolor}
\documentclass[hyper]{JHEP}
\usepackage{graphicx,amssymb,bm,latexsym,amsmath}
\usepackage{epstopdf}
\usepackage{caption}
\usepackage{mathtools}
\usepackage{rotating}
\usepackage{enumitem}
\usepackage[normalem]{ulem}
\usepackage{caption,subcaption}
\newcommand{\bej}[1]{ \begin{equation}\label{#1} }
\newcommand{\eej}{\end{equation}}
\newcommand{\beaj}[1]{\begin{eqnarray}\label{#1} }
\newcommand{\eeaj}{\end{eqnarray}}
\newcommand{\eq}[1]{(\ref{#1})}
\def\ZZZ{{\hskip-3pt\hbox{ Z\kern-1.6mm Z}}}
\def\zzz{{\hskip-3pt\hbox{ z\kern-1mm z}}}

\newcommand{\bd}{\bar{\rm D}}

\newcommand{\N}{\frac{m_{2}}{k_{2}}-\frac{m_{1}}{k_{1}}}

\newcommand{\be}{\begin{equation}}
\newcommand{\ee}{\end{equation}}
\newcommand{\ben}{\begin{eqnarray}\displaystyle}
\newcommand{\een}{\end{eqnarray}}

\def\one{{\hbox{ 1\kern-.8mm l}}}
\def\zero{{\hbox{ 0\kern-1.5mm 0}}}

\def\be{\begin{equation}}       
\def\ee{\end{equation}}         

\def\bea{\begin{eqnarray}}      
\def\eea{\end{eqnarray}}
                                
\def\ba{\begin{array}}
\def\ea{\end{array}}
\def\bd{\begin{displaymath}}
\def\ed{\end{displaymath}}

\def\eq{\begin{equation}}
\def\eqe{\end{equation}}
\def\eqa{\begin{eqnarray}}
\def\eqae{\end{eqnarray}}
\def\ena{\end{eqnarray}}

\def\nn{\nonumber}
\def\Tr{{\rm Tr}}
\def\tr{{\rm tr}}

\def\unit{1 \hskip-.3em \raise2pt\hbox{$ \scriptstyle |$ } }

\def\bd{\begin{displaymath}}
\def\ed{\end{displaymath}}

\def\6{\partial}
\def\N4{{\cal N}=4}

\def\bop#1{\setbox0=\hbox{$#1M$}\mkern1.5mu
        \vbox{\hrule height0pt depth.04\ht0
        \hbox{\vrule width.04\ht0 height.9\ht0 \kern.9\ht0
        \vrule width.04\ht0}\hrule height.04\ht0}\mkern1.5mu}
\def\>{\rangle} %right angle

\def\<{\langle} %left angle
\def\Dsl{D \hskip-.6em \raise1pt\hbox{$ / $ } }
\def\to{\rightarrow}

\def\+{\oplus}

\def\Tr{{\rm Tr}\, }

\def\as2{AdS_3\times S^3_1 \times S^3_2}

\title{On interpolating anomalous dimension of twist-two operators with general spins}

\author{Aritra Banerjee$^{a}$, Abhishek Chowdhury$^{b}$, Somyadip Thakur$^{a}$ and Gang Yang$^{a}$ \\
$^a$CAS Key Laboratory of Theoretical Physics,\\ Institute of Theoretical Physics, Chinese Academy of Sciences,\\  Beijing 100190, P.R.China\\
$^b$Institute for Theoretical Physics,\\
TU Wien,
Wiedner Hauptstr. 8-10,\\
A-1040 Vienna,
Austria\\

\vskip .2in
Email: \email{aritra, somyadip, yangg@itp.ac.cn; abhishek.chowdhury@tuwien.ac.at}}

\abstract{We study non--perturbative interpolating functions to probe the physics of anomalous dimensions associated with twist--two operators in ${\cal N}=4$ SYM of finite and infinite spin. Compared to previous studies, the novel result of this paper is to introduce single multivariate functions of \emph{both} coupling $g$ and spin $j$ to approximate such anomalous dimensions. 
	We provide a unified framework to study such operators in interim ranges of the parameters which so far has eluded previous results. 
	Explicitly, we consider twist--two anomalous dimensions in two distinct scenarios using  interpolating functions. For the large $N$ case, we stick to simple Pad\'{e} approximants and its generalizations . For the finite $N$ case, ${\cal N}=4$ SYM is expected to be S--dual invariant, hence the observables are expected be modular invariant. To probe the finite  $N$ physics, we take into account the non--planar and instanton contributions by constructing modular invariant interpolating functions to approximate the cusp and twist--two anomalous dimensions. We also consider interpolating functions for the twist--four operators and study level crossing phenomenon between the twist--two and twist--four operators.

}


\begin{document}

%%%%%%%%%%%%%%%%%%%%%%%%%%%%%%%%%%%%%%%%%%%%%%%
%%%%%%%%%%%%%%%%%%%%%%%%%%%%%%%%%%%%%%%%%%%%%%%
\section{Introduction }
\label{intro}

As a close cousin of quantum chromodynamics (QCD), ${\cal N}=4$ super Yang--Mills (SYM) theory has drawn significant attention in the past twenty years. Unlike QCD which is analytically under control only in the perturbative ultraviolet (UV) regime, in ${\cal N}=4$ SYM, both weak and strong coupling behavior can be understood quantitively, thanks to the AdS/CFT correspondence \cite{Maldacena:1997re, Gubser:1998bc, Witten:1998qj}. In the `t Hooft large $N$ limit \cite{tHooft:1973alw}, the integrability makes it even possible to study various physical observables non--perturbatively, see e.g.~\cite{Beisert:2010jr} for a review. This makes ${\cal N}=4$ SYM an ideal model to study the non--perturbative dynamics of a four dimensional interacting gauge theory, which hopefully may lead to better understanding in generic gauge theories, including realistic QCD.

A particularly interesting class of observables are the anomalous dimensions of twist--two Wilson operators, which will be the central objects of our study. 
In ${\cal N}=4$ SYM, the simplest twist--two operator consists of two scalars written as ${\rm tr}(\phi D^j \phi)$, which has classical dimension $\Delta_0 = j+2$ and spin $j$, and the twist is defined by the difference $\Delta_0 - j=2$. The classical dimension receives quantum corrections which is the anomalous dimension $\gamma(g,j)$, where $g$ is the gauge coupling.
In QCD, twist--two operators appear in the operator product expansion description of deep inelastic scattering processes, and they play an important role for determining the parton distribution functions, see e.g.~\cite{Korchemsky:1988si}.

At least in this UV regime where  the twist--two operators contribute to high energy QCD processes, one expects that the two theories, QCD and ${\cal N}=4$ SYM, have many features in common.
Let us point out  two interesting connections between the two theories. Firstly, in the large spin limit, the twist--two anomalous dimensions in both theories develops a logarithmic behaviour with respect to the large spin dependence, and the coefficient is given by the cusp anomalous dimension (CAD) $\gamma_{\rm cusp}$ \cite{Polyakov:1980ca, Korchemsky:1985xj}, i.e.
\begin{equation}
\gamma(g, j) \sim \Gamma_{\rm cusp}(g) \log(j)\,, \qquad \textrm{for}\ j \gg 1 \ \textrm{and} \ j \gg g \,.
\end{equation}

Another intriguing connection is the so--called ``maximal transcendentality principle" which was first observed in \cite{Kotikov:2002ab, Kotikov:2004er}. It says that the anomalous dimensions of twist--two operators in ${\cal N}=4$ SYM can be obtained from the maximally transcendental part of the QCD results \cite{Moch:2004pa}. (Here, ``transcendentality" refers to transcendentality degree which characterizes the ``complexity" of mathematical numbers or functions, for example, the Riemann zeta value $\zeta_n$ and the polylogarithm ${\rm Li}_n$ have degree $n$.) This is a conjecture expected to be true perturbatively to all orders. Further evidences of these correspondences were also found for other observables such as amplitudes and Wilson lines \cite{Brandhuber:2012vm,Brandhuber:2017bkg,Jin:2018fak,Li:2014afw,Li:2016ctv}. These little understood mystical correspondence between ${\cal N}=4$ SYM and QCD indicates there is a deep connection between the two theories.

While the non--perturbative QCD is an extremely hard question to address, the goal of the present paper is to study non--perturbative interpolating functions to approximate the twist--two anomalous dimensions in ${\cal N}=4$ SYM. We hope such a study might help us to probe the real physics in QCD, as also indicated by the aforementioned connections.
Similar studies of using interpolating functions have been considered in \cite{Sen:2013oza,Beem:2013hha,Alday:2013bha,Honda:2014bza, Honda:2015ewa, Chowdhury:2016hny}.
Compared to previous studies, the main new aspect of this work is that under a single multivariate function we can analytically consider the anomalous dimension as a function of \emph{both} the coupling $g$ and spin $j$. 
This provides a unified framework for many previous results.  For example, the dependence on spin allows us to incorporate the information of the cusp anomalous dimension as the large spin limit of the constructed functions.
Explicitly, we will consider twist--two anomalous dimensions in two distinct scenarios, using two distinct classes of interpolating functions. 

The first scenario is the planar large $N$ limit, where ${\cal N}=4$ SYM is integrable and one can employ the tools of AdS/CFT duality to explore non--trivial gauge invariant objects. This case is relatively well understood due the significant progress made in past years, and in principle, non--perturbative results can be obtained via integrability techniques \cite{Beisert:2010jr}. In ${\cal N}=4$ SYM, one could think of twist--two operators $\tr (\phi D^j \phi)$ as adding large number of derivative impurities to the protected half--BPS vacuum $\tr(\phi^2)$, which forms the backbone of various integrability related investigations for such operators and their dual string states. While such a closed form for twist--two anomalous dimensions for arbitrary spin is not yet known, perturbative expansion in various limits are known to higher orders. These data allows us to construct relatively reliable interpolating functions, from which one can study the non--perturbative properties in regimes unreachable herein-before.  One interesting feature is that the anomalous dimensions in the small spin ($j\ll g$) and large spin ($j\gg g$) limits have very different behaviours. In this work we will use the framework of interpolating functions to explicitly capture such disparate physical phenomena and discuss the consequences. 

The second scenario is the case of finite $N$, which physically is more closely related to realistic QCD (where $N=3$).  This case is much harder to study, because the theory is no longer integrable and much less data are available. On the other hand, an important new ingredient in this case is the S--duality property of ${\cal N}=4$ SYM \cite{Osborn:1979tq}. (Note that we do not expect S--duality for twist--two anomalous dimensions in the large $N$ limit.) This suggests the use of interpolating functions that are invariant under $SL(2,Z)$ modular transformations provides strong constraints on the result. Furthermore, since we will be using a basis of modular invariant Eisenstein series (instead of polynomials) for our interpolating functions, both the instanton corrections as well as the non--planar data can be incorporated in our unified framework. Compared to the large $N$ case, a subtle new physical feature expected to appear at finite $N$ is the level crossing phenomenon, see e.g.~\cite{Korchemsky:2015cyx}. 
We will try to address this issue based on the interpolating function. Since the available finite $N$ data is limited, the main goal here is to have a qualitative or even speculative understanding of the physical picture. Hopefully this can provide new insight for (or be tested by) further studies in this regime. Besides, we will also encounter several technical challenges, such as getting the correct coupling powers in the strong coupling expansion and encoding the non--planar and instanton contributions in the modular invariant functions, which we will explicitly address in our construction.

The structure of the paper is planned as follows. 
We first review the construction of various classes of interpolating functions in section \ref{intfunc}.
Then we explicitly compile the data available in the literature for cusp and finite spin twist--two anomalous dimensions in section \ref{sec:dataofAD}, which will serve as input constraints for the interpolating functions in the subsequent sections. 
In section \ref{sec:planarcase}, we construct the interpolating function for the twist--two anomalous dimensions in the planar limit. We elaborate on both large and small spin physics here based on construction of two--point Pad\'e type approximants.
In section \ref{sec:finiteNcase}, we turn to the theory with finite value of $N$. We first discuss the interpolating function of cusp anomalous dimension with S--duality, which is related to the large spin physics. Then we discuss the interpolating function for the anomalous dimensions with finite spin.
We discuss the results and present an outlook for our constructions in section \ref{sec:discussion}. Finally, we give several appendices covering the technical details of the construction of interpolating functions.

%%%%%%%%%%%%%%%%%%%%%%%%%%%%%%%%%%%%%%%%%%%%%%%%%%%%%%%%%%%%%%%

%%%%%%%%%%%%%%%%%%%%%%%%%%%%%%%%%%%%%%%%%%%%%%%%%%%%%%%%%%%%%%%

%%%%%%%%%%%%%%%%%%%%%%%%%%%%%%%%%%%%%%%%%%%%%%%
%%%%%%%%%%%%%%%%%%%%%%%%%%%%%%%%%%%%%%%%%%%%%%%
\section{ Interpolating functions: Construction}
\label{intfunc}
	
Before going into the details of our interpolating functions for the anomalous dimensions of twist--two operators, let us first briefly discuss some well known techniques employed to study non--perturbative interpolations via approximants.
	In literature there exists a variety of sophisticated techniques to resum perturbative expansions  (both strong coupling and weak coupling) to get non--perturbative answers in quantum mechanics and quantum field theory. If the  perturbative series of the theory under consideration is Borel summable \cite{Kleinert:2001ax}, powerful techniques can be employed to resum the perturbative series. So it should be possible to obtain a great deal of insight into the behaviour of the resummed function associated with an observable at any finite coupling with some additional information about the strong coupling expansion along with the perturbative series. For us, the physical observable would be the anomalous dimensions of twist--two operators. 
	
	Although it is always theoretically possible to write a non--perturbative function which encodes the perturbative expansions of such anomalous dimensions, in general it becomes quite hard to encode extra symmetries such as those implied by $duality$ in ${\cal N}=4$ SYM on the functions themselves. It would be interesting to find appropriate resummation methods which are not only compatible with the asymptotic behaviour of perturbation theory, but also clearly encodes the features of duality symmetry.
	% so we can combine the resulting symmetry constraints with a more detailed consideration of the analytic properties of the anomalous dimensions.
	 It should be noted that such dualities are extremely valuable to constrain the structure of the anomalous dimensions. As a starting point to discuss such non--perturbative duality invariant results, one could explicitly construct interpolating functions which are inherently invariant under such dualities.
	
	Motivated by this, in this work we construct interpolating functions to study the non--perturbative properties of twist--two anomalous dimension in ${\cal N}=4$ SYM both at finite $N$ and large $N$ limits. In the former case where it is expected that anomalous dimensions will be invariant under the action of the full SL$(2,\mathbb{Z})$ modular group, we construct  interpolating functions that are invariant under modular transformations. 

In the latter case, i.e., in the large $N$ limit where the observables are not invariant under the symmetries imposed by S--duality,\footnote{For example, the instanton corrections, which are important for preserving the S--duality, are exponentially suppressed in the large $N$ limit, thus effectively breaking the modular invariance.} we would construct simple interpolating functions consistent with the weak coupling and the strong coupling expansions.

	To this effect, we will make use of distinct classes of interpolating functions constructed in the literature \cite{Sen:2013oza,Beem:2013hha,Alday:2013bha,Honda:2014bza,Chowdhury:2016hny}.
	
Broadly speaking, we will be employing two different approaches to approximate the anomalous dimensions for leading twist two operators both at finite and large spin limits.
	\begin{itemize}
		\item Interpolating functions without S--duality 

		\item S--duality invariant interpolating functions
		
	\end{itemize}
	In what follows we will briefly describe various classes of such interpolating functions which will be important to us and build up to the machinery we extensively use in the later sections.
	
	\subsection{Interpolating functions without S--duality }\label{wthoutS}
	As mentioned before, in the large $N$ limit it is expected that the observables i.e, the anomalous dimensions of twist--two operators in ${\cal N}=4$ SYM are not invariant under the full modular group. We provide the methodology of constructing such interpolating functions which  are consistent with both the weak coupling and the strong coupling expansions without invoking any modular invariance. 
	
	The idea of an interpolating function is to broadly know the perturbative results at strong and weak coupling and match the results from both ends. Suppose we want to interpolate a function $F(g)$ which  has the weak coupling  expansion $F_w^{(N_w )}(g)$ up to $(a+N_w)$th order around $g=0$ 
	\begin{equation}
	F_w^{(N_w )}(g) = g^a \sum_{k=0}^{N_w} w_k g^k \,,
	\end{equation}
	and strong coupling expansion $F_s^{(N_s )}(g)$ up to $(b-N_s)$th order around $g=\infty$ 
	\begin{equation}
	F_s^{(N_s )}(g) = g^b \sum_{k=0}^{N_s} s_k g^{-k} \,.
	\end{equation}
	Then for a consistent interpolation we expect that the Taylor expansion of the interpolating function will match these two expansions around the weak and strong coupling,
	\begin{equation}
	F(g) = F_w^{(N_w )}(g) +\mathcal{O}(g^{a+N_w +1}) = F_s^{(N_s )}(g) +\mathcal{O}(g^{b-N_s -1}) .
	\end{equation}
	In terms of these expansions, 
	we would like to construct smooth interpolating function
	which coincides with the small--$g$ and large--$g$ expansions up to the given orders.

%%%%%%%%%%%%%%%%%%%%%%%%%%%%%%%%%%%%%%%%%%%%%%%
%%%%%%%%%%%%%%%%%%%%%%%%%%%%%%%%%%%%%%%%%%%%%%%
\subsection*{Pad\'e approximant:}

	A simple possibility for an interpolating function is the two--point Pad\'e approximant. 
	Let us construct the Pad\'e approximant $\mathcal{P}_{m,n}(g)$  for the function $F(g)$,
	with $m\leq N_w$ and $n\leq N_s$.
	The formal definition of the Pad\'e approximant interpolating function  for $b-a \in \mathbb{Z}$ is given by
	\begin{equation}
	\mathcal{P}_{m,n}(g)
	= w_0 g^a \frac{ 1 +\sum_{k=1}^p c_k g^k}{1 +\sum_{k=1}^q d_k g^k } ,
	\label{eq:Pade}
	\end{equation}
	where 
	\begin{equation}
	p = \frac{m+n+1 +(b-a)}{2}  ,\qquad q = \frac{m+n+1 -(b-a)}{2}  .
	\end{equation}
	The coefficients $c_k$ and $d_k$ in  \eqref{eq:Pade} can be fixed such that
	power series expansions of the  Pad\'e approximation, $\mathcal{P}_{m,n}(g)$ around $g=0$ and $g=\infty$ agrees with
	the weak coupling expansion and strong coupling  expansions up to the given order in the perturbation  up to $\mathcal{O}(g^{a+m+1})$ and $\mathcal{O}(g^{b-n-1})$ respectively. 
	Thus by  construction the Pad\'e approximant would satisfy $$F(g) - \mathcal{P}_{m,n}(g) =\mathcal{O}(g^{a+m +1}, g^{b-n -1}).$$
	For this construction we  need
	\begin{equation}
	\frac{m+n-1+b-a}{2} \in \mathbb{Z} .
	\label{eq:Pade_constraint}
	\end{equation}
	Pad\'e approximants are widely used to construct non--perturbative answers for perturbatively known functions in diverse areas of physics. However in some situations the denominator in \eqref{eq:Pade} runs into zeroes in physically interesting regions 
	and poles show up in the total function.
	This situation signals limitation of approximation by the Pad\'e, and except for cases where $F(g)$ itself has poles, it would become necessary to investigate the radius of convergence of the Pad\'e approximant.

%%%%%%%%%%%%%%%%%%%%%%%%%%%%%%%%%%%%%%%%%%%%%%%
%%%%%%%%%%%%%%%%%%%%%%%%%%%%%%%%%%%%%%%%%%%%%%%
\subsection*{Fractional Power of Polynomial (FPP):}

	In \cite{Sen:2013oza} the author constructed  a new type of interpolating function, 
	which we refer to as the fractional power of polynomial method (FPP), having the  following skeleton  structure,
	\begin{equation}
	F_{m,n}(g)
	= w_0 g^a \Biggl[ 1 +\sum_{k=1}^m c_k g^k +\sum_{k=0}^n d_k g^{m+n+1 -k} \Biggr]^{\frac{b-a}{m+n+1}} .
	\label{eq:FPP}
	\end{equation}
	We can  determine  the coefficients $c_k$ and $d_k$ in a similar fashion as explained in the case of Pad\'e approximant.
	By construction, the  FPP would  satisfy  $$F(g) - F_{m,n}(g) =\mathcal{O}(g^{a+m +1}, g^{b-n -1}).$$
	
	The FPP does not have constraint on the parameters such as \eqref{eq:Pade_constraint} in the Pad\'e approximant.
	But as in the previous case, these functions are also not free from running into non--analytic regions. One can encounter cases where the polynomial itself can become negative in physically important regions. Consequently, when the power $(b-a)/(m+n+1)$ is not an integer, 
	the FPP takes complex value and signals a breakdown of the approximation.
	
	Although we will use this FPP in its current form only to approximate functions which are not expected to be invariant under duality symmetries,  it is interesting to note that this method has been applied to capture improvements of string perturbation theory via S--duality  \cite{Sen:2013oza,Pius:2013tla}.\footnote{For other curious applications of this method in the 4d $\mathcal{N}=4$ super Yang-Mills theory, the reader may consult \cite{Beem:2013hha,Alday:2013bha}.}

%%%%%%%%%%%%%%%%%%%%%%%%%%%%%%%%%%%%%%%%%%%%%%%
%%%%%%%%%%%%%%%%%%%%%%%%%%%%%%%%%%%%%%%%%%%%%%%
\subsection*{Fractional Power of Rational function (FPR):}

	Based on the earlier two constructions, one can construct a more general class of interpolating function \cite{Honda:2014bza} with  the basic structure as follows
	\begin{equation}
	F_{m,n}^{(\alpha )} (g)
	= w_0 g^a \Biggl[ \frac{ 1 +\sum_{k=1}^p c_k g^k}{1 +\sum_{k=1}^q d_k g^k } \Biggr]^\alpha ,
	\label{eq:FPR}
	\end{equation}
	where the parameters are as following
	\begin{equation}
	p = \frac{1}{2} \left( m+n+1 -\frac{a-b}{\alpha} \right) ,\quad
	q = \frac{1}{2} \left( m+n+1 +\frac{a-b}{\alpha} \right).
	\end{equation}
	We can easily deduce that the Pad\'e and FPP both are special cases of the above interpolating function, by taking the following
	special limits of \eqref{eq:FPR}:
	\begin{itemize}
		\item If $2\ell+1=a-b$ for $a-b \in \mathbb{Z}$ and $m+n$ is even, (\ref{eq:FPR}) reduces to Pad\'e approximant.
		\item If $2\ell+1=m+n+1$ ($2\ell=m+n+1 $) for even (odd) $m+n$, (\ref{eq:FPR}) reduces to   FPP.
	\end{itemize}

	The coefficients $c_k$ and $d_k$ could be determined in a similar way as explained in the  Pad\'e and FPP.
	We refer to this interpolating function as ``fractional power of rational function method" (FPR).
	To construct this interpolating function we require that 
$p,q \in \mathbb{Z}_{\geq 0} \,,$
	which gives us the following constraint on $\alpha$
	\begin{equation}
	\alpha = \left\{ \begin{matrix}
	\frac{a-b}{2\ell +1}  & {\rm for} & m+n:{\rm even} \cr
	\frac{a-b}{2\ell}  & {\rm for} & m+n:{\rm odd}
	\end{matrix} \right. ,\quad
	{\rm with}\ \ell \in\mathbb{Z} .
	\end{equation}
		In analogy to the case of FPP, when the rational function in the parenthesis has
	poles or takes negative values for non--integer $\alpha$,
	we cannot trust approximation of the function $F(g)$ by the FPR.

%%%%%%%%%%%%%%%%%%%%%%%%%%%%%%%%%%%%%%%%%%%%%%%
%%%%%%%%%%%%%%%%%%%%%%%%%%%%%%%%%%%%%%%%%%%%%%%
\subsection{Interpolating functions with S--duality}
\label{intdualitysect}

In the previous section we have introduced a class of interpolating functions which are not inherently invariant under any duality transformations. As mentioned earlier, one can use, for example FPP, to approximate functions invariant under S--duality \cite{Sen:2013oza,Pius:2013tla}. The strategy in these cases is to demand S--duality invariance for the whole function, which in turn gives strong constraints on the coefficients of the polynomials. However, this is not enough to handle functions with all possible non--perturbative effects (like instantons for example) since these contributions cannot be taken care of using polynomials. In our approach, we will be using a better guiding principle to ascertain S--duality in the function.  

 Now to study the non--perturbative properties of the anomalous dimensions at finite $N$ case, it is expected that S--duality would play an important role.
This would   imply  that observables should transform appropriately under the full modular transformations of coupling parameters
\begin{equation}
\mathbf{h}\cdot \tau = \frac{a\tau +b}{c\tau +d} ,\quad {\rm where }\ ad-bc=1,\ a,b,c,d\in\mathbb{Z},
\label{eq:duality}
\end{equation}
which is a combination of $S$-- and $T$--transformations:
\begin{equation}\label{tst}
\mathbf{S}\cdot \tau = -\frac{1}{\tau} ,\quad
\mathbf{T}\cdot \tau = \tau +1 , 
\end{equation} 
where $\tau$ is the complex gauge coupling 
\begin{equation}
\tau = \frac{\theta}{2\pi} +\frac{i}{g} ,\quad 
{\rm with}\ \  g=\frac{g_{\rm YM}^2}{4\pi} .
\end{equation}

In the case of finite $N$, we will mostly focus on the anomalous dimensions of leading--twist operators and construct 
interpolating functions which  satisfy full S--duality invariance.
To construct such a function, the basic philosophy is to choose inherently modular invariant building blocks, instead of polynomials in $g$, as expansion basis. One such natural choice is the real or non--holomorphic Eisenstein series\footnote{
	Note that $s$ can be non--integer and $E_s (\tau )$ has a pole at $s=1$.
	Hence we take $s>1$.
}
\begin{equation}\label{eisenexp}
E_s (\tau ) =\frac{1}{2} \sum_{m,n\in\mathbb{Z}-\{0, 0\}}
\frac{1}{|m+n\tau |^{2s}} ({\rm Im}\tau )^s .
\end{equation}
Since  the Eisenstein series are invariant under the duality transformation \eqref{eq:duality}, by construction interpolating functions constructed out of Eisenstein series as the basic building blocks are invariant under the full S--duality.

The non--holomorphic Eisenstein series $E_s (\tau )$
has the following expansion for small argument (see e.g. section 5.3 of \cite{Klevang})
\begin{eqnarray}
E_s (\tau )
&=& \zeta (2s) ({\rm Im}\tau )^s
+\frac{\sqrt{\pi}\Gamma (s-1/2)}{\Gamma (s)} \zeta (2s-1) ({\rm Im}(\tau ))^{1-s} \\\nn
&&+\frac{4\pi^s}{\Gamma (s)} \sqrt{{\rm Im}(\tau )}
\sum_{k=1}^\infty \sigma_{1-2s}(k) k^{s-\frac{1}{2}} 
K_{s-\frac{1}{2}} \left( 2\pi k {\rm Im}(\tau ) \right)  \cos{\left( 2\pi k {\rm Re}(\tau )\right)} ,
\label{eq:eisenN}
\end{eqnarray}
where $\sigma_s (k)$ is the divisor function defined by\footnote{The sum of positive divisors function $\sigma_s (k)$, for a real or complex number $s$, is defined as the sum of the $s^{th}$ powers of the positive divisors of $k$, with $d|k$ is shorthand for d divides k. }
$\sigma_s (k) =\sum_{d|k} d^s.$
In terms of $(g,\theta )$,
it is written as,
\begin{eqnarray}\label{eisendef}
E_s (\tau )
&=& \zeta (2s) g^{-s}
+\frac{\sqrt{\pi}\Gamma (s-1/2)}{\Gamma (s)} \zeta (2s-1) g^{s-1} \\\nn
&&+\frac{4\pi^s}{\Gamma (s)} g^{-\frac{1}{2}}
\sum_{k=1}^\infty \sigma_{1-2s}(k) k^{s-\frac{1}{2}} 
K_{s-\frac{1}{2}} \left( \frac{2\pi k}{g} \right)  
\cos{\left( k\theta\right)} .
\end{eqnarray}
Here the $\Gamma$ is the usual gamma function and $K$ is the modified bessel function. Below we present a brief account of the interpolating function methods developed using such Eisenstein series in \cite{Alday:2013bha} and \cite{Chowdhury:2016hny}.

We note an important feature of the Eisenstein series $E_s(\tau)$, it contains both a perturbative and a non--perturbative part. The non--perturbative part of the Eisenstein series contains power of $q=e^{2\pi i \tau }$, which would play a crucial role in reproducing the correct instanton contributions, as we will use later for the cusp anomalous dimension.

%%%%%%%%%%%%%%%%%%%%%%%%%%%%%%%%%%%%%%%%%%%%%%%
%%%%%%%%%%%%%%%%%%%%%%%%%%%%%%%%%%%%%%%%%%%%%%%
\subsection*{FPP--like interpolating function involving Eisenstein series:}

In \cite{Alday:2013bha}  
the following type of interpolating function has been constructed 
\begin{equation}
\bar{F}_m^{(s)} (\tau ) = \left( \sum_{k=1}^m c_k E_{s+k} (\tau ) \right)^{-\frac{1}{s+m}} ,
\label{eq:AB}
\end{equation}
where the coefficient $c_k$'s are determined such that
expansion of $\bar{F}_m^{(s)}$ around   $g=0$ agrees with the weak coupling expansion of the anomalous dimension, $F (\tau )$ up to $\mathcal{O}(g^{m+1} )$. One could see that the above function has structural similarities with FPP like interpolating functions \eqref{eq:FPP}, where instead of a polynomial in $g$, the Eisenstein series has been used. 
Thus an appropriate choice of $c_k$ correctly gives the weak coupling expansion of $F (\tau )$. However, since this function is not actually constrained by strong coupling expansion, it is only natural to consider generalisations of it where strong coupling data has significant role to play.

%%%%%%%%%%%%%%%%%%%%%%%%%%%%%%%%%%%%%%%%%%%%%%%
%%%%%%%%%%%%%%%%%%%%%%%%%%%%%%%%%%%%%%%%%%%%%%%
\subsection*{FPR--like duality invariant interpolating function:}
\label{fprsdual:}

 In section \ref{cinfnc} we will construct  FPR--like duality invariant interpolating functions to study the cusp anomalous dimensions and the anomalous dimensions for finite spin operators at finite $N$.
Here, we give a brief methodology of building such generalised function and will provide a more detailed construction procedures in the later section.
	 The FPR--like duality invariant interpolating functions constructed in \cite{Chowdhury:2016hny} has the following structure:
\begin{equation}
\tilde{F}_m^{(s,\alpha )} (\tau ) 
= \Biggl[ \frac{\sum_{k=1}^p c_k E_{s+k} (\tau )}{\sum_{k=1}^q d_k E_{s+k} (\tau )} \Biggr]^\alpha ,
\label{eq:FPR_wo_gravity}
\end{equation}
where we can  determine the coefficients $c_k$ and $d_k$ ({\it except} $d_1$) such that
expansion of $\tilde{F}_m^{(s,\alpha )}$ around $g=0$ agrees with $F (\tau )$ up to $\mathcal{O}(g^{m+1} )$.\footnote{
	Note that $m$ should be $m\geq 2$
	since we need two coefficients at least 
	for this interpolating function.
}
Matching at $\mathcal{O}(g)$ leads to
\begin{equation}
\alpha (-p +q ) =1,\quad
\left( \frac{c_{p}\zeta (2s+2p)}{d_{q}\zeta (2s+2q)} \right)^\alpha = w_1 .
\end{equation}
Now,  the interpolating function is invariant 
under the scaling $c_k ,d_k \rightarrow \lambda c_k ,\lambda d_k $,
so without any loss of generality we can take
\begin{equation}
d_{q} =1 .
\end{equation}
Matching at other orders leads to the constraint
$p+q-1 = m ~,$and hence we find the relation between the parameters 
\begin{equation}
p = \frac{1}{2}\left( m+1-\frac{1}{\alpha }\right) ,\quad
q = \frac{1}{2}\left( m+1+\frac{1}{\alpha }\right) .
\end{equation}
We also require the condition $p,q \in \mathbb{Z}_{\geq 1}$,
which implies
\begin{equation}
\alpha = \left\{ \begin{matrix}
\frac{1}{2\ell }  & {\rm for} & m:{\rm odd} \cr
\frac{1}{2\ell +1}  & {\rm for} & m:{\rm even} \end{matrix} \right. ,\quad
{\rm with}\ \ell \in\mathbb{Z} .
\end{equation}

The above interpolating function can be further constrained if we impose the matching of coefficients from either the strong coupling or the finite $N$ results. Let us discuss the constraints imposed by strong coupling results with $\mathcal{O}(\frac{1}{N^2})$ corrections on the interpolating functions. 
We start with 't Hooft expansion (i.e. in $\lambda$, see equation (\ref{eq:thooft})) of the interpolating function:\footnote{In principle the supergravity results can be a function of the spin-$j$ of the operator and the genus correction \cite{Goncalves:2014ffa,Alday:2018pdi}. }
\begin{equation}
\tilde{F}_m^{(s,\alpha )} \left( \frac{iN}{\lambda} \right) 
= f_0 (\lambda,j ) +\frac{f_2 (\lambda,j )}{N^2} +\frac{f_4 (\lambda,j )}{N^4} +\cdots .
\end{equation}
Then we determine the yet unknown coefficient $d_1$  to satisfy
\begin{equation}
\lim_{\lambda\rightarrow\infty} \left( f_0 (\lambda,j ) +\frac{f_2 (\lambda,j )}{N^2}+\frac{f_4 (\lambda ,j)}{N^4} \right)
= \gamma_M^{\rm SUGRA} (N) ,
\end{equation}
where $\gamma_M^{\rm SUGRA}$ is the result in the supergravity limit
given by \eqref{eq:result_gravity}.
Imposing matching of other orders leads us to
$
p+q-2 = m ,
$
and therefore we get
\begin{equation}
p = \frac{1}{2}\left( m+2-\frac{1}{\alpha }\right) ,\quad
q = \frac{1}{2}\left( m+2+\frac{1}{\alpha }\right) .
\end{equation}
We also require $p,q \in \mathbb{Z}_{\geq 1}$,
which constrains $\alpha$ as
\begin{equation}
\alpha = \left\{ \begin{matrix}
\frac{1}{2\ell +1 }  & {\rm for} & m:{\rm odd} \cr
\frac{1}{2\ell }  & {\rm for} & m:{\rm even} \end{matrix} \right. ,\quad
{\rm with}\ \ell \in\mathbb{Z} .
\end{equation}
The constraints coming from the supergravity, where the data is of the same form as above, will be important to construct the interpolating function for the finite spin twist--two operators. 

Note that  there are clearly three parameters $(m,s,\alpha )$ driving the interpolating function. There could be infinite choices for this set of parameters, leading to infinite number of possible interpolating functions.  This {\it{``landscape problem of interpolating functions"}} was studied earlier in 
\cite{Honda:2014bza}. It is a priori unclear which set of values of $(m,s,\alpha )$ would give us the 
best approximation. We will briefly discuss the procedure to choose optimal values of  $(m,s,\alpha )$ in our construction, see also \cite{Honda:2014bza,Chowdhury:2016hny}.
\begin{itemize}
	\item {\it{Choice of $m$}}.\\ By construction the interpolating function should reproduce the correct weak coupling expansion up to the given $m$--th loop order. The best choice for $m$ depends on the details of the weak coupling expansion. One of the important criteria would be the convergence property of the weak coupling expansion. Suppose the weak coupling expansion is convergent series then we can take $m$ as large as possible otherwise we have to judiciously select $m$.
	\item  {\it{Choice of $s$}}.\\
	Since the weak coupling expansion of the function of interest only contains positive integer powers of $g$, however  in principle the interpolating function  \eqref{eq:FPR_wo_gravity} can contain fractional power of $g$ for any arbitrary value of $s$. In order to guarantee
	absence of such fractional powers, we should take\footnote{Note that $s$ can be non-integer and $E_s (\tau )$ has a pole at $s=1$	hence we take $s>1$.}
	$
	2s\in\mathbb{Z} .
	$
	As discussed in \cite{Chowdhury:2016hny}, most of the construction has little dependence on $s$ and thus we get an infinite class of interpolating functions with extremely close numerical values but different structures.

	\item {\it{Choice of $\alpha$}}.\\
	The parameter $\alpha$ determines the type of branch cuts of the interpolating functions. Hence a correct choice of $\alpha$ would be related to the analytic properties of the interpolating functions.
	\end{itemize}
We will provide more details of the choice of such parameters in section \ref{sec:finiteNcase} where we construct such functions explicitly. 

Finally, let us mention an issue of matching strong coupling expansion which would require a further generalization of the above function. An important feature of the above function is that in the limit $\lambda\gg 1$ (planar limit), since the non--perturbative part of the Eisenstein series is suppressed  \eqref{eisendef}, the coefficients $c_k$ and $d_k$ are determined only in terms of the perturbative part  i.e. $\mathcal{O}(g^{-s})$ and $\mathcal{O}(g^{s-1})$ terms. Furthermore, in the regime discussed in \cite{Chowdhury:2016hny}, at large $s$ only the $\mathcal{O}(g^{-s})$ part is relevant and in the planar limit the total function have a schematic form\footnote{We would expect it to be $\mathcal{O}(N^{-(s+k)})$ but since we normalize $d_{s+q}=1$, we multiply each coefficient by $N^{s+q}$.}
\begin{equation}
\label{eq:power-issue1}
\Biggl[ \frac{\sum_{k=1}^p \bar{c}_k \lambda^{-(s+k)} }
{\sum_{k=1}^q \bar{d}_k \lambda^{-(s+k)} } \Biggr]^\alpha \quad  \text{with}  \; c_k \sim d_k \sim \mathcal{O}(N^{q-k} ) \,,
\end{equation}
where we have
\begin{equation}
\label{eq:power-issue2}
\bar{c}_k = \lim_{N\rightarrow\infty} \zeta (2s+2k)N^{k-q} c_k  \,,\quad
\bar{d}_k = \lim_{N\rightarrow\infty} \zeta (2s+2k)N^{k-q} d_k  \,.
\end{equation}
In the large $\lambda$ limit, $\bar{c}_k$ and $\bar{d}_k $ are essentially 
$\mathcal{O}(1) $. Furthermore, the function in--principle can't generate any fractional powers in $\lambda$ in the strong coupling limit. In order to take into account any fractional powers of $\lambda$ in the strong coupling limit we have to further generalize this function, which we defer for a details discussion in section \ref{cinfnc}.

%%%%%%%%%%%%%%%%%%%%%%%%%%%%%%%%%%%%%%%%%%%%%%%%%%
%%%%%%%%%%%%%%%%%%%%%%%%%%%%%%%%%%%%%%%%%%%%%%%%%%
\section{Data on anomalous dimensions}
\label{sec:dataofAD}

There has been many well known investigations on  twist--two operators in the relevant literature. In this section, we will very briefly review some aspects of these investigations and mention the main results of such works, which will serve as effective input data in our work. 

We summarize our convention for the coupling constants here:
\begin{eqnarray}
\label{eq:thooft}
\frac{1}{2\pi\alpha\prime}= \frac{\sqrt{g_{YM}^2 N}}{2\pi}=\frac{\sqrt{\lambda}}{2\pi}=2\tilde g \,,
\end{eqnarray}
where the $g_{YM}$ is the Yang-Mills coupling constant and $\lambda$ is the `t Hooft coupling. 
The `t Hooft large $N$ limit is taken by $N\to \infty$, keeping $\lambda$ or $\tilde g$ constant.
For the study of finite $N$ case, we also introduce 
\begin{equation}
g=\frac{g^2_{YM}}{4\pi}=\frac{4\pi\tilde g^2}{N} \,.
\end{equation}

%%%%%%%%%%%%%%%%%%%%%%%%%%%%%%%%%%%%%%%%%%%%%%%%%%
%%%%%%%%%%%%%%%%%%%%%%%%%%%%%%%%%%%%%%%%%%%%%%%%%%
\subsection{Results on cusp anomalous dimension}
\label{data1}

Cusp anomalous dimension is an important observable that governs the universal scaling behavior of various gauge invariant quantities. 
As we already mentioned, it governs the large spin scaling behavior of twist--two anomalous dimensions.
By definition it is also the anomalous dimension of Wilson loop with a light--like cusp singularity \cite{Polyakov:1980ca, Korchemsky:1985xj}.
Furthermore, it provides the leading infrared divergences of on--shell amplitudes and is an essential ingredient in constructing amplitudes, such as in \cite{Bern:2005iz}.
In AdS/CFT correspondence, it is related to dual description in terms of spinning strings \cite{Gubser:2002tv} or cusped minimal surface \cite{Kruczenski:2002fb} in the AdS background.

In the planar limit, in principle we can find the weak and strong coupling expansions up to any loop order for cusp anomalous dimension using the BES formula \cite{Beisert:2006ez}. 
This is an integral equation derived from all--loop Bethe Ansatz equation with a mathematically complicated kernel structure. 
At weak coupling, the planar expansion for cusp anomalous dimension has been obtained up to four loops from rigorous perturbative analysis \cite{Bern:2006ew,Cachazo:2006az,Henn:2013wfa}. From semiclassical computations in string theory the strong coupling expansion has been explicitly computed up to two loops from the analysis of quantum string sigma model in $AdS$ \cite{Roiban:2007dq, Roiban:2007ju}.

On the other hand, non--planar corrections to quantities like cusp anomalous dimension is so far hard to compute within a framework like $AdS/CFT$, where these corrections correspond to string loop corrections. Also, the power of integrability fails here since its role beyond planar limit is yet to be fully uncovered. 
Recently, progress in computing the non--planar corrections to cusp anomalous dimension has been made  via a numerical calculation of Sudakov form factor \cite{Boels:2017skl, Boels:2017ftb}, where the non--planar part enters into the result first in the fourth loop order in weak coupling. 

Let us summarize the result of cusp anomalous dimension up to four loops:
\begin{eqnarray}
\label{cuspdef1}
\Gamma_{{\rm cusp,w}} = 4\tilde g^2-\frac{4}{3}\pi^2 \tilde g^4+\frac{44}{45}\pi^4 \tilde g^6 + \left( -\frac{292}{315}\pi^6 - 32\zeta(3)^2+\frac{\Gamma^{\rm np}}{N^2}\right)\tilde g^8+\mathcal{O}(\tilde g^{10})
\end{eqnarray}
where the non--planar four--loop cusp anomalous dimension, $\Gamma^{\rm np}$, is given by\footnote{We use the central value of the non--planar result \cite{Boels:2017skl}. Note that the definition of cusp anomalous dimension is different from that in \cite{Boels:2017skl} by an overall factor $2$.}
\begin{equation}
\label{nonplan}
\Gamma^{\rm np} \sim -2400.
\end{equation}
At strong coupling, we quote the result \cite{Roiban:2007dq}
\begin{eqnarray}
\label{cuspst}
\Gamma_{{\rm cusp,s}}= 2\tilde g-\frac{3\log 2}{2\pi}+\mathcal{O}(\frac{1}{\tilde g}) \,.
\end{eqnarray}

As mentioned before, higher order data can be extracted from the BES equation recursively. For example, in the next orders of weak and strong coupling expansions, one can read off, 
\begin{eqnarray}\label{higherloop}
\Gamma_{{\rm cusp,w}}^{(5)} &=&16 \left( \frac{887}{14175}\pi^8+\frac{4}{3}\pi^2\zeta(3)^2+40\zeta(3)\zeta(5)   \right) =10601.9 \,, \nn \\ 
\Gamma_{{\rm cusp,s}}^{(3)} &=& \frac{\mathbf K}{8\pi^2} =-0.0116\, ,
\end{eqnarray}
where $\mathbf K$ is the Catalan constant.
This allows us in principle to predict the planar cusp anomalous dimension perturbatively to arbitrary loop order \cite{Basso:2007wd}.

One important motivation of this paper is to consider truly non--perturbative corrections to the anomalous dimensions. In a recent work \cite{Korchemsky:2017ttd}, leading instanton contribution to the light--like cusp anomalous dimension 
has been computed. We briefly review this below. In general quantum corrections to four point functions of half--BPS operators have the following form in the weak coupling limit,
\begin{equation}
\mathcal{G}(u,v) = \Phi_0(u, v, g^2)+\sum_{n\geq1}\left( e^{2\pi i n\tau}+e^{-2\pi i n \bar\tau }\right)\Phi_n(u, v, g^2),
\end{equation}
where the first term is the perturbative part and the other term is the non--perturbative correction and $u, v$ are the two cross ratios.
In  \cite{Korchemsky:2017ttd} it is mentioned that in large spin limit the leading instanton contribution scales as 
\begin{equation}\label{weakinst}
\gamma(j) \sim g_{YM}^8 e^{2\pi i \tau}\log(j) \,, \quad j \gg1.
\end{equation}
This result is valid for the SU(2) gauge group. Remember that in the light--like limit the cross ratios $u,v\to 0$ and this reproduces the contribution for the cusp anomalous dimension, as have been quoted above. For the one instanton correction, the contribution has the following form 

\begin{equation}
\label{inss2}
\Gamma_{{\rm cusp, inst}}=-\frac{4}{15}\left( \frac{g_{YM}^2}{4\pi^2} \right)^4\left( e^{2\pi i \tau}+e^{-2\pi i  \bar\tau }\right).
\end{equation}
Now we may generalize these instanton corrections to the $SU(N)$ group as follows.
The non--perturbative  correction appears in the non--planar sector so there is a explicit factor of $1/2^2$ for the $SU(2)$ case. We can rewrite \eqref{inss2}\footnote{Here we have assumed that the leading instanton corrections starts at $1/N^2$ order\cite{Alday:2016tll}.} as follows
\begin{equation}
\label{inssex2}
\Gamma_{{\rm cusp, inst}}=-\frac{1}{2^2}\frac{16}{15}\left( \frac{g_{YM}^2}{4\pi^2} \right)^4\left( e^{2\pi i \tau}+e^{-2\pi i  \bar\tau }\right) .
\end{equation}
In the large $N$ limit, the result for the leading instanton correction gets multiplied by a $N$ dependent factor \cite{Korchemsky:2017ttd}  due to contribution from all bosonic and fermionic modes arising from embedding the $SU(2)$ instanton into $SU(N)$. The appearance of such a factor has been worked out in details in \cite{Dorey:1998xe} and in our case it has a form

\begin{equation}
c(N)= \frac{(2N-2)!}{2^{2N-3}(N-1)!(N-2)!}.
\end{equation}
For large $N$, one can see that the total factor before the instanton contribution then has a factor $\mathcal{O}(N^{-3/2})$.

%%%%%%%%%%%%%%%%%%%%%%%%%%%%%%%%%%%%%%%%%%%%%%%%%%
%%%%%%%%%%%%%%%%%%%%%%%%%%%%%%%%%%%%%%%%%%%%%%%%%%
\subsection{Data for finite spin anomalous dimensions}

Next, we consider the anomalous dimension of twist--two operator with generic spins,
$\mathcal{O}_{j} = \Tr(\phi D^j \phi)$.
The scaling dimension of these operators can be written schematically as,
\be
\Delta(\tilde g, j, N)  = 2 + j +\gamma(\tilde g, j, N),
\ee
where $\gamma(\tilde g)$ is the anomalous dimension.
Using conventional methods, one can study these functions only up to first few loops in the strong/weak coupling regimes. With integrability techniques, employing the TBA or Y--system \cite{Alday:2010vh}, one gets analytically very complicated integral equations. With the advent of techniques associated with Quantum Spectral Curve (QSC) \cite{Gromov:2013pga,Gromov:2014bva,Gromov:2014caa,Gromov:2017blm}, in principle one can extract such data at any value of the spin and coupling constant. However, such an explicit computation in particular for generic spin dependence still seems to be absent in the present literature. 
As mentioned earlier, the results for anomalous dimensions in $\mathcal{N}=4$ SYM can be extracted from QCD results by isolating the maximally transcendental part perturbatively at different orders in the coupling constant \cite{Kotikov:2002ab,Kotikov:2001sc, Kotikov:2007cy}. 

The internal symmetries of $\mathcal{N}=4$ SYM makes it evident that the basic building block of such anomalous dimension are sums of the form $\sum_{i}\frac{1}{j^i}$, where $i$ is the level of transcendentality, indicating that these anomalous dimensions are polynomials in Riemann Zeta value or it multi--index generalisations. The basis for the results is formed from these $harmonic~sums$ defined as follows,
%\be
%S_a(j) = \sum_{m=1}^{j} \frac{(sgn(a)^m)}{m^{|a|}},~~~S_{a_1, a_2...,a_n}(j)=\sum_{m=1}^{j} \frac{(sgn(a_1)^m)}{m^{|a_1|}}S_{a_2,...,a_n}(m).
%\ee
%Below we give an explicit form of the harmonic numbers  that will be used in defining the anomalous dimension and used in our computation.
\begin{eqnarray}
&&\hspace*{-1cm} S_{a}(j)\ =\ \sum^j_{m=1} \frac{1}{m^a},
\ \ S_{a,b,c,\cdots}(j)~=~ \sum^j_{m=1}
\frac{1}{m^a}\, S_{b,c,\cdots}(m),  \label{ha1} \nonumber \\
&&\hspace*{-1cm} S_{-a}(j)~=~ \sum^j_{m=1} \frac{(-1)^m}{m^a},~~
S_{-a,b,c,\cdots}(j)~=~ \sum^j_{m=1} \frac{(-1)^m}{m^a}\,
S_{b,c,\cdots}(m),  \nonumber \\
&&\hspace*{-1cm} \overline S_{-a,b,c,\cdots}(j) ~=~ (-1)^j \, S_{-a,b,c,...}(j)
+ S_{-a,b,c,\cdots}(\infty) \, \Bigl( 1-(-1)^j \Bigr).  \label{ha3}
\end{eqnarray}
Due to this remarkable structure, we can define the anomalous dimension for finite spin twist--two operators using a basis of \emph{harmonic sums}. 
We can now write the anomalous dimension up to three loops as 
\begin{equation}\label{finspindata}
\gamma(j)\equiv \tilde g^2 \gamma^{(1)}(j)+\tilde g^4
\gamma^{(2)}(j) +\tilde g^6 \gamma^{(3)}(j) + ...  \,,
\end{equation}
where $\tilde g$ has been defined as before. This $\gamma$ is the anomalous dimension defined at finite values of $j$.  \\
Using the above assumptions, the leading order (LO) and the next--to--leading order (NLO) anomalous dimensions for twist--two operators were found in \cite{Kotikov:2000pm, Kotikov:2003fb}. The three loop expressions were obtained in \cite{Kotikov:2004er} by extracting the most complicated contributions from the three loop non--singlet anomalous dimensions in QCD \cite{Moch:2004pa}. To keep our considerations simpler, we will use the expression for first few loop data as follows,\footnote{Our conventions are the same as of \cite{Kotikov:2004er}. Higher loop data for twist two operators have been explored in \cite{Lukowski:2009ce,Velizhanin:2010cm,Marboe:2014sya,Marboe:2016igj}. We note that the non--planar corrections were computed for spin 2, 4, 6 and 8 \cite{Velizhanin:2009gv,Velizhanin:2010ey,Velizhanin:2014zla}, but for generic spin the results are still not available.}
\begin{eqnarray}
\label{finspinres}
\, \gamma^{(1)}(j+2) &=&  4 S_1,  \label{uni1.1} \\
\, \gamma^{(2)}(j+2) &=& -8\left[\Bigl(S_{3} +
\overline S_{-3} \Bigr) - 2\, \overline S_{-2,1} + 2\,S_1\Bigl(S_{2} +
\overline S_{-2} \Bigr)\right],  \label{uni1.2} \\
\, \gamma^{(3)}(j+2) &=& -32 \Bigg[2\, \overline S_{-3}\,S_2 -S_5 -
2\, \overline S_{-2}\,S_3 - 3\, \overline S_{-5}
+24\, \overline S_{-2,1,1,1} 
\label{uni1.5}\\
&&+ 6\biggl( \overline S_{-4,1} +  \overline S_{-3,2} +
\overline S_{-2,3}\biggr)
- 12\biggl( \overline S_{-3,1,1} +  \overline S_{-2,1,2} +
\overline S_{-2,2,1}\biggr)\nonumber \\
&&   -
\biggl(S_2 + 2\,S_1^2\biggr) \biggl( 3 \, \overline S_{-3} + S_3
- 2\,  \overline S_{-2,1}\biggr)
- S_1\biggl(8\, \overline S_{-4} +  \overline S_{-2}^2\nonumber \\
&&  + 4\,S_2\, \overline S_{-2} +
2\,S_2^2 + 3\,S_4 - 12\,  \overline S_{-3,1} - 10\,  \overline S_{-2,2}
+ 16\,  \overline S_{-2,1,1}\biggr)\Bigg] \,. \nonumber
\end{eqnarray}
The $j\to\infty$ results are important here, especially for matching to the cusp anomalous dimension. From each of these terms we will get a $\log(j)$ contribution in the $j\to \infty$ limit, the coefficients of which can be exactly matched to the ones written in equation (\ref{cuspdef1}). 
So, in the weak coupling case, the finite spin data smoothly connects to the infinite spin case and gives rise to the \textit{log} behaviour.

On the other hand, we are also interested in the coefficients of the strong coupling expansion of the anomalous dimension at finite spin values. In planar limit, this can be (partially) computed from AdS/CFT prescription and in principle via integrability. Let us call the strong coupling anomalous dimension $\mathcal{G}(\tilde g)$,
\begin{equation}
\label{deltaki}
\mathcal{G}(\tilde g) = \mathcal{G}^{(1)} (4\pi\tilde g)^\frac{1}{2} + \mathcal{G}^{(2)} (4\pi\tilde g)^{-\frac{1}{2}}  + \mathcal{G}^{(3)}  (4\pi\tilde g)^{-\frac{3}{2}} + \mathcal{G}^{(4)}  (4\pi\tilde g)^{-\frac{5}{2}} 
+\dots .
\end{equation}
For the twist--$\ell$ operators in the $SL(2)$ sector, there are analytical predictions for the first four coefficients of (\ref{deltaki}).
The coefficients for twist--two operators can be written as functions of $j$ and is known to take the following form via Quantum Spectral Curve calculation \cite{Gromov:2014bva}: 
\begin{equation}\label{D01}
\mathcal{G}^{(1)}=\sqrt{2 \, j}, \qquad \mathcal{G}^{(2)}=\frac{\, 8+j(3 \, j-2)}{4 \, \sqrt{2 j}},  
\end{equation}
\begin{equation}\label{D2}
\mathcal{G}^{(3)} = \frac{-21\,j^4 +(24-96\,\zeta_3) j^3+68 j^2+32 j -64}{64 \sqrt{2}\,j^{3/2}},
\end{equation}
\begin{eqnarray}
\mathcal{G}^{(4)} &=& \frac{187\,j^6 + 6\,(208\,\zeta_3 + 160\,\zeta_5-43)\,j^5 +\left(-584 - 4\,(336\,\zeta_3-41)\right)j^4 }{512 \sqrt{2}\,j^{5/2}} + \nonumber \\
& & + \frac{\left(128\,(6\,\zeta_3+7)\,-88\right)j^3 -288 j^2 - 384 j + 512}{512 \sqrt{2}\,j^{5/2}}. \label{D3}
\end{eqnarray}
The first two coefficients in (\ref{D01}) can be determined either from Basso's slope function \cite{Basso:2011rs} or from
semi--classical computations in string theory \cite{Gromov:2011de,Roiban:2011fe,Vallilo:2011fj}. The next two coefficients were determined by matching
the $O(j^2)$ term of the small spin expansion with classical and semi--classical results \cite{Gromov:2014bva}.

 In the small spin regime the data can be compared with the dispersion relation of a small--spin circular string moving near the centre of AdS. The other regime of the strong coupling data, i.e. at $j\to \infty$ limit is given by the folded string dispersion relation. It is important to note here that unlike weak coupling, it is not possible to reproduce the $\log(j)$ behaviour from the strong coupling data \eqref{D01}-\eqref{D3} at large spin limit. This would indicate a non-trivial ``phase-transition'' --like physics in large spin regime when we go to the dual AdS string picture. We will come back to this picture in section \ref{transition}.

At finite $N$, another important ingredient that will enter our calculations at strong coupling is the double--trace twist--four operator from operator mixing effect.
At planar large 't Hooft coupling, the anomalous dimension of  
the twist--two operator $\mathcal{O}_M$ \eqref{deltaki}
grows without any bound \cite{Gubser:1998bc},
and mixing effect with double--trace operators kicks in. 
We will discuss more in section \ref{sec:finiteNcase} when constructing interpolating functions at finite $N$.
Here, let us introduce the double--trace operators 
which has a schematic form
\begin{equation}\label{Dtrace}
{\rm tr}(\phi^{(i} \phi^{j)} ) D^M {\rm tr}(\phi^{(i} \phi^{j)} ) ,
\end{equation}
where ${\rm tr}(\phi^{(i} \phi^{j)})$ is the  
symmetric traceless part of ${\rm tr}(\phi^{i} \phi^{j})$
and a chiral primary operator belonging to $\mathbf{20'}$ representation of $SU(4)_R$.
There are known results for  the anomalous dimension of such operators from  supergravity 
\cite{Dolan:2001tt,DHoker:1999mic,Arutyunov:2000ku} computation in $AdS_5\times S^5$.
Below we note down 
the anomalous dimension for spin $j$ double trace operator to ${\cal{O}}(1/N^4)$ order as considered in  \cite{Alday:2017xua} and \cite{Aprile:2017bgs}
\begin{equation}\begin{split}\label{eq:result_gravity}
\gamma(j)=&2-\frac{96}{(j+1) (j+6) N^2}\\&-\frac{96 \left(j^6+21 j^5+123 j^4+103 j^3+2348 j^2+19196 j+13488\right)}{(j-1) (j+1)^3 (j+6)^3 (j+8) N^4}.
\end{split}
\end{equation} 
Here we have defined anomalous dimension of the double trace operator as $\gamma(j)= \Delta_s-(j+2),$ where $\Delta_s$ is the  scaling dimension of the double trace operator.
%%%%%%%%%%%%%%%%%%%%%%%%%%%%%%%%%%%%%%%%%%%%%%%%%%
%%%%%%%%%%%%%%%%%%%%%%%%%%%%%%%%%%%%%%%%%%%%%%%%%%
\section{The $planar$ case: Finite spin twist--two operators}
\label{sec:planarcase}
In this section, we would try to understand the physics of anomalous dimension of  ``short'' (small spin) and ``long'' (large spin) twist--two operators from the interpolating function point of view in the large $N$ limit.
In what follows, we would try to address this issue using simple Pad\'{e} type approximants as toy models. Later we will propose betterment over such simple approximants and try to see whether this improves the interpolation. As we will see, the underlying proposed physical picture is fairly independent of the construction itself. 
The 't Hooft coupling is denoted by ${\tilde g}$ or equivalently $\lambda$, and we again recall our convention: $${\tilde g} = \frac{\sqrt{\lambda}}{4\pi} = \frac{\sqrt{g_{YM}^2 N}}{4\pi} = \frac{1}{4\pi\alpha\prime}.$$

%%%%%%%%%%%%%%%%%%%%%%%%%%%%%%%%%%%%%%%%%%%%%%%%%%
%%%%%%%%%%%%%%%%%%%%%%%%%%%%%%%%%%%%%%%%%%%%%%%%%%
\subsection{Results from Pad\'{e} approximant}
\label{padeconst}

In the last subsection, we discussed various regimes of available data for  anomalous dimensions of twist--two operators. Let us list them in one place in the following simplistic  way,
\begin{enumerate}[label=\Alph*.]
\item \textbf{Small $\tilde{g}$ and small $j$ (can also connect to large $j$):} Data given in terms of Harmonic Sums as function of $j$ as in (\ref{finspinres})-\eqref{uni1.5}.

\item \textbf{Large $\tilde{g}$ and small $j$:} Combined data from semi--classical string computations and Quantum Spectral Curve (\ref{D01})-\eqref{D3}. 

\item \textbf{Small $\tilde{g}$ and large $j$:} Data given by cusp anomalous dimension (coefficient of $\log(j)$) in the weak coupling expansion i.e. planar version of (\ref{cuspdef1}).

\item \textbf{Large $\tilde{g}$ and large $j$:} Data given by cusp anomalous dimension (coefficient of $\log(j)$) in the strong coupling expansion (\ref{cuspst}), also from AdS/CFT via folded strings.

\item \textbf{Large $\tilde{g}$ and small $j$ (which can also connect to large $j$):} Data from results of string computations which take care of small spins, i.e. ``short strings''.

\end{enumerate}
%%%%%%%%%%%%%%%%%%%%%%%%%%
\begin{figure}[!tb]
		\centering
		\includegraphics[width=8cm]{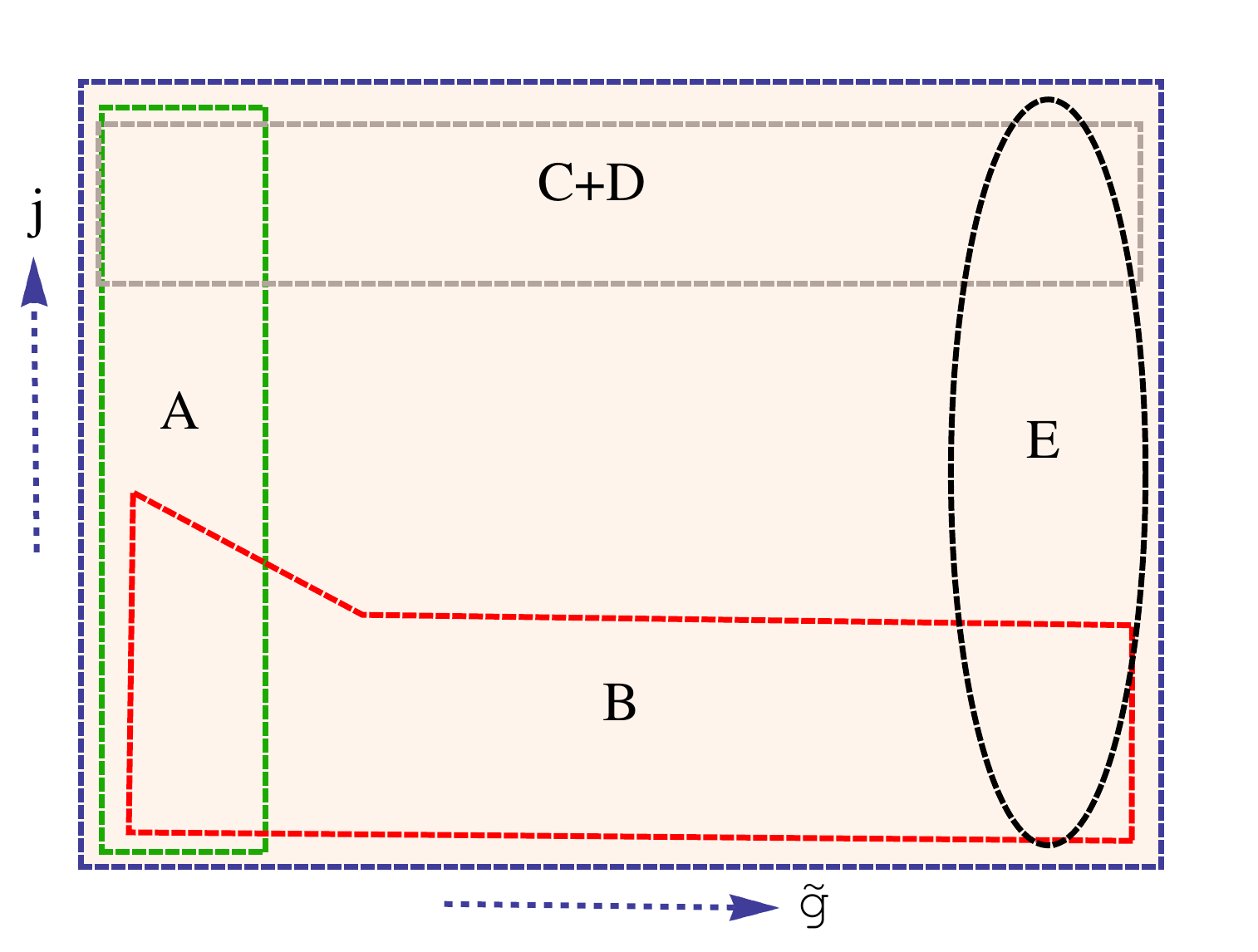}
		\caption{A visualisation of the data in the $\tilde g$--$j$ plane as presented above.}
		\label{viz}
\end{figure}
%%%%%%%%%%%%%%%%%%%%%%%%%%

In short, we now know input data for four corners in a $\tilde g$--$j$ parameter space but we don't know the expressions for anomalous dimensions in the interim regime of parameters. In principle, one might be able to construct a master interpolating function along both $g$ and $j$ direction to extract data at any point in this space, but that turns out to be a 
difficult job. In what follows, we would try to take a different route. By constructing two interpolating functions in the coupling constant, one for the 
small $j$ region and one for the large $j$ region, we would try to piecewise fit the total parameter space.  It is quite evident that the intersection line for such two functions would signify a change of physical regimes.

\begin{figure}[!tb]
\centering
\begin{minipage}{.5\textwidth}
\centering
  \includegraphics[width=0.93\linewidth]{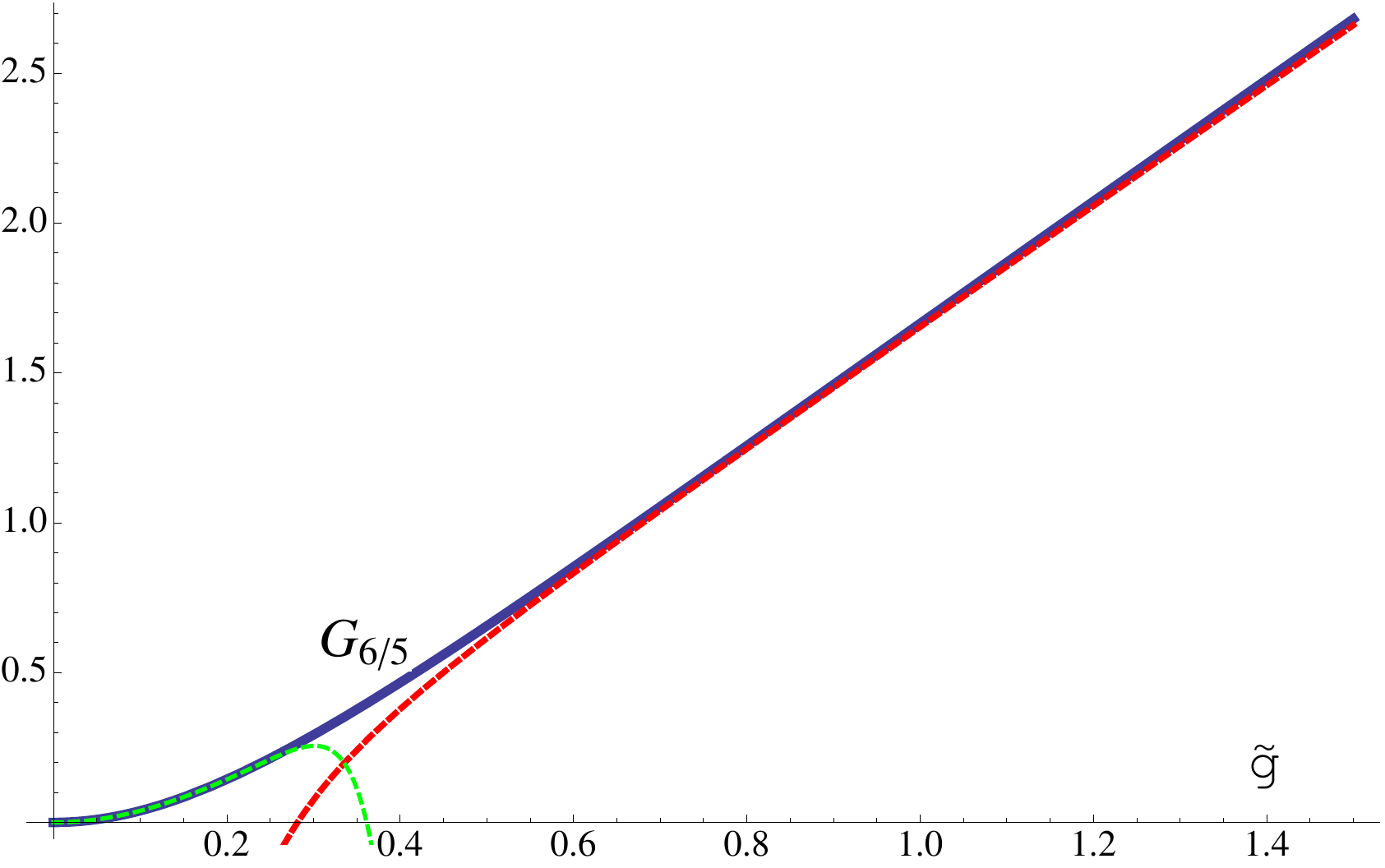}\label{intp}
  \caption*{(a)}
\end{minipage}%%%%%%
~~
\begin{minipage}{.5\textwidth}
  \centering
  \includegraphics[width=.93\linewidth]{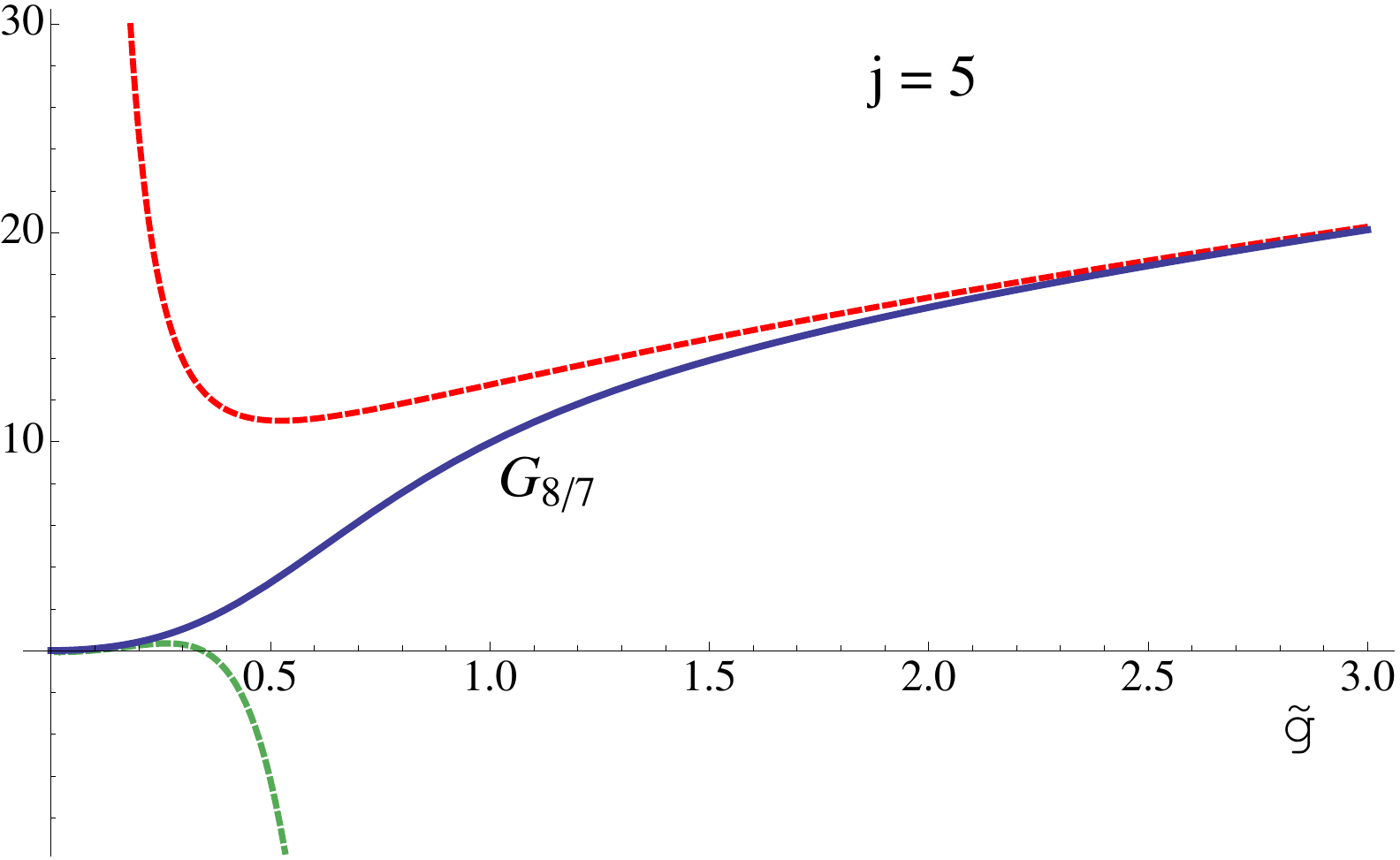}
    \caption*{(b)}
\end{minipage}
%%%%%%%%%%%%%%%%%%%%%%%%%%

\caption{(a)The weak coupling expansion for the cusp is given by the Green dashed line and the strong coupling expansion is given by the Red dashed line. The interpolating function $\Gamma_{\rm cusp}^{(G_{6/5})}$ is given by the solid Blue line. (b) The weak coupling expansion for the anomalous dimension at finite spin is given by the Green dashed line and the strong coupling expansion via QSC is given by the Red dashed line. The interpolating function $G_{8/7}$ is given by the solid Blue line.}
\label{fig:testG65}
\end{figure}

We start with constructing an interpolating function in the large $j$ region. In this case, we can approximate the anomalous dimension by $\Gamma_{\rm cusp}(\tilde g)\log j$, and it suffices to construct an approximant for only $\Gamma_{\rm cusp}(\tilde g)$.  Using the data given in (\ref{data1}), we can construct the following two--point minimal Pad\'e approximant in the tune of eq.~\eqref{eq:Pade},\footnote{Here, only the first four loops in the weak coupling and first two loops in the strong coupling side has been taken into account. For related construction, see for example \cite{Banks:2013nga}.}
\be
\Gamma_{\rm cusp}^{(G_{6/5})} = \frac{87.4384 \tilde g^6+97.1024 \tilde g^5+57.0406 \tilde g^4+11.1317 \tilde g^3+4 \tilde g^2}{43.7192 \tilde g^5+55.7857 \tilde g^4+33.4311 \tilde g^3+17.55 \tilde g^2+2.78293 \tilde g+1} \,.
\ee
In figure \ref{fig:testG65}(a) , we show the smooth interpolation resulting from this approximant.

 We next move on to construct another Pad\'{e} approximant for the small $j$ region using the anomalous data up to three loops (weak coupling as in (\ref{finspinres})) and QSC data up to four loops (strong coupling (\ref{deltaki})).  Let us call this function $G_{8/7}$ for simplicity whose explicit is given in the Appendix \ref{padfing}. In figure \ref{fig:testG65}(b), we compare the interpolating function with the perturbative data for $j=5$ to showcase the efficiency of interpolation.

As we have seen earlier, in general the small $j$, small $\tilde g$ data (Section A  in figure \ref{viz}) can also be interpolated to the large $j$ region via the property of the Harmonic Sums. But our function $G_{8/7}$ also includes strong coupling data from semi--classical and QSC computations, so we can't a priori assume that the whole function will also be valid into the large $j$ region. We can however expect that up to finite values of $j$, the approximant works quite well. However, for larger values  of $j$ and $\tilde g$ there is a competition between the two regimes of data as a dependence in order of limits kicks in. Physically, it seems plausible that in $j\gg \tilde g$ region the $\log j$ dominates and in  $\tilde g \gg j$ region the finite spin behaviour (starting with $\sqrt{j}$) dominates. We will discuss this more in section \ref{transition}.

Now, we have two different models to fit two different $j$ regimes in the $\tilde g$--$j$ parameter space, i.e. $\Gamma_{\rm cusp}^{(G_{6/5})}\log(j)$ and $G_{8/7}$. We can plot them together and discuss the implications. It is indeed notable that there is a well defined `sharp' transition region between the two functions, given by the equation 
\begin{equation}
\label{eq:planarIntersec}
\Gamma_{\rm cusp}^{(G_{6/5})}(g) \log(j)-G_{8/7}(j, g)=0 \,.
\end{equation}
In figure 3(a), we show the real solution of $j = j(\tilde g)$  from this equation (apart from $\tilde g = 0$). 
 A comparison with figure 1 can assure the reader that this line roughly separates regions (A+B) and (C+D), while maintaining a sharp transition along a $j\propto \tilde g$ line for higher values of $\tilde g$.  One can see from the figure that this curve splits the parameter space into two, and it can be checked that in large $j$, in the left region value of $G_{8/7}$ dominates while in right region $\Gamma_{\rm cusp}^{(G_{6/5})}(g) \log(j)$ dominates. The line that appears in the small $j$ region is, however, not trustworthy.

We also plot the whole parameter space in figure 3(b)  to show clearly the features of  two dimensional surfaces corresponding to the functions and their intersection region. In the figure, it is evident that the two regions are separated by the almost linear transition region i.e. the transition occurs along a $j \propto \tilde g$ curve, although in the plot we can mostly see this line in large enough $j$ and $\tilde g$ region. 

We must note here that by our construction $G_{8/7}$ should be well suited for small $j$ region for all $\tilde g$ and $\Gamma_{\rm cusp}^{(G_{6/5})}\log(j)$ should be valid for predominantly large $j$ region. But it seems that due to the logarithmic term, the latter takes over in the parameter space sooner than expected and hence the intersection becomes important. In figure 3(b), we can clearly see that around the intersection, our expectation of $\log j$ dominating in $j\gg \tilde g$ region and $\sqrt{j}$ dominating in the $\tilde g \gg j$ region is fulfilled by the subdominant branches of the two--dimensional surfaces, i.e. the ones drawn in solid colour. 

Now due to the extra line in the small $j$ region, it seems $\Gamma_{\rm cusp}^{(G_{6/5})}(g) \log(j)$ is the right choice in the region with low $j$ but high $\tilde g$ (analogous to section B in figure \ref{viz}), but this appears to be counter--intuitive. We conclude that in small $j$ region, our approximation may not be trustable. Before we go into the physical implications of the above constructions, let us first see how we can in principle improve this interpolating function and show that $j\propto \tilde g$ transition behaviour is universal even with improvements.
\begin{figure}[!tb]
\centering
\begin{minipage}{.48\textwidth}
\centering
  \includegraphics[width=.95\linewidth]{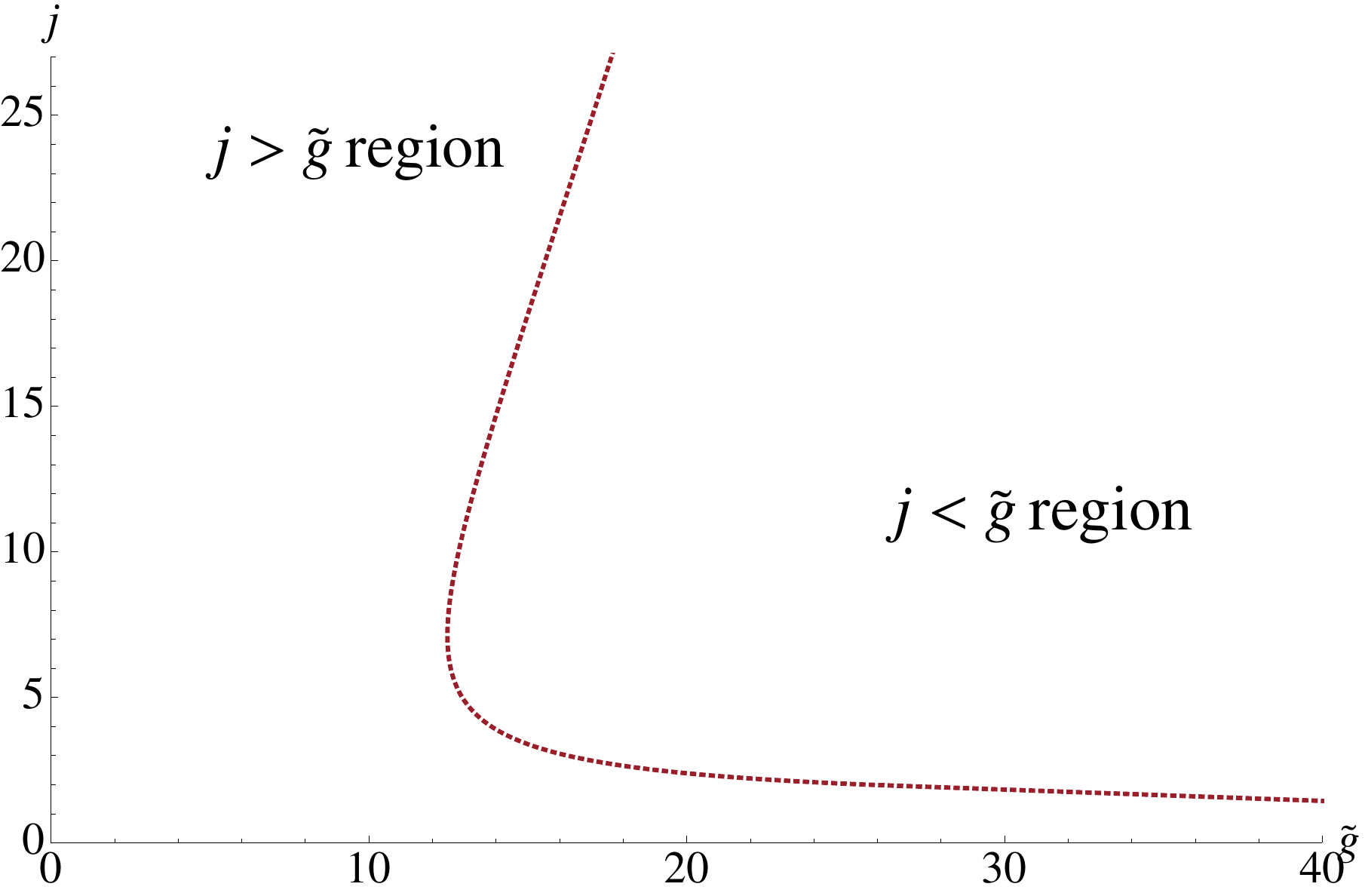}
  \caption*{(a)}
\end{minipage}%%%%%%
~~
\begin{minipage}{.52\textwidth}
  \centering
  \includegraphics[width=.95\linewidth]{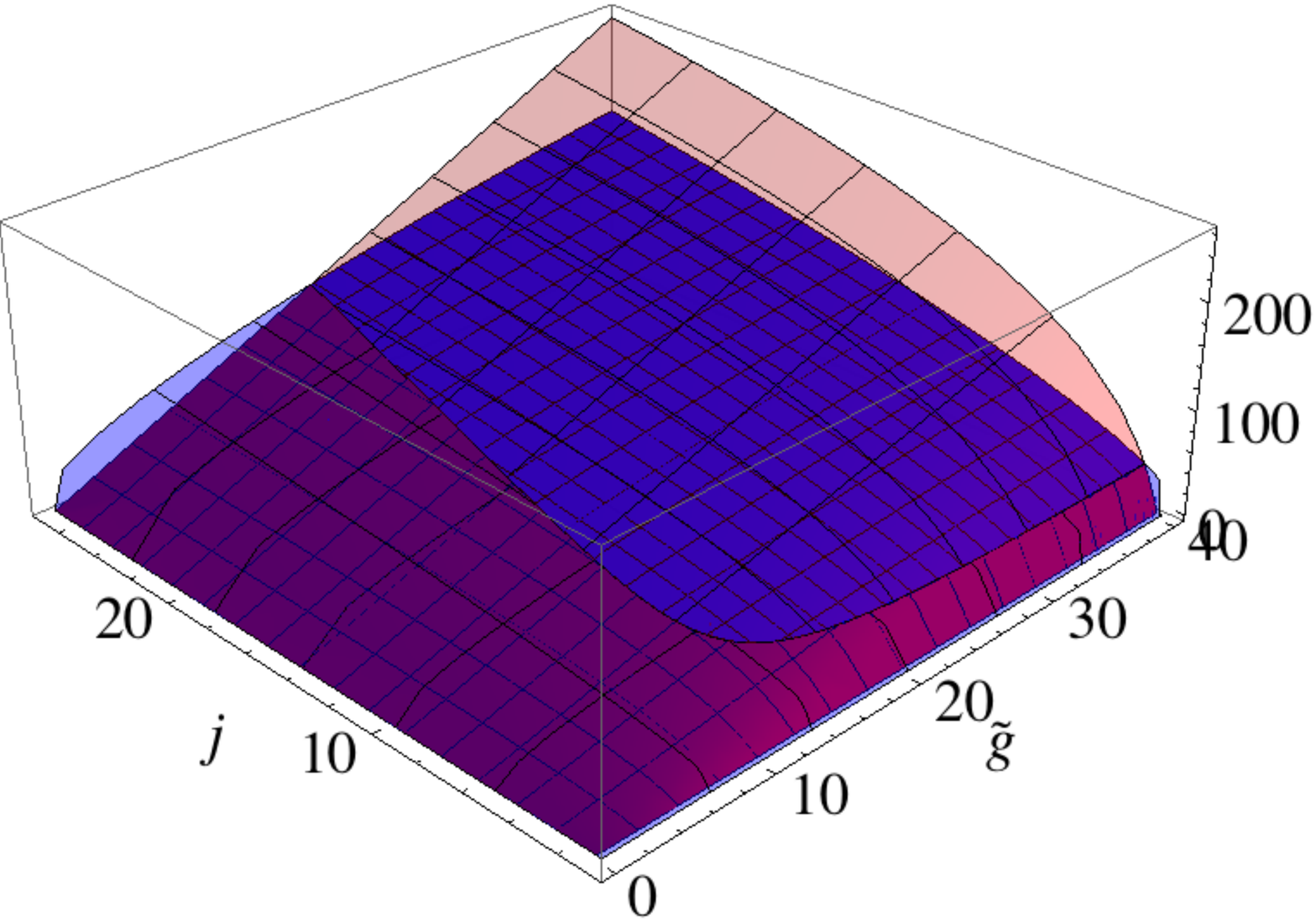}
  \caption*{(b)}
    \label{fig:planartwist2crossing2}
\end{minipage}

\caption{(a) Real solutions of the equation $\Gamma_{\rm cusp}^{(G_{6/5})}\log(j)-G_{8/7}=0$, showing the intersection curve between the two functions. (b)  Relevant parts of the surfaces $\Gamma_{\rm cusp}^{(G_{6/5})}\log(j)$ (red solid and transparent surface) and $G_{8/7}$ (blue solid and transparent surface) plotted together. Dominant branches are in transparent colour while subdominant ones are in solid colour. The transition line has the distinctive shape we got from the analytical solution and the total surface represents the anomalous dimension over the $\tilde g-j$ plane. }
 
\end{figure}

\subsection{Improving the interpolating functions}
\label{fppconst}

A notable issue with construction of interpolating functions for a physical object is that there could be various ways one could improve the behaviour of such a function.\footnote{In general, the structure of large spin expansion of the cusp anomalous dimension follows the expression
\be
\gamma(j) = A(\lambda)\log(j) + B + C(\lambda)\frac{\log(j)}{j}+D(\lambda)\frac{1}{j}+...
\ee
Now from the available literature one could read of the values of the functions $A,B,C,D$ depending on the gauge coupling. In principle one could improve the total interpolating function by constructing individual interpolating functions for $B,C,D...$ etc.} 
For example, one could take data up to more loops or even construct a different approximant that gives correct expansion at two ends. We have already encountered a number of different interpolating functions that could do the trick. However, our skeleton construction can be dubbed sufficient if, with various improvements, the physical properties captured by the functions do not change much. In what follows we will briefly talk about possible improvements to the construction we presented in the last section.

%%%%%%%%%%%%%%%%%%%%%%%%%%%%%%%%%%%%%%%%%%%%%%%%%%
%%%%%%%%%%%%%%%%%%%%%%%%%%%%%%%%%%%%%%%%%%%%%%%%%%
\subsubsection*{Improvement via FPP construction}

For the sake of completeness we should try to construct better interpolating functions for our small spin and large spin data. One example could be the Fractional Power Polynomial (FPP) type of interpolating function, which has been briefly introduced in section \ref{wthoutS}. We remind the reader again the structure of such function. For our case with small spin data, we can see that the parameters of FPP can be fixed as following,
\be
b-a=-3, ~m+n+1 = 15.
\ee
Then the total interpolating function for the small spin case will have the form
\be
F_{\rm small}(x,j)=F_{8,6}^{(-1/5)}(x,j)=s_0 x^4 \left(1+\sum_{k=1}^8 c_k(j) x^k+\sum _{k=0}^6 d_k(j) x^{15-k} \right)^{-1/5},~~~\tilde{g} = x^2.
\ee
Similarly, for the large spin approximation we have
\be
F_{\rm large}(\tilde{g},j) = \Gamma_{\rm cusp}^{(F_{6,4})}(\tilde{g})\log(j) \,,
\ee
where we write an FPP for cusp anomalous dimension as
\be
\Gamma_{\rm cusp}^{(F_{6,4})}(\tilde{g})=F_{6,4}^{(-1/11)}(\tilde g)= l_0 \tilde{g}^2 \left(1+\sum_{k=1}^6 e_k \tilde{g}^k+\sum _{k=0}^4 f_k \tilde{g}^{11-k} \right)^{-1/11}.
\ee

\begin{figure}[!tb]
		\centering
		\includegraphics[width=9cm]{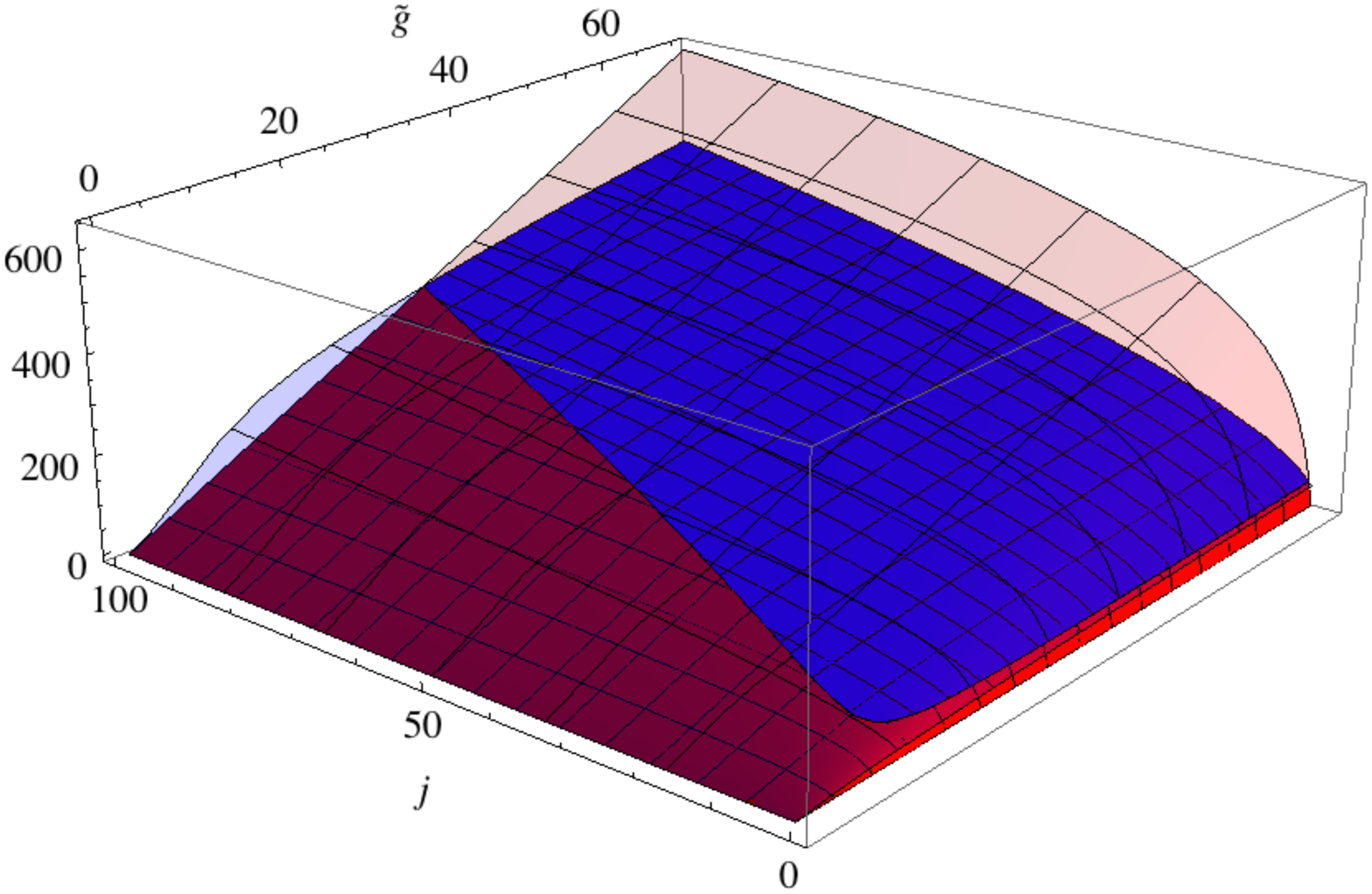}
		\caption{$F_{\rm large}$ (red solid and transparent surface) and $F_{\rm small}$ (blue solid and transparent surface) plotted together in the $\tilde g- j$ plane. Dominant branches are in transparent colour while subdominant ones are in solid colour. The transition line between them is clearly visible.}
		\label{fig:planartwist2crossingfp}
\end{figure}

The problem with such a function is that the denominator has a fractional power, so depending on the coefficients it can run into complex values, which will signal the breakdown of our approximation. One can in principle solve the denominator for negative values to find the breakdown parameters. Careful investigation shows that such breakdown of the approximations in $F_{\rm small}$ occur here not before $j\sim 300$ which is a improvement over our model with Pad\'{e} approximants. A plot of the two functions covering the $\tilde g$--$j$ plane is shown in figure \ref{fig:planartwist2crossingfp}. One may note the qualitative similarities between this picture and figure 3(b).
%%%%%%%%%%%%%%%%%%%%%%%%%%
%%%%%%%%%%%%%%%%%%%%%%%%%%

%%%%%%%%%%%%%%%%%%%%%%%%%%%%%%%%%%%%%%%%%%%%%%%%%%
%%%%%%%%%%%%%%%%%%%%%%%%%%%%%%%%%%%%%%%%%%%%%%%%%%
\subsubsection*{Improvement via nonlinear variable transformations}

Another problem with both Pad\'{e} and FPP type approximants lie in the appearance of anomalous powers in the expansion along weak or strong coupling points. This becomes particularly clear as we expand our Pad\'{e} approximants or FPP's beyond the orders up to which we have known data. For example, if one fixes the expansion of such a function upto $\tilde g^8$, and tries to expand upto higher orders, anomalous powers like $\tilde g^9$ can also creep in. 

These unwanted powers can be in principle managed by adding ``offset'' interpolating functions to cancel them up to arbitrary orders, however it does not actually provide a permanent solution. A tentatively better way is to tweak the variables in the interpolating functions so that it only spews out the right powers in both weak and strong coupling expansion. 
As an example let us consider the variable change\footnote{See \cite{Gromov:2011bz} for related non--linear transformations in interpolating functions to generate correct powers in expansions.}
\be
y(\tilde{g})=\frac{\tilde{g}^2}{\sqrt{1+\tilde{g}^2}} \,.
\label{eq:nonlinearvar}
\ee
Then, we can consider an interpolating function of the form
\be
P_1(y)=\sqrt{\frac{\sum _{i=1}^4 A_i y^i}{\sum _{i=1}^3 B_i y^i+1}},
\ee
the expansion of which only gives terms of order $\{\tilde{g}^2,\tilde{g}^4,\tilde{g}^6...\}$ in the weak coupling and $\{\tilde{g}^{1/2},\tilde{g}^{-1/2},\tilde{g}^{-3/2}...\}$ in the strong coupling expansion, i.e exactly reproduces our small spin data. Similarly, the following function can also be considered,
\be
P_2(y)=\frac{\sum _{i=1}^4 C_i y^i}{\sum _{i=1}^3 D_i y^i+1}.
\ee
This gives $\{\tilde{g}^2,\tilde{g}^4,\tilde{g}^6...\}$ in the weak coupling and $\{\tilde{g},\tilde{g}^{0},\tilde{g}^{-1}...\}$ in the strong coupling expansion, i.e exactly reproduces our cusp data. One should note that, similar problem with generation of right powers will reappear in the case of finite $N$ (section \ref{sec:finiteNcase}) too, but will be dealt with in a different way.

%%%%%%%%%%%%%%%%%%%%%%%%%%%%%%%%%%%%%%%
%%%%%%%%%%%%%%%%%%%%%%%%%%%%%%%%%%%%%%%
\subsection{Towards an interpretation of the ``transition region''}
\label{transition}

As we mentioned in the last section, in the weak coupling, the twist--two anomalous dimension has a distinct $\log(j)$ scaling behaviour in the large spin limit to all perturbative orders. On the other hand, the story at strong coupling has two different cases, making it more subtle. When the spin is very large 
the $\log(j)$ scaling is still true. However, when we have small spins,
 the scaling becomes $\gamma \sim \sqrt{j}$, see \eqref{D01}.  Intriguingly, the dependence on the `t Hooft coupling is also different in these two cases. The former case has $\sqrt{\lambda}$ dependence at the leading order, while the latter is $\lambda^{1/4}$ dependent. To summarize, we have
\begin{eqnarray}
& \gamma(j, \lambda) \sim \sqrt{\lambda} \log(j), & \quad \textrm{Large spins} ,  
\label{eq:large-j-case}
\\
& \gamma(j, \lambda) \sim \lambda^{1/4} \sqrt{j}, & \quad \textrm{Small spins}.
\label{eq:small-j-case}
\end{eqnarray}
Obviously, there is an order of limits issue when taking large $\lambda$ and large $j$ together. The region where such a transition from one behaviour to another happens can be characterized by the intersection line from eq. \eqref{eq:planarIntersec} obtained in the previous section. Below we will try to give a physical account for the dependence of this two limits and comment on the physics of the transition. The best way to describe this physics seems to be in the realm of classical strings in $AdS/CFT$, where large $\lambda$ finds a natural existence. The general expansion of string energy calculated from such prescription then appears as a dispersion relation between conserved charges from the sigma model.

On the classical string side, the appearance of $\lambda^{1/4}$ scaling is not that surprising, even though one would expect that leading order energy contribution from string sigma model would start at $\sqrt{\lambda}$.
Since spin of string roughly counters the contracting effect due to the string tension ($T\sim \sqrt{\lambda}$), strings with smaller spins actually correspond to smaller lengths for strings with a centre of mass near $AdS$ centre ($\rho = 0$).  In \cite{Tirziu:2008fk, Roiban:2009aa}, it has been argued that one could then define an effective parameter for the case of the spinning folded string as $\mathcal{J} = j/\sqrt{\lambda}$, which actually measures this interplay of spin and string tension and remains fixed even when $j\to \infty$ and $\lambda \to \infty$. 

So, in general there is a case of order of limits involved in this energy expansion. Formally, a string with $\mathcal{J}\ll 1~ (j\ll \sqrt{\lambda})$ can still be described to be in the `short string' phase. Following the computation of \cite{Tirziu:2008fk, Roiban:2009aa} for energies of such string states, we can write the desired dispersion relation
\be
\mathcal{E} = a \mathcal{J}^{1/2}+b \mathcal{J}^{3/2}+c \mathcal{J}^{5/2}+...
\ee
with $\mathcal{E} = E/\sqrt{\lambda}$ and $a,b,c...$ are pure constants. For this case the `t Hooft coupling dependence to above expansion can be restored  in the form 
\be
E = A(j) \lambda^{1/4}+\frac{B(j)}{\lambda^{1/4}}+... \,,
\ee
where evidently $A(j)$  goes as $\sqrt{j}$. This expression can easily be checked on physical grounds to be low lying excited string states and always behave as $m^2\sim \frac{1}{\alpha'}$, i.e., they have the flat space Regge trajectory behaviour and hence the dimension of dual operators go as $\lambda^{1/4}$ in the leading order.\footnote{We may also recall that $\lambda =  {R^4 \over \alpha'^2} \sim {R^4 \over l_s^4}$,
where $R$ is the AdS radius, and $l_s$ is the string length. For large $\lambda$ and for small spin states, it is equivalent to think that the AdS radius (which we set to be one) is much large comparable to the string length. Thus we can approximate the close string as moving in the flat background which gives the well--known Regge trajectory behavior between mass and spin, $m \sim \sqrt{j}/\alpha'$.} One has to include subleading corrections starting from $\mathcal{O}(\lambda^{-1/4})$ due to curvature of the target space. 
 
\begin{figure}[!tb]
		\centering
		\includegraphics[width=9cm]{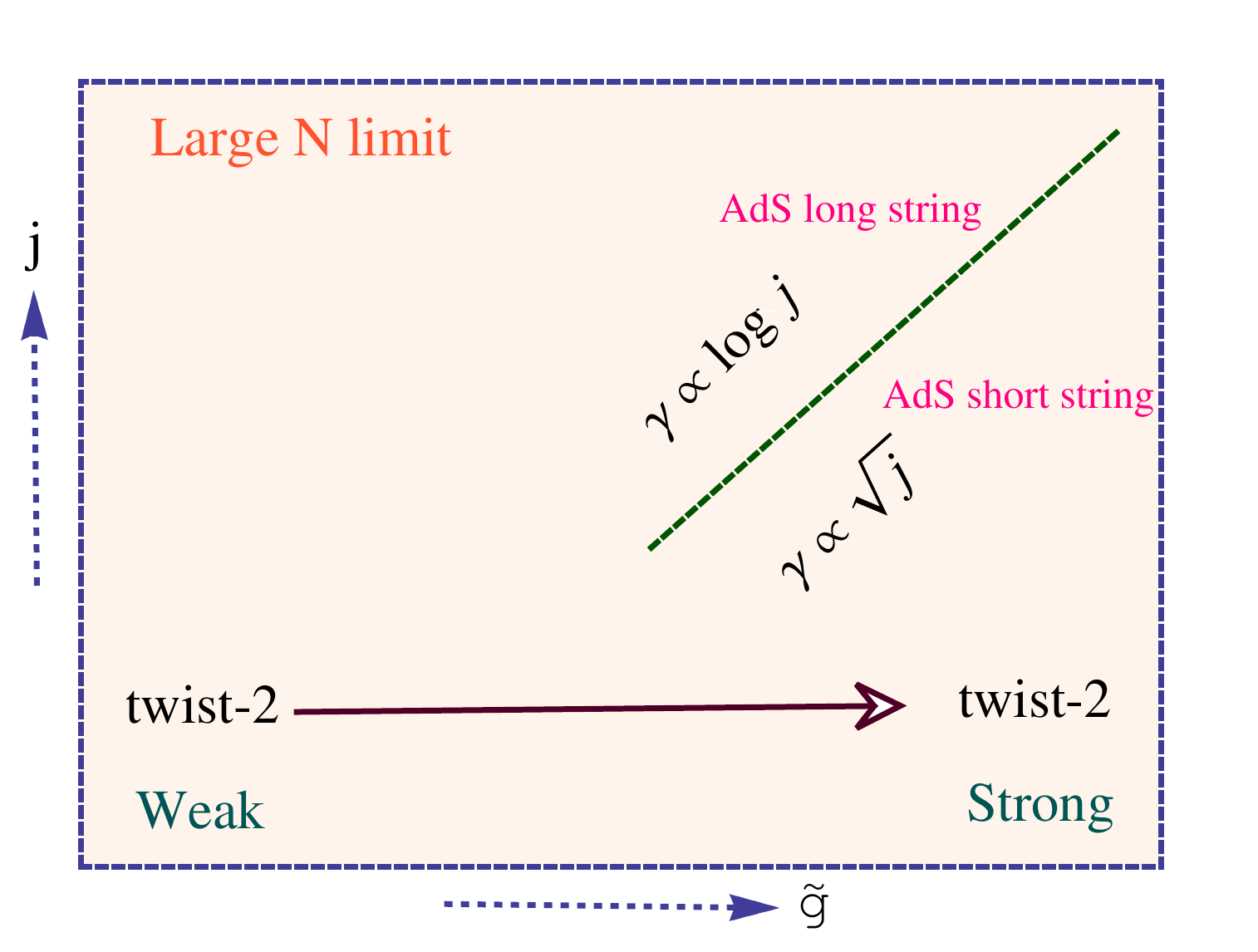}
		\caption{The transition between different $j$ dependence for twist--two anomalous dimensions in the large $N$ case.}
		\label{phase}
\end{figure}

 On the other hand, it is well known that to capture the physics of cusp anomalous dimensions we need to take the large spin limit, where $j\gg \sqrt{\lambda}\gg 1$ \cite{Gubser:2002tv}, and the dispersion relation clearly gives rise to the $\sqrt{\lambda}\log j$ term in the leading order.  So, it can be clearly seen that the parameter space is divided into two regions along the line $j\sim \sqrt{\lambda}$. But we must remind ourselves that this works only for large enough value of the coupling, where the string description is still valid. It seems that at least in the large enough coupling region, the transition region between our two interpolating functions are in tune with this description. The situation has been summarised in figure \ref{phase}. Comparing with figure 3(b) and figure 4, we could see that the solid colour surfaces, i.e. the subdominant ones are the physically trustable regions. 
 %Since the picture is not clear in small $j$, large $\tilde g$ region, we refrain to comment on it.
 
 Physically, these two regions on the parameter space signify two ends of the $AdS$ spinning string spectrum, the `long' string that almost touches the boundary and the `short' string which is not stretched much
compared to the radius of curvature of $AdS$ and stays near the centre. Both of these pictures are defined in the strong coupling regime, but as we mentioned earlier, the physics depends on the order of limits between the coupling and the spin. The two regimes of classical strings are discriminated by the appearance of ``cusps''  or ``folds'' on the string profile. As we increase the spin and go towards $j\sim \sqrt{\lambda}$, the string starts to get more and more stretched along the $AdS$ radius, and proper length of this highly excited string starts becoming comparable to $AdS$ radius. 

%%%%%%%%%%%%%%%%%%%%%%%%%%
%%%%%%%%%%%%%%%%%%%%%%%%%%

A nice way to see this would be to consider the profile and conserved charges associated with spinning strings in the $AdS_3\in AdS_5$
spaces. For example the profile of  a string spinning with angular velocity $\omega$ is given by\footnote{ See \cite{Floratos:2013cia} for a recent take on this issue.}
\be\label{profile}
\frac{d\sigma}{d\rho} \sim \frac{1}{\sqrt{(\omega^2-1)(\frac{1}{\omega^2- 1}-\sinh^2\rho)}} \,,
\ee
where $\rho$ is the radial direction in the $AdS_3$ and $\sigma$ is the spacelike worldsheet coordinate. We parameterize the AdS coordinates in the form $t\sim\tau,~\phi\sim\omega \tau,~\rho = \rho(\sigma)$ and consider a spinning folded closed string whose center lies at rest at $\rho=0$. The extent of the string $\rho_0$ along AdS radial direction is roughly given by $\coth\rho_0= \omega$.

 It is evident that for a closed string with $\omega^2>1$ and $\sinh^2\rho < \frac{1}{\omega^2-1}$, we can have two cases:
\begin{enumerate}
\item $\omega \to \infty$: No derivative discontinuities appear on $\rho(\sigma)$, i.e. no cusps appear on the string,  which corresponds to the ``short'' string phase.

\item $\omega \to 1$: $\rho(\sigma)$ develops a derivative discontinuity at particular point(s), i.e cusps begin to appear on the string, which corresponds to the ``long'' string phase.
\end{enumerate}
For actual $\omega=1$, the cusps on the string touches the AdS boundary. So, in general the information of both the  short and long strings can be explored from the spinning string setup at large coupling. The Noether charges associated with the string can be calculated using (\ref{profile}) and is given by \cite{Gubser:2002tv}
\be
E  =  \frac{2\sqrt{\lambda}}{\pi}\frac{\omega}{\omega^2 -1}\mathbb{E}\left(\frac{1}{\omega^2}\right),~~~j = \frac{2\sqrt{\lambda}}{\pi}\left[ \frac{\omega}{\omega^2 -1}\mathbb{E}\left(\frac{1}{\omega^2}\right) - \mathbb{K}\left(\frac{1}{\omega^2}\right)   \right],
\ee
where $\mathbb{E}$ and $\mathbb{K}$ are the usual complete Elliptic integrals. One can see here that the charges implicitly depend on each other, and hence the dispersion relation is given by $E= E(j)$. A way to look at the anomalous dimension for any $j$ is to consider the quantity $\gamma = \frac{\pi}{2\sqrt{\lambda}}(E-j)$. Although it is quite complicated to evaluate this quantity for different $j$ and $\lambda$, one can try to do so numerically. In figure 6 we plot this quantity from the equations above. We see that  no such sharp transition occurs in this case as in the last, and the small $j$ and large $j$ regimes interpolate smoothly into each other. From the interpolating function point of view, if one could properly construct a master interpolating function in both $j$ and $g$ then it is expected that the physics of this transition would become more clear and one could predict data at any non--trivial point on the surface which could later be checked against analytics.
%%%%%%%%%%%%%%%%%%%%%%%%%%
\begin{figure}[!tb]
		\centering
		\includegraphics[width=9cm]{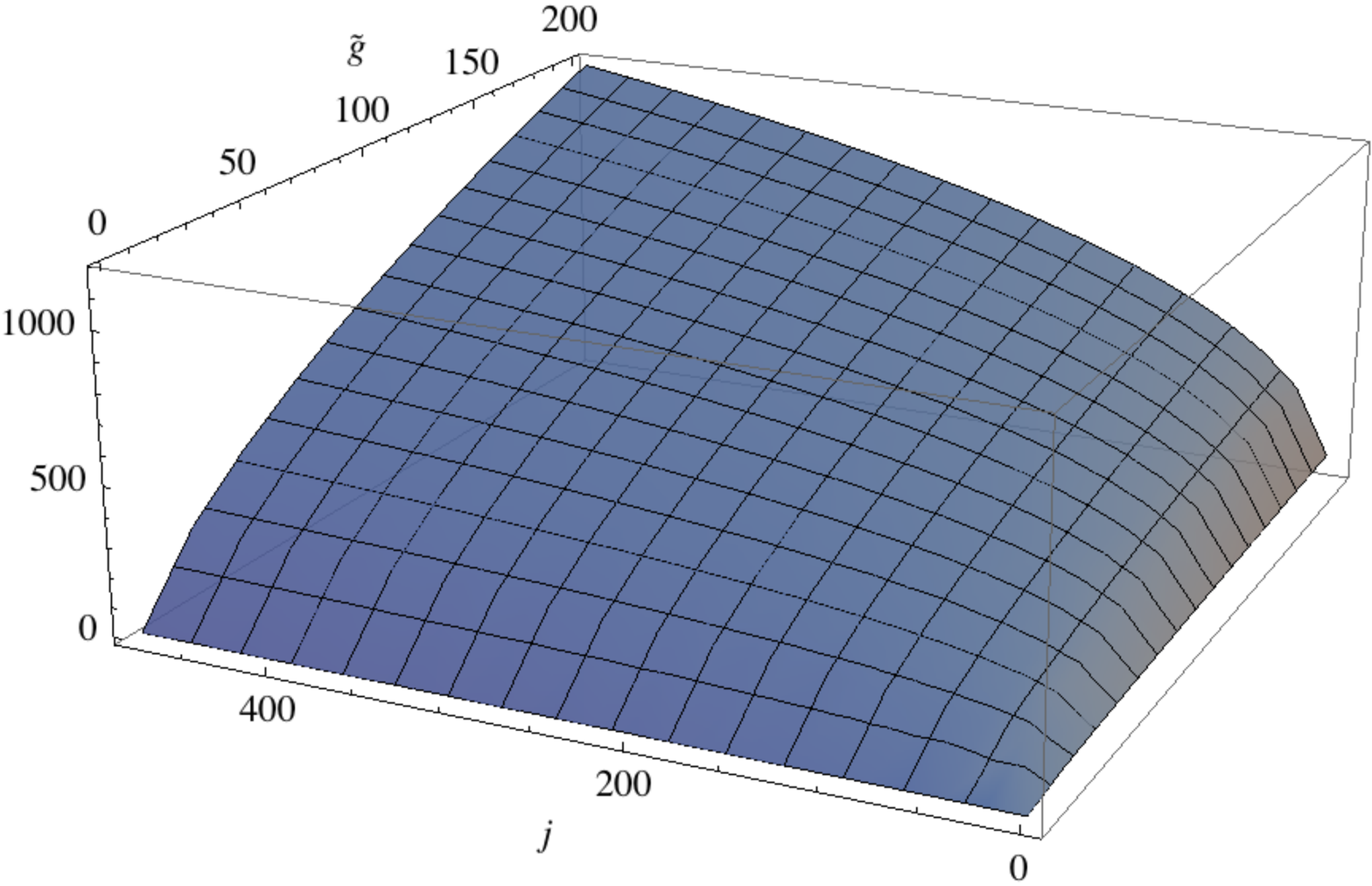}
		\caption{Numerical data for twist--two anomalous dimension plotted for various values of $\tilde g$ and $j$ using the classical string solution. }
\end{figure}
%%%%%%%%%%%%%%%%%%%%%%%%%%

%%%%%%%%%%%%%%%%%%%%%%%%%%%%%%%%%%%%%%%%%%%%%%%%%%
%%%%%%%%%%%%%%%%%%%%%%%%%%%%%%%%%%%%%%%%%%%%%%%%%%
\section{Probing the finite $N$ case: Modular invariant interpolating functions}
\label{sec:finiteNcase}

So far in this work, we have focused mainly on the planar (large $N$) limit where ${\cal N}=4$ SYM is expected to be integrable and there exists a well defined  AdS/CFT dictionary to study various gauge--invariant observables. However, moving away from the large $N$, ${\cal N}=4$ SYM is a much harder model to study yet interesting to explore due to close relation to realistic QCD (where $N$=3). For finite $N$, it is expected 
that  ${\cal N}=4$ SYM   possesses S--duality \cite{Osborn:1979tq} which connects weak and strong coupling regimes of the theory. It is a well known fact for gauge invariant observables in $\mathcal{N}=4$ SYM that one has to include all non--planar  and non--perturbative corrections to restore full S--duality to the result. In particular, the instanton contributions are very important in the S--duality context, but they are supposed to be exponentially small at large $N$, see e.g.~\cite{Korchemsky:2017ttd}. The main challenge in studying S--duality invariant object is that one has to look at truly non--planar and non--perturbative data, which is hard to find in many cases.

With this new symmetry  at hand, we would attempt to take a step towards constructing interpolating functions on the same line of modular invariant interpolating forms presented in \cite{Chowdhury:2016hny} to approximate the anomalous dimensions of twist--two operators. As before, we pay special attention to the dependence on the spins. 
Below we will construct the modular invariant interpolating function first for the cusp anomalous dimension and then for the twist--two anomalous dimension with general spin dependence.

We recall our convention again for convenience, 
\begin{equation}
g=\frac{g^2_{YM}}{4\pi}=\frac{4\pi\tilde g^2}{N} = \frac{\lambda}{4\pi N} \,, \qquad \tau = \frac{\theta}{2\pi} +\frac{i}{g}\,.
\end{equation}

%%%%%%%%%%%%%%%%%%%%%%%%%%%%%%%%%%%%%%%%%%%%%%%%%%
%%%%%%%%%%%%%%%%%%%%%%%%%%%%%%%%%%%%%%%%%%%%%%%%%%
\subsection{Cusp anomalous dimension}
\label{cinfnc}

As we mentioned before, the cusp anomalous dimension is related to the UV singularity of a cusped Wilson loop \cite{Korchemsky:1985xj}, and since a Wilson loop is expected to have a S--duality completion, it is natural to expect that there is a non--perturbative definition for the cusp anomalous dimension that also satisfies the modular invariance.
We will thus construct the interpolating function by using Eisenstein series as the building block as introduced in section \ref{intfunc}. 
Concretely, we will use the FPR--type S--duality invariant interpolating functions as eq.~\eqref{eq:FPR_wo_gravity} for our construction.

Let us explain the physical data we will use for the construction. 

\begin{itemize}
\item 
In the weak coupling expansion, we will use the perturbative data up to four loops given in \eqref{cuspdef1}--\eqref{nonplan}, including the non--planar result that appears first at the fourth loop.
\item
 At strong coupling,  under the S--duality it is expected that a Wilson loop is related to a `t Hooft loop, and in this regard one should use corresponding quantity associated to the \emph{cusped} 't Hooft loop, a construction which seems to be still missing in the literature.
Instead, here we assume that the cusp anomalous dimension has `weak' mixing effect, and we will (naively) use the holographic data of cusp anomalous dimension at strong coupling in our construction.

It would be very interesting to study this quantity, for example, by considering the D1 string coupled to \textit{cusped} 't Hooft loop at AdS boundary \cite{Drukker:2005kx,Chen:2006iu}. 
\item
Finally, another important additional structure that we will take into account is the instanton contribution \cite{Korchemsky:2017ttd}.

\end{itemize}

%%%%%%%%%%%%%%%%%%%%%%%%%%%%%%%%%
%%%%%%%%%%%%%%%%%%%%%%%%%%%%%%%%%
\subsubsection*{Resolving power issue at strong coupling}

For the FPR type interpolating function  \eqref{eq:FPR_wo_gravity}, there is a problem of matching the correct power at strong coupling as explained around \eqref{eq:power-issue1}-\eqref{eq:power-issue2}; the constraints on the parameters of the function does not allow for fractional powers of $\lambda$.\footnote{ We will shift our convention of coupling constant to the t' Hooft coupling $\lambda$, to study the finite $N$ scenario. This will be essential in defining the holographic limit in a systematic way. The data collected in section (\ref{data1}) should be appropriately multiplied/divided by factors of $4 \pi$ to interpret the series as expansions in t' Hooft coupling $\lambda$.} 
Note that the change of variable methods used in section \ref{fppconst} will not work here in a straightforward way, since the new non--linear variable as in eq.~\eqref{eq:nonlinearvar} won't meet the requirement of modular invariance directly. Instead, we would have to take an alternative approach.

To take into account the failure of \eqref{eq:FPR_wo_gravity} to encode the fractional power in the strong coupling expansion, let us consider the following generalization of the usual construction of $F_{m}^{(s,\alpha)}$ by introducing
\begin{equation}
I_m^{(s,t,\alpha )} (\tau ) 
= \Biggl[ \frac{ c_1 E_{s+t+1}(\tau ) +\sum_{k=2}^p c_k E_{s+k} (\tau )}
{\sum_{k=1}^q d_k E_{s+k} (\tau )} \Biggr]^\alpha ,
\label{eq:inter_t}
\end{equation}
where $c_1 \sim \mathcal{O}( N^{q-t-1})$ 
and
the only difference from \eqref{eq:FPR_wo_gravity}
is the presence of a new integer parameter $t$ in the first term of the numerator. The planar limit of this interpolating function for large $s$ is given by
\begin{equation}
\left. I_m^{(s,t,\alpha )} (\tau ) \right|_{\rm planar}
= \Biggl[ \frac{ \bar{c}_1 \lambda^{-(s+t+1)} +\sum_{k=2}^p \bar{c}_k \lambda^{-(s+k)} }
{\sum_{k=1}^q \bar{d}_k \lambda^{-(s+k)} } \Biggr]^\alpha 
= \Biggl[ \frac{ \bar{c}_1 \lambda^{-t} +\sum_{k=2}^p \bar{c}_k \lambda^{-k+1} }
{\sum_{k=1}^q \bar{d}_k \lambda^{-k+1} } \Biggr]^\alpha ,
\end{equation}
where the changed coefficients have a form 
$\bar{c}_1= \lim_{N\rightarrow\infty} \zeta (2s+2)N^{t+1-q} c_1$,  $\bar{c}_k=\lim_{N\rightarrow\infty} \zeta (2s+2k)N^{k-q} c_k$ and
$\bar{d}_k = \lim_{N\rightarrow\infty} \zeta (2s+2k)N^{k-q} d_k$.

For a negative $t$ and large $\lambda$,
the leading order for the above expression is $\mathcal{O}(\lambda^{-\alpha t})$.  We can now match any fractional power in the strong coupling expansion of the form $\mathcal{O}(\lambda^c)$, where $c$ is any fractional power by solving 
\begin{equation}
-\alpha \, t = c \,. 
\end{equation} 
Note that if the leading power in \eqref{eq:inter_t} is $\mathcal{O}(\lambda^c) $ the subsequent powers are $\mathcal{O}(\lambda^{c-1}), $
$\mathcal{O}(\lambda^{c-2}) $ and so on. So essentially, our problem with fractional powers is solved by tweaking the previous construction and yet it is consistent with  the weak coupling data \eqref{cuspdef1} and the strong coupling data \eqref{cuspst}.

As is evident from the above discussion, we now need to fix $\alpha t=-1/2$ to get the right strong coupling expansion. But, there is a caveat as the subsequent order in the expansion of the function is  $\mathcal{O}(\lambda^{-1/2})$, rather than $\mathcal{O}(1)$. To reproduce all the correct powers of $\lambda$ in \eqref{cuspst} we would need to construct an interpolating function which is a linear combination of the basic interpolating functions.
	The total interpolating function that has the correct properties to generate strong coupling powers like $\{\lambda^{1/2}, \lambda^0, \lambda^{-1/2}....\}$  is  
	
	\begin{equation}	\label{interstrongq}
I= w_1 I^{(10,-1,1/2)}_5 + w_2 I^{(10,0,1/2)}_5 \,, \quad \text{where} \ \  w_1 + w_2 =1 \,,
	\end{equation}
where all the coefficients including $w_1  $, $ w_2$
	are fixed by respectively matching
	to the $\mathcal{O}(\lambda^{\frac{1}{2}})$ and $\mathcal{O}(1)$ coefficients in the strong coupling data.  Essentially, while the first function generates terms of the order $\{\lambda^{1/2}, \lambda^{-1/2}....\}$, the second function complements with the orders $\{\lambda^{0}, \lambda^{-1}....\}$ and together they can explicitly match the strong coupling power series.  For a detailed account on how to fix the coefficients of the function, the reader is directed to Appendix \ref{resultsdcp}.
	
	Note that, in general the linear combination of functions could involve different number of basic functions depending on which powers we want to have in the strong coupling expansion, i.e. the general schematic ansatz would be $I_{total} = \sum_{i=1}^{n}w_i I^{(i)}$ with $\sum_{i=1}^{n}w_i = 1$.\footnote{ For example, in \cite{Chowdhury:2016hny} we have studied the anomalous dimension of the Konishi constrained on  the strong coupling by supergravity. However if we assume that Konishi operator doesn't level cross on the strong coupling to double trace operator and used the naive gauge theory result at the strong coupling which starts at $\mathcal{O}(\lambda^{1/4})$, one would have to start with $n=3$.}

%%%%%%%%%%%%%%%%%%%%%%%%%%%%%%%%%
%%%%%%%%%%%%%%%%%%%%%%%%%%%%%%%%%
\subsubsection*{Non--planar corrections}	

To take into account the non-planar piece of data (\ref{nonplan}), let us consider weak coupling data up to four loops and strong coupling up to two loops. As explained before, our minimal ansatz then should have a form  of (\ref{interstrongq}), i.e. we will need a linear combination of two interpolating functions. We should note that in the large $N$ limit the coefficients for a fixed order of $g$ or $\lambda$ in the weak or strong coupling expansions has $ \left \{\mathcal{O}\left(\frac{1}{N^2}\right)\, , \mathcal{O}\left(\frac{1}{N^4}\right) \cdots \right \} $ non--planar corrections which induces a similar non--planar contribution to the solved coefficients of the interpolating function. Hence, the planar parts of the interpolating function receives no corrections from the non--planar data and the construction of the previous section goes through with systematic non--planar corrections. 

	However, the effect of adding such a term shows up starting from $\mathcal{O}({\lambda^4})$ and subsequent orders in weak coupling expansion receive non--planar corrections, so does the strong coupling expansion.\footnote{Here, we would like to stress the fact that strong coupling limit of the interpolating function is first the $N \rightarrow \infty $ limit and then the $\lambda \rightarrow \infty $ limit. It allows us to drop the exponentially suppressed $\mathcal{O}\left(e^{-\frac{8 n \pi^2 N}{\lambda}}\right)$ terms arising from the non--perturbative parts of the Eisenstein series and focus only on the perturbative expansion.} From the weak coupling expansion \eqref{cuspdef1}, we can see that the numerical values of the planar and non--planar terms are of the  same order.
As a check of the construction, we can predict the data for the 5th--loop $\mathcal{O}\left(\lambda^{5} \right)$ order,\footnote{We mention that the $1/N^4$ correction appears only from six loops, where higher order group theory invariants enter, see e.g. Appendix B of \cite{Boels:2012ew}.}
\begin{equation}
\label{eq:fiveloopNPprediction}
\Gamma^{(5)}_{{\rm cusp,w}}=\left (6580.7+\frac{21308.8}{ N^2}\right) \times \left(\frac{1}{4 \pi}\right)^{10}.
\end{equation} 
The $\gamma^{(5)}_{{\rm cusp,w}}$ predicted from our construction can be checked against the result from the BES equation\eqref{higherloop} and our result is within $35\% $ error bar \footnote{Though $35 \%$ seems high, we would like to point out that in terms of $\lambda$--expansion the  BES equation \eqref{higherloop} i.e the 5th--loop data is of the order $10^{-7}$ i.e. very small. Therefore, it is better to minimise the error on the strong coupling side to the maximum extent possible.}. We can also predict the next order in the strong coupling limit i.e. $\mathcal{O}\left(\sqrt{\lambda}\right)$ and we found that our result is only $0.6\%$ off from the result obtained from BES equation \eqref{higherloop} which reads,\footnote{
	For completeness, let us give an example of parameter counting here. Lets start with combination of two $I_5$'s as in \ref{interstrongq}, each of which contains five unknown coefficients. After matching the weak coupling data up to four loops, three unknown variables remain, of which two can be solved using the two loop strong coupling data. The remaining one variable should be thought of as an adjustable free parameter and we can fix it by demanding that the absolute relative error in predicting the weak coupling planar $\mathcal{O}(\lambda^5)$ and strong coupling $\mathcal{O}\left(\lambda^{-1/2}\right)$  terms is minimised. }
\begin{equation}
\Gamma^{(3)}_{{\rm cusp,s}}= (-0.011535) \times (4 \pi) \,.
\end{equation} 
In figure 7, we plot the modular invariant interpolating functions of cusp anomalous dimension at $N=2$ and $N=20$. We also compare the effect of adding the non-planar correction to the data for each of these cases.
	
%%%%%%%%%%%%%%%%%%%%%%%%%%
\begin{figure}[!tb]
		\label{finite_weak}
		\centering
		\begin{minipage}{.48\textwidth}
		\centering
			\includegraphics[width=1.05\linewidth]{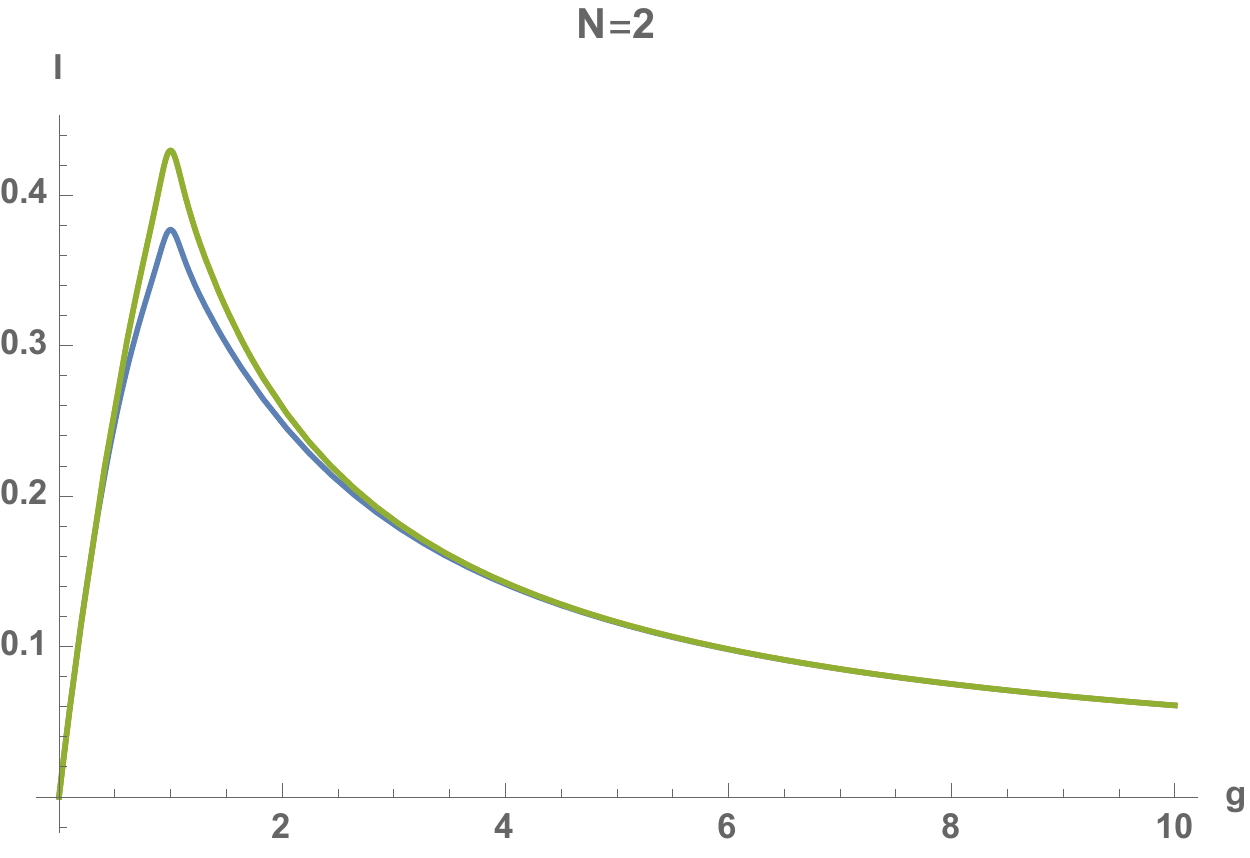}
			
		\end{minipage}%%%%%%
		~~
		\begin{minipage}{.5\textwidth}
			\centering
			\includegraphics[width=1\linewidth]{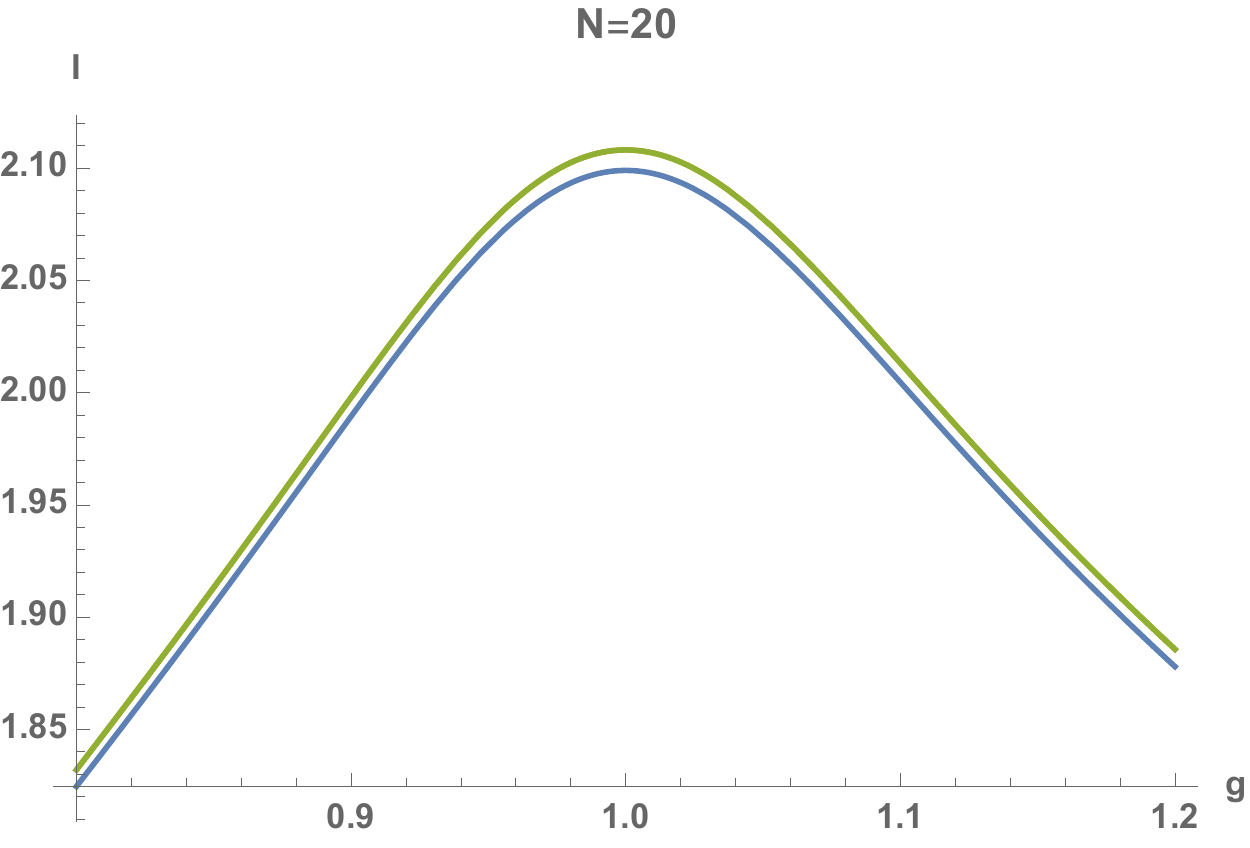}
			
		\end{minipage}
		\caption{In the above figures the blue and green lines respectively denote the modular invariant interpolating functions for the cusp anomalous dimension ($\Gamma_{\rm cusp}(\lambda)$) with zero and non-zero value of the non-planar term $\Gamma^{np}$. The non--planar  correction decreases with increasing $N$ and hence the maximum difference is visible at $N=2$  and $g=1$. The same is plotted for $N=20$ zooming around $g=1$, which shows a decrease in the difference with increasing $N$.}	
		\label{weak_nonplanar}
\end{figure}
%%%%%%%%%%%%%%%%%%%%%%%%%%

%%%%%%%%%%%%%%%%%%%%%%%%%%%%%%%%%%
%%%%%%%%%%%%%%%%%%%%%%%%%%%%%%%%%%	
\subsubsection*{Adding Instanton corrections}	
\label{sdulaity_inst}

The next hurdle in the process is to include instanton corrections at the weak coupling expansion of the function. Such a correction would first occur at $\mathcal{O}(g^4e^{-\frac{2\pi}{g}})$  \eqref{weakinst} . Since the weak coupling expansion does not have any fractional powers of $\lambda$, we could start with a single interpolating function which gives the right powers up to four loop expansion and solve the coefficients accordingly. For example, taking the FPR--like function (see (\ref{eq:FPR_wo_gravity})) $F_{4}^{(2,\frac{1}{3})}$ with $p=1,q=4$, one could generate a $\mathcal{O}(g^4e^{-\frac{2\pi}{g}})$  term from the non-perturbative part of the Eisenstein series. However, this does not serve our purpose as the $\mathcal{O}(g^4e^{-\frac{2\pi}{g}})$ term of the interpolating function which goes like $\sim N^{-3/2}$ and $\mathcal{O}(g)$ term which goes like $\sim N$  shares the same coefficient and hence is impossible to satisfy.
 
To incorporate all $N$ powers and $g$ powers at the right places it is better to follow with a construction discussed previously in this section and use a linear combination of interpolating functions ($I_{m}^{(s,t,\alpha)}$) like 
\begin{equation}
I = I_{4}^{(15,0,1)}+I_{2}^{(2,0,1)}-I_{14}^{(10,0,1)}.
\end{equation}
The choice of $t=0$ simply points to the fact that we do not constrain our function from the strong coupling data and as we discussed such functions for this choice are nothing but the FPR--like duality invariant functions as in (\ref{eq:FPR_wo_gravity}). Here the first function simply generates the weak coupling perturbative terms, the second one generates the $\mathcal{O}(g^4e^{-\frac{2\pi}{g}})$ with right power of $N$ and the third one compensates the extra terms generated in the process, for details and exact forms we refer the readers to  Appendix \ref{inst}. It is the combined effect of the last two functions that enables us to get the right instanton contribution at the weak coupling.\footnote{Although the numerical value of the instanton correction  is exponentially small in the large $N$ limit, nevertheless it plays a crucial role in restoring the S--duality and studying AdS/CFT beyond the planar limit  \cite{Korchemsky:2017ttd}.  In principle one could also constrain the interpolating function with the non--planar and strong coupling data over this basic construction including the instanton corrections, but we find  that  the leading instanton sector  gives very small correction to the interpolation (see Appendix \ref{inst} for details).}

%%%%%%%%%%%%%%%%%%%%%%%%%%%%%%%%%%%%%%%%%%%%%%%%%
%%%%%%%%%%%%%%%%%%%%%%%%%%%%%%%%%%%%%%%%%%%%%%%%%
\subsection{Finite spin twist--two anomalous dimensions}

In this subsection we consider the twist--two anomalous dimension with finite spin dependence. There is an essential difference between large $N$ and finite $N$ cases. 
In the planar limit the single--trace twist--two operators have well--defined anomalous dimension to all range of 't Hooft coupling.
On the other hand, at finite $N$, we have to be cautious in our program at strong coupling. For Konishi, it is clearly known that while the weak coupling anomalous dimension starts at $2+\mathcal{O}(\lambda)$ and grows to $\mathcal{O}(\lambda^{1/4})$ in the strong coupling, the leading twist operator in strong coupling are the double--trace twist--four operator as \eqref{Dtrace}, which is protected from corrections in the planar limit. This indicates that their anomalous dimensions will cross each other at certain finite coupling. The operator mixing effect would indicate that Konishi should not be \textit{self~S--dual} since the strong coupling description changes the operator altogether.  The behaviour for higher spin operators could be more complicated. Thus for modular invariant interpolating function at finite $N$, one expects that there is an operator mixing behaviour such that there is change of operators from weak to strong coupling, such as from single--trace twist--two operators to double--trace twist--four operators.

The above consideration leads us to the following choice of physical data for constructing the modular invariant interpolating function.
\begin{itemize}
\item 
At weak coupling,  the operator corresponds to (single--trace) twist--two operators, and the anomalous dimensions are given by the gauge theory computation as \eqref{finspindata}. We use the perturbative twist--two result up to three loops from (\ref{finspinres}). Note that the four--loop non--planar corrections have been computed for twist--two operators up to spin 8.  
\item
At strong coupling, we will assume the dominant contribution comes from the double--trace twist--four operators of the schematic form  $[\textrm{tr}(\phi^2)D^j \textrm{tr}(\phi^2)](x)$, and we use the results from supergravity dual picture as quoted in eq.~(\ref{eq:result_gravity}). It is indeed true that one could have other possible tower of multi trace operator eigenstates in the strong coupling side, but their contributions are expected to be suppressed in $\mathcal{O}(\frac{1}{N})$. We explicitly focus on this set of double trace operator as we hope this approximation may provide some qualitative picture for the physics.\footnote{In general for a multi trace operator with twist $2t$ and spin $j$,  one could have $(t-1)$ degenerate double trace operators of the form $\mathcal{O}_t \square^n D^j \mathcal{O}_t$ with $n=t-2,t-3...0$. For our case, we restrict to twist-four double trace operators.} 
\item
The non--zero instanton correction is known  for Konishi operator \cite{Alday:2016tll}. For operators with higher spin $(j>2)$ the leading instanton correction vanishes \cite{Alday:2016jeo}. In our study of interpolating functions for operators with finite spin we will not consider the instanton corrections. 

\end{itemize}
%Note that in this picture, we have neglected the mixing effect between different spins, by assuming the mixing between different spin is weak.
 Now, with all these ingredients, we go forward to study the modular invariant function as a function of both spin and gauge coupling and expect this provides a qualitative approximation to the true physics.

From three loops in weak coupling and the supergravity result, we can see that while the weak coupling starts at $\mathcal{O}(g)$, there is no $g$ dependence in the strong coupling. So following the construction we presented in section \ref{cinfnc}, we can again restrict ourselves to interpolating functions with $t=0$, i.e. which reduces to FPR--like functions of (\ref{eq:FPR_wo_gravity}). Note here, that in contrast to cases in the previous subsection, we have  $\mathcal{O}(\frac{1}{N^2})$ corrections in the strong coupling side. Here, our discussion presented in section \ref{intdualitysect} on $\mathcal{O}(\frac{1}{N^2})$ correction  comes into play. For this case  the interpolating function does not need to match non--trivial powers on both sides and is simply given by 
	%	The finite spin interpolating function for $s=10, \, m=4 $ and $ \alpha =1 $ reads, 
	\begin{equation}\label{intfins}
	I_4^{(10,0,1)}=F_4^{(10,1 )} (\tau ) 
	= \Biggl[ \frac{\sum_{k=1}^2 c_k E_{10+k} (\tau )}{\sum_{k=1}^3 d_k E_{10+k} (\tau )} \Biggr] ,
	\end{equation}
	where the coefficients $c_k$ and $d_k$ are functions of spin $j$ and the details can be found in appendix \ref{finsd}. 
	
Features of this function, namely the presence of extremas in $\tau$ plane, has been plotted in figures 8 and 9 for simplicity. One could actually see here that the peak values of the function on the whole $\tau$ plane appears respectively at  two points $\tau = i$ and $\tau = e^{i\pi/3}$, which turn out to be two special values of $\tau$ in \eqref{tst} invariant under $\mathbf{S}$--transformation and $(\mathbf{T}\cdot\mathbf{S})$--transformation.
	
%%%%%%%%%%%%%%%%%%%%%%%%%%
\begin{figure}[!tb]\label{sugra11}
	\centering
		\includegraphics[width=.8\linewidth]{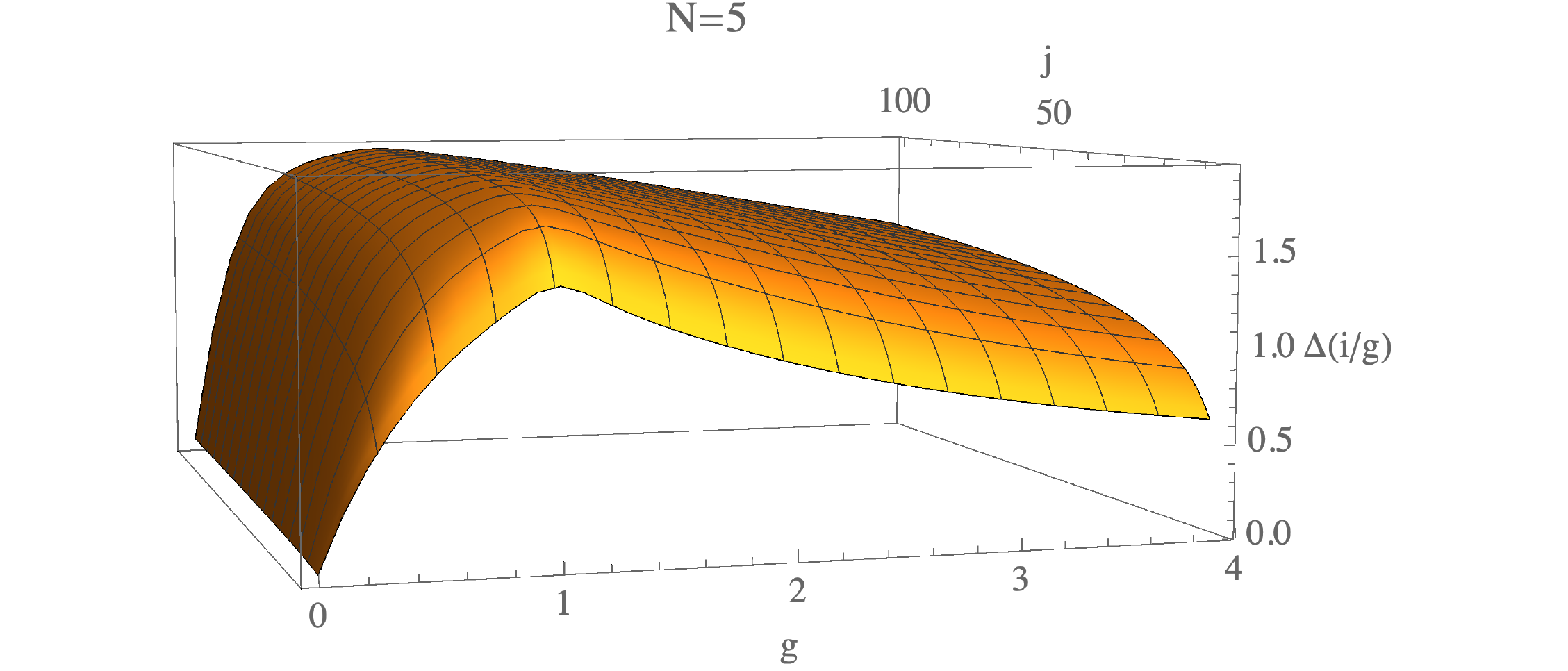}
	\caption{$g-j$ plot for the interpolating function \eqref{intfins} for $N=5$ for $\theta=0$ and $\tau=i/g$. One can notice here, since the supergravity data falls like $\mathcal{O}(\frac{1}{j^2})$ for a fixed $N$, the shape of the curve almost does not change after some initial values of $j$. The interpolating function has peak value at $g=1$ which corresponds to one of the duality invariant point $\tau=\tau_{S}$.  }\label{sfinj}
\end{figure}
%%%%%%%%%%%%%%%%%%%%%%%%%%
%%%%%%%%%%%%%%%%%%%%%%%%%%
\begin{figure}[!tb]\label{sugra22}
	\centering
		\includegraphics[width=.8\linewidth]{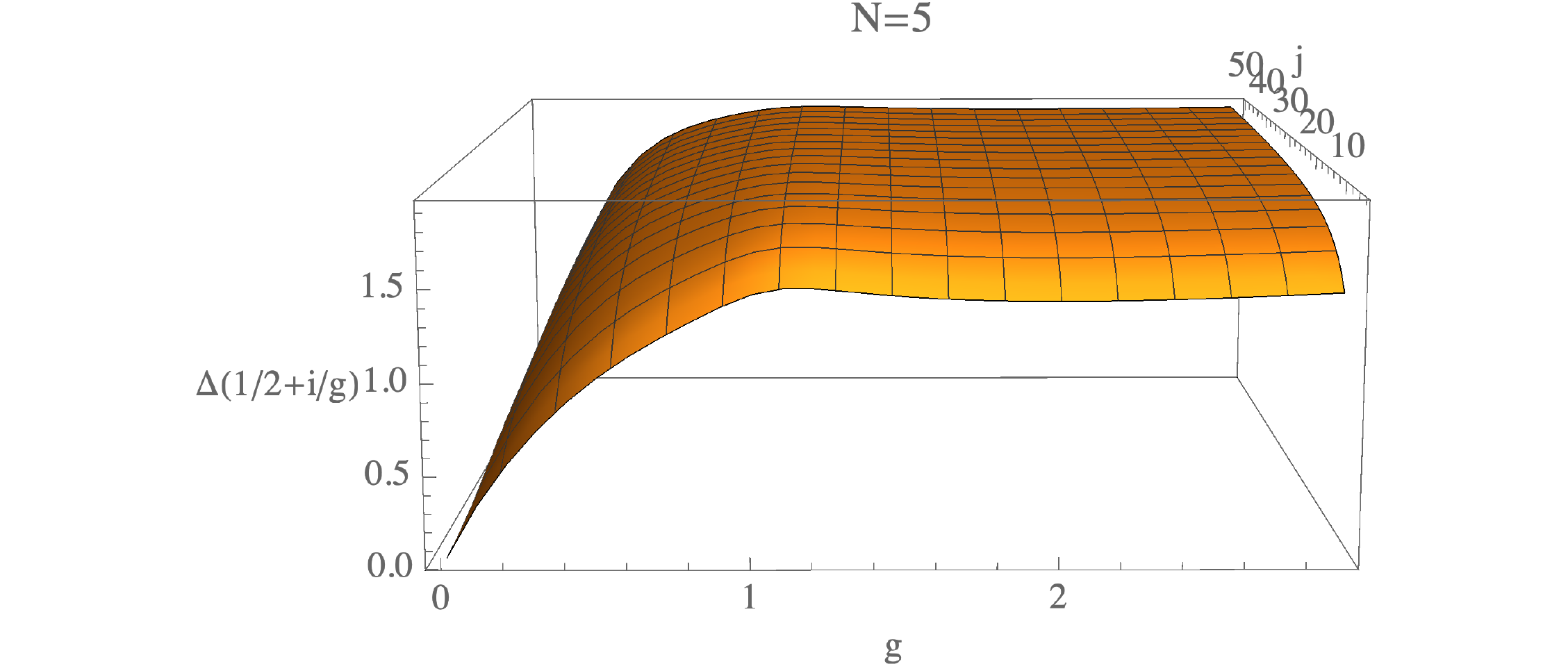}
	
	\caption{$g-j$ plot for the interpolating function \eqref{intfins} for $N=5$ for $\theta=\pi$ and $\tau=1/2+i/g$. The interpolating function has peak value at $g=\frac{2}{\sqrt{3}}$ which corresponds to one of the duality invariant point $\tau=\tau_{TS}$. }\label{tsfinj}
\end{figure}
%%%%%%%%%%%%%%%%%%%%%%%%%%

%%%%%%%%%%%%%%%%%%%%%%%%%%%%%%
%%%%%%%%%%%%%%%%%%%%%%%%%%%%%%
\subsection*{On level crossing} 
\label{sec:levelcross}

We can consider the phenomena of level crossing between the single--trace and the double--trace operators of the same spin using our formalism. This in particular means that  as we increase the coupling $g$ the dimension of the leading twist operator increase while the dimension of the subleading twist operator decreases. Therefore, it is possible that for some finite value of the coupling the dimension of both the leading and subleading operators becomes equal and they cross over. 

As an attempt to probe this, we consider two interpolating functions, for purely single--trace operators and purely double--trace operators respectively, and we study their crossing.

At first, consider the function which gives out weak and strong coupling data for twist--four double trace operators in the finite $N$ case.  The strong coupling data for such an operator with leading order corrections in $\mathcal{O}(\frac{1}{N})$ can be simply written from the supergravity approximation,
\begin{equation}
\Delta_{{\rm DT,sugra}}(j,N)=2-\frac{96}{(j+1) (j+6) N^2} \,.
\end{equation}
In the weak coupling, since not much is known for leading order correction to anomalous dimensions of double trace operators of general spin,\footnote{To our best knowledge, anomalous dimension of only spin--0 operator to the one--loop correction exist in literature \cite{Beisert:2003tq,Arutyunov:2002rs}. This one-loop effect was taken into account in the study of interpolating functions for crossing involving only spin-0 operators in \cite{Chowdhury:2016hny}.}
we will naively just consider the bare dimension of such operators and construct a simple interpolating function which reproduces the correct leading order results
\begin{equation}\label{subtw}
F_{{\rm DT}}(g,j,N)=\frac{g \left(2-\frac{96}{(j+1) (j+6) N^2}\right)+2}{g+1} \,.
\end{equation}

The other case we need is a function defined in finite $N$ that mimics the behaviour of finite spin twist--two operators at both ends of the spectrum. We had already constructed such a function in section \ref{sec:planarcase} using Pad\'e approximants for large $N$ case. As further naive approximation, one could demand that such a function could be valid also for smaller $N$ as in large $N$ one would not expect any crossing.

\begin{figure}[!tb]
	\centering
	\begin{minipage}{.5\textwidth}
		\centering
		\includegraphics[width=1.0\linewidth]{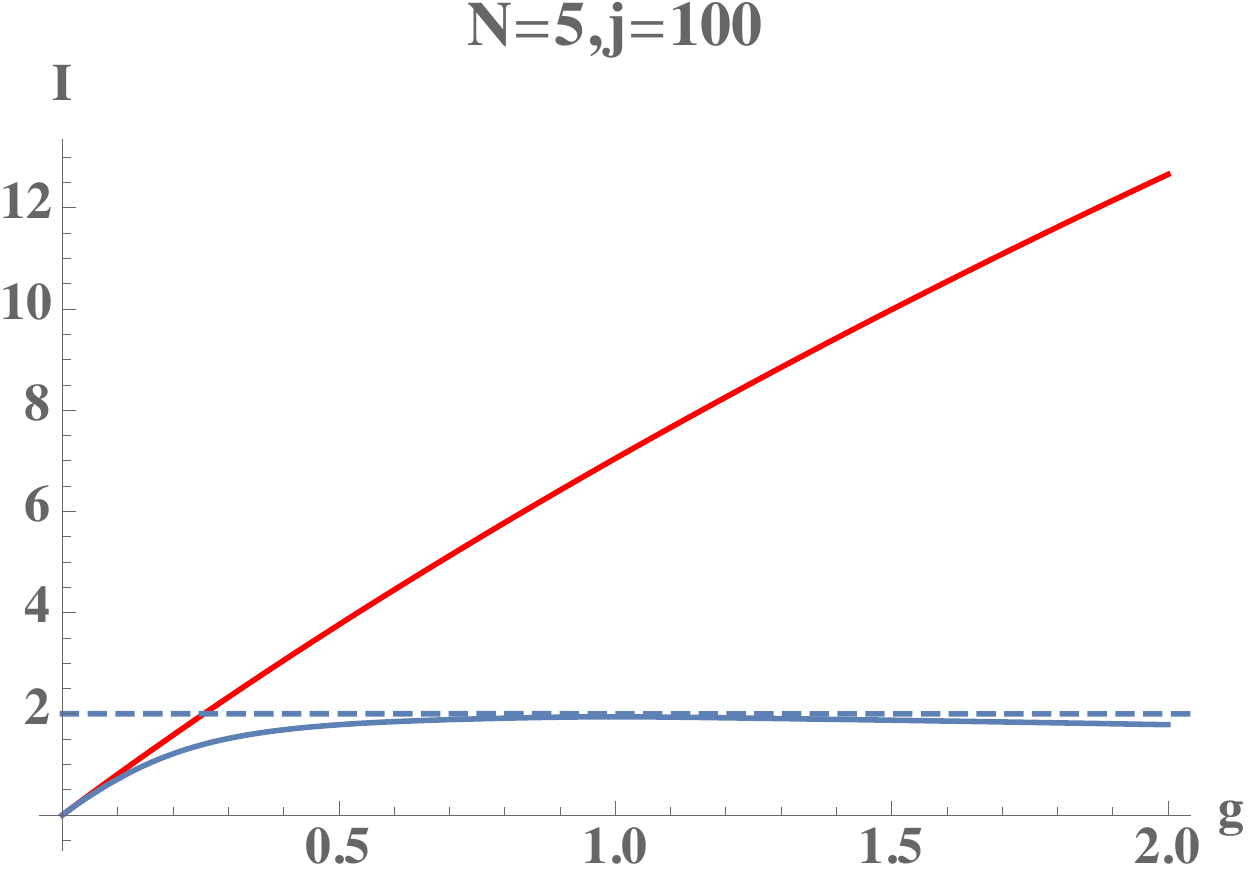}
		
	\end{minipage}%%%%%%
	~~
	\begin{minipage}{.5\textwidth}
		\centering
		\includegraphics[width=1\linewidth]{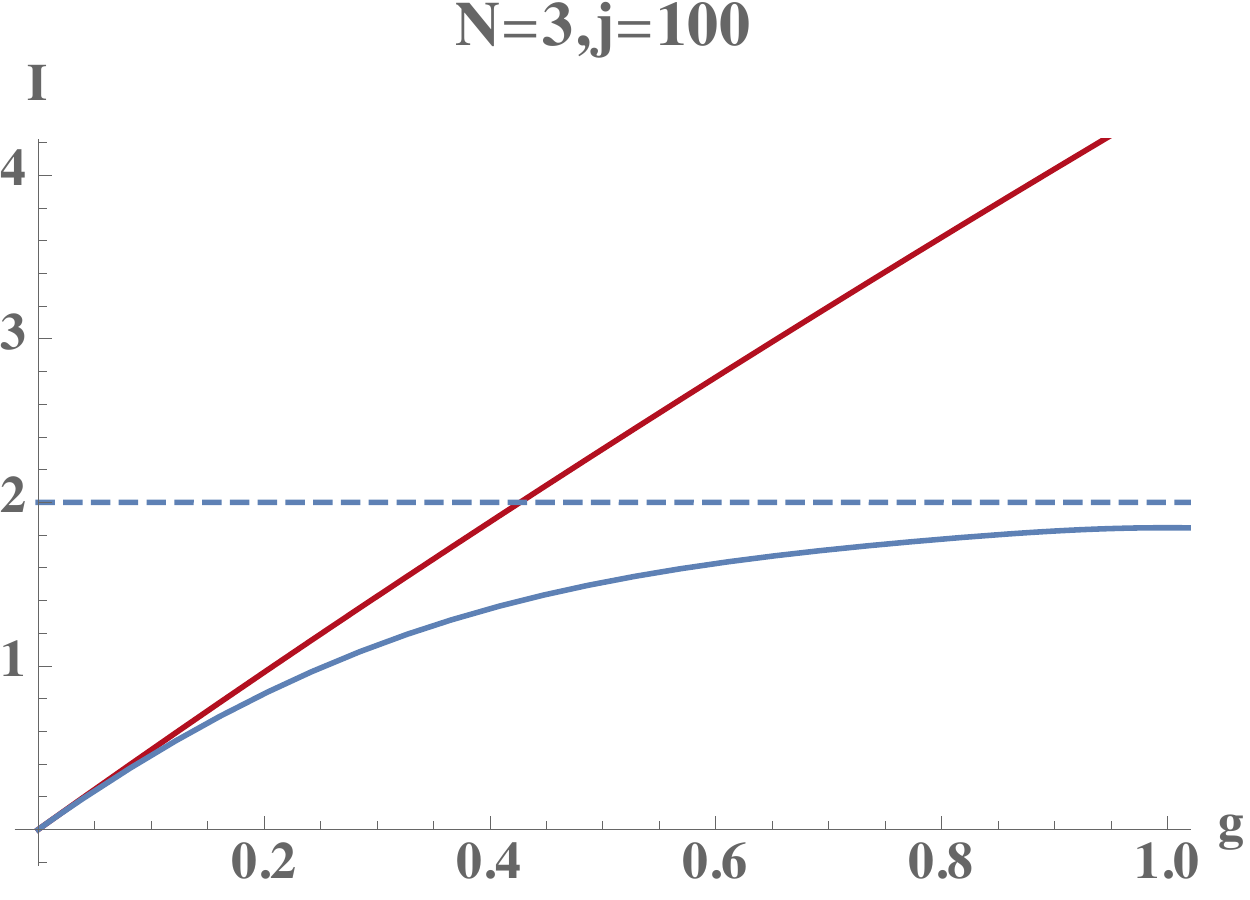}
		
	\end{minipage}
	\captionof{figure}{Dimensions of leading and subleading twist operators at finite  values of $N=5$ and $N=3$. The crossing point is given by the intersection of the red curve (Pad\'e for finite spin twist--two) and the blue dashed curve (Pad\`e for twist--four). The solid blue line indicates the modular invariant interpolating function (\ref{intfins}).}	\label{fig:levelcross}
\end{figure}

We can compare the anomalous dimension of the single--trace and double--trace operators, as shown in figure \ref{fig:levelcross}. We observe that the anomalous dimensions of the twist--two and twist--four operators do cross--over at some finite value of the coupling $g$ (red and the blue dashed curve in figure  \ref{fig:levelcross}) which marks the onset of level crossing in this case.  Hence, in such a physical crossing over between the dimensions, the interpolating functions are reliable only up to the crossing region as the dimensions themselves change their behaviour in the vicinity of such point.

%%%%%%%%%%%%%%%%%%%%%%%%%%
%%%%%%%%%%%%%%%%%%%%%%%%%%

Our modular invariant interpolating function \eqref{intfins}, on the other hand, explicitly takes single--trace twist--two anomalous dimension in weak coupling to double--trace twist--four anomalous dimension in strong coupling, thus avoiding the crossing of anomalous dimensions. This non--crossing of the dimensions is consistent with the Wigner--von Neumann no--crossing rule, where the dimensions of the new eigenstate (due to operator mixing) would repel each other. 
The mechanism for such non--crossing is precisely due to the non-planar corrections as studied in  \cite{Korchemsky:2015cyx}.

\begin{figure}[!tb]
	\centering
	\begin{minipage}{.5\textwidth}
		\centering
		\includegraphics[width=1.0\linewidth]{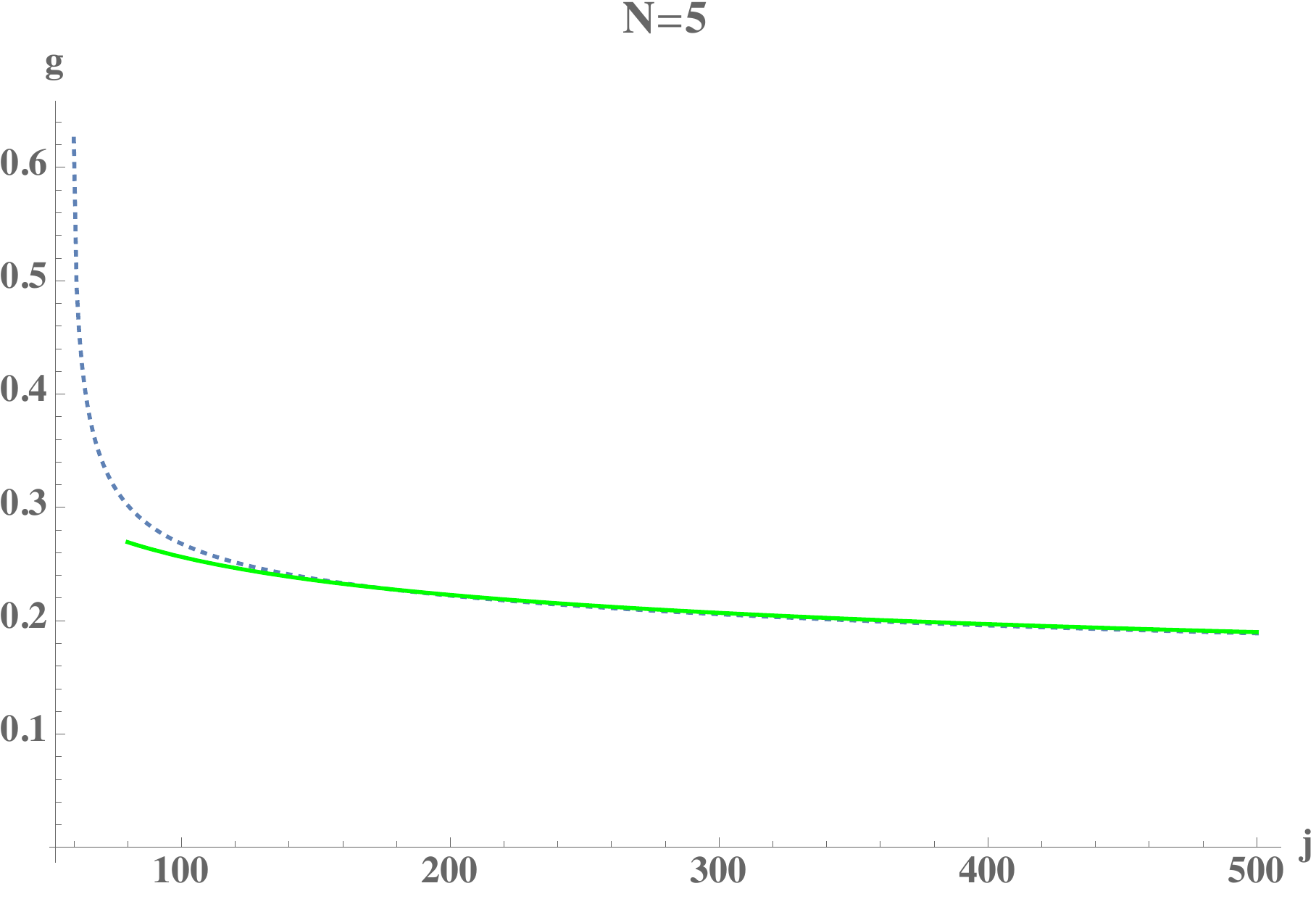}
		
	\end{minipage}%%%%%%
	~~
	\begin{minipage}{.5\textwidth}
		\centering
		\includegraphics[width=1\linewidth]{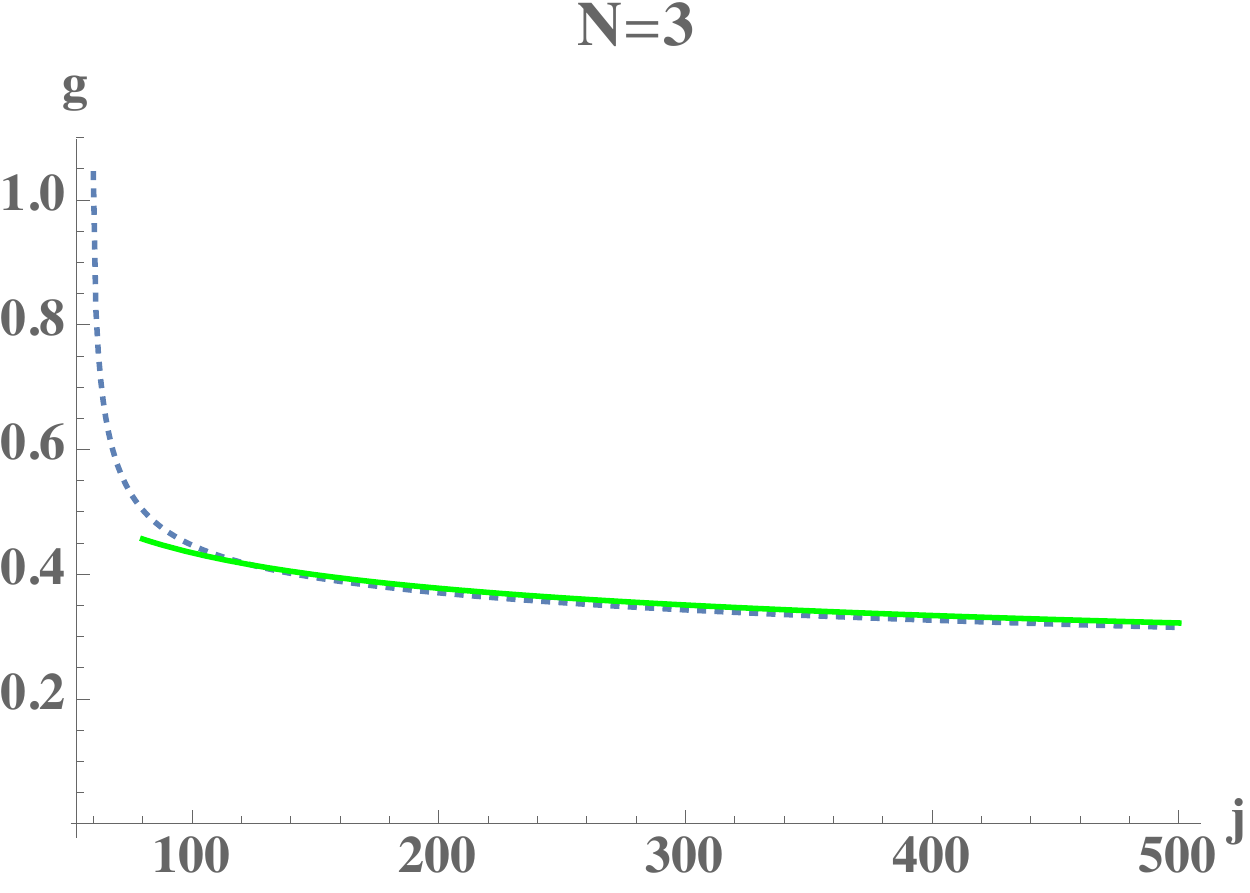}
		
	\end{minipage}
	\captionof{figure}{Crossing point for the dimensions of leading and subleading twist operators at finite  values of $N=5$ and $N=3$ as a function of $j$. The crossing point is given by the intersection of the (Pad\'e for finite spin twist--two) and the  (Pad\`e for twist--four). The solid curve is for the approximate fitting function $\frac{A}{\log j}$ for $ A=1.18 \, (N=5)$ and $A=2 \, (N=3)$.}	\label{fig:lvllr}
\end{figure}

Let us consider the spin dependence of the crossing points. At large spin the  anomalous dimension the crossing is expected to happen at lower values of the coupling. In large spin the anomalous dimensions  of leading twist operators grows logarithmically as
\begin{equation}
\gamma(j)\sim \Gamma_{\rm cusp,w}\log (j) \,.
\end{equation}
On the other hand, for sufficiently large spin the  anomalous dimension of subleading twist operator goes as
\begin{equation}
\Delta_{DT,sugra}(j,N)= 2-d(N)\frac{1}{j^2} \,.
\end{equation}
Hence, at some value of the coupling constant of order 
\begin{equation}
\label{crosg}
g \sim \frac{2\pi}{N\log j} \,,
\end{equation}
the dimensions of the leading and the subleading twist operators will cross over. As we increase the spin of the operator the cross--over will happen at lower values of the coupling \cite{Alday:2013bha}. In figure \ref{fig:lvllr} we have plotted the crossing point for large spin limit and fitted with approximate fitting function $$g\sim\frac{A}{\log j}.$$ We observe that as we increase the spin of the operator the crossing happens at lower values of coupling and the coefficient of the approximate fitting function is close to the $\frac{2\pi}{N}$.
For small spin ($j\sim \rm{coupling}$) the cross over between the 
 anomalous dimension of the single trace operator \eqref{deltaki} and  the double trace operator in the strong coupling(supergravity limit) occurs at the strong coupling. As we increase the spin of the operator the cross over shifts towards the weak coupling end.

\begin{figure}[!tb]
	\centering
	\begin{minipage}{.5\textwidth}
		\centering
		\includegraphics[width=1.0\linewidth]{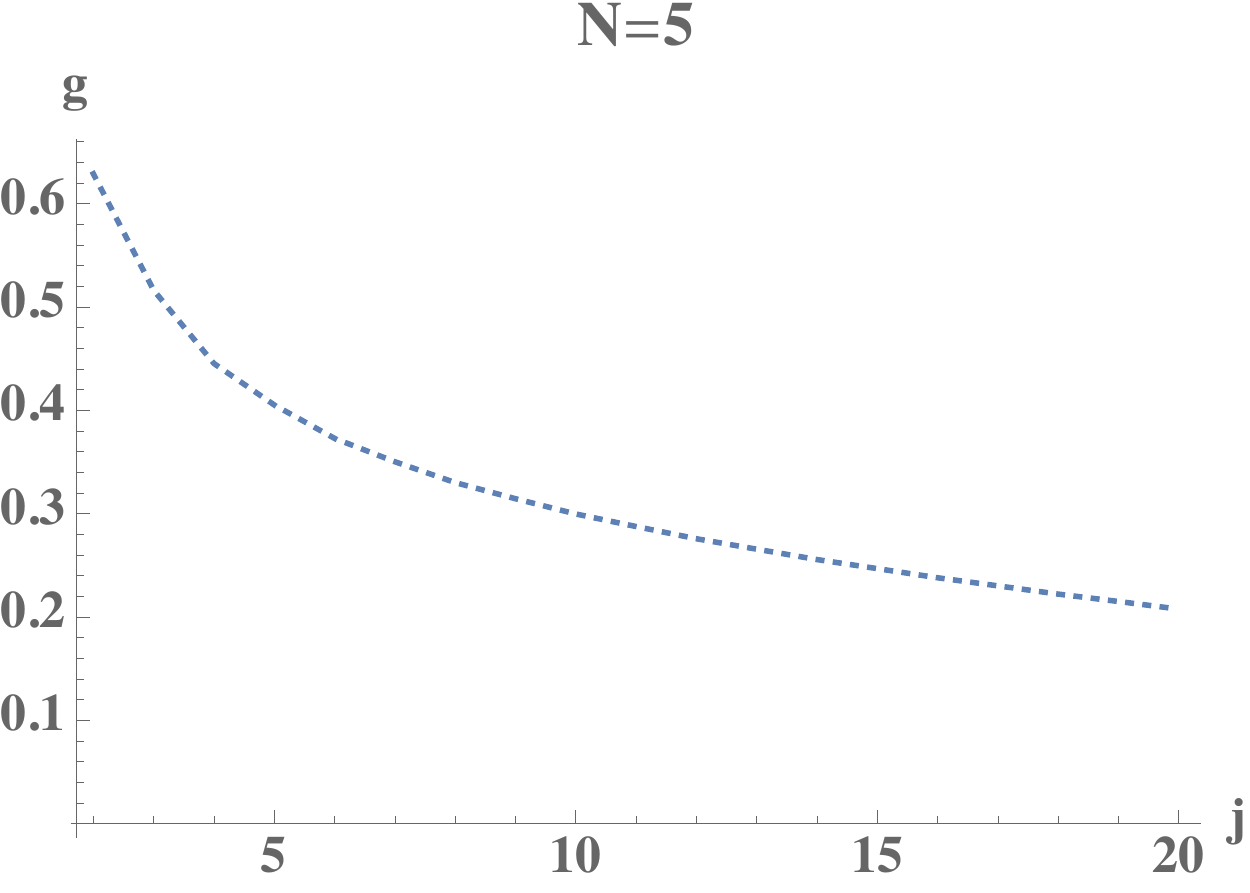}
		
	\end{minipage}%%%%%%
	~~
	\begin{minipage}{.5\textwidth}
		\centering
		\includegraphics[width=1\linewidth]{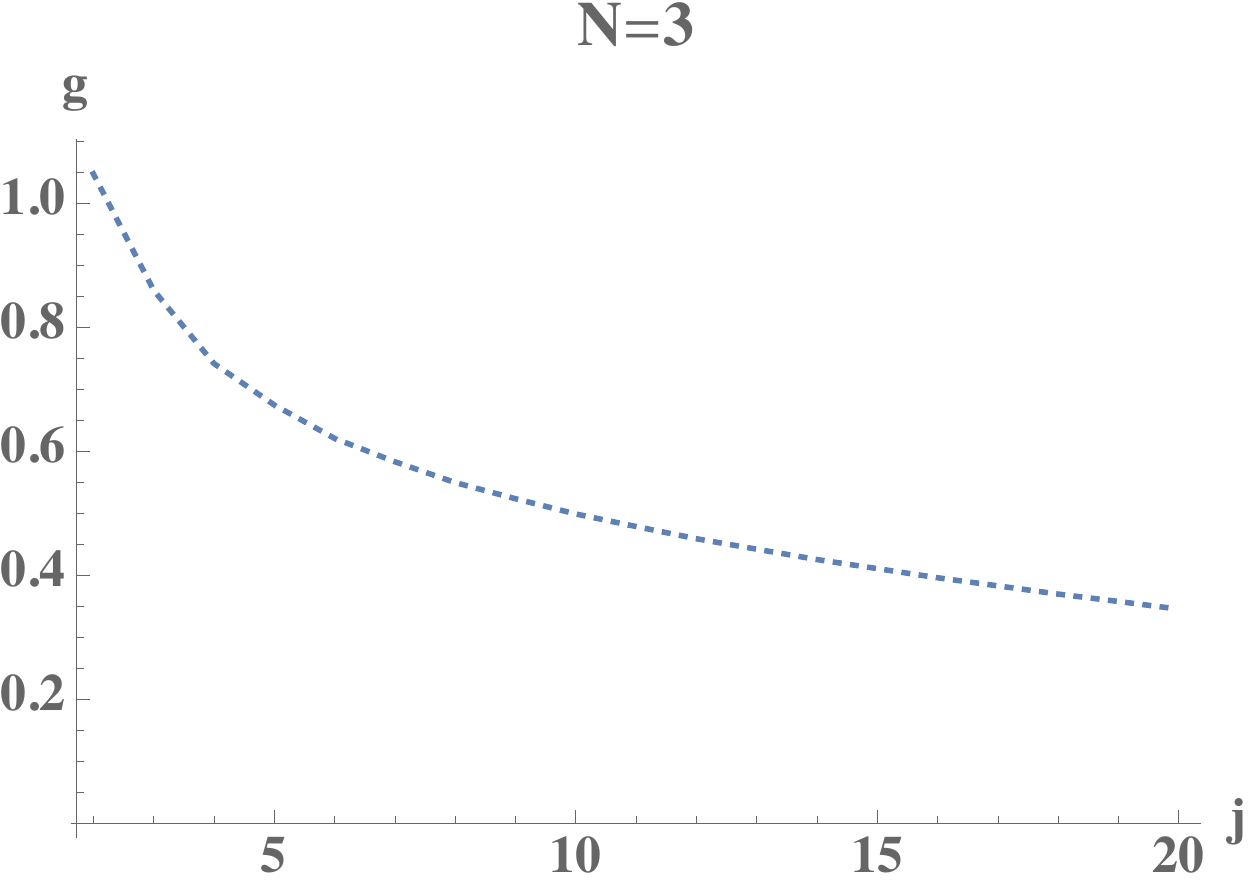}
		
	\end{minipage}
	\captionof{figure}{Crossing point for the dimensions of leading and subleading twist operators at finite  values of $N=5$ and $N=3$ as a function of $j$. The crossing point is given by the intersection of the (Pad\'e for finite spin twist--two) and the  (Pad\`e for twist--four). }	\label{fig:levelcrossgj}
\end{figure}

In figure \ref{fig:levelcrossgj} we have presented  plot of the crossing point for the anomalous dimension of the leading and the subleading twist operators as a function of $j$ for finite but smaller spin. From figure  \ref{fig:levelcrossgj} we observe  that as we increase the spin of the operators the crossing happens for lower value of the coupling. Hence, for very large spin of the operator the crossing occurs in the weak coupling and  there is maximal mixing of the operator in the large spin limit. 

%%%%%%%%%%%%%%%%%%%%%%%%%%%%%%%%%%%%%%%%%%%%%%%%%
%%%%%%%%%%%%%%%%%%%%%%%%%%%%%%%%%%%%%%%%%%%%%%%%%	
\subsection*{Upper bounds on anomalous dimensions and conformal bootstrap}

One immediate application of the finite $N$ interpolating function could in principle be in constraining the upper bound on the anomalous dimensions of leading twist operators for any arbitrary spin $j$. The $\mathcal{N}=4$ superconformal bootstrap approach \cite{Beem:2013qxa}
	has obtained some 
	upper bounds on the dimensions of the unprotected leading twist operators
	by studying the four-point function 
	(see also \cite{Alday:2013opa,Alday:2014qfa,Alday:2014tsa})
	\begin{equation}
	\langle \mathcal{O}_{\bf 20'}^{I_1}(x_1 ) \mathcal{O}_{\bf 20'}^{I_2}(x_2 )
	\mathcal{O}_{\bf 20'}^{I_3}(x_3 ) \mathcal{O}_{\bf 20'}^{I_4}(x_4 )\rangle ,
	\end{equation}
	where $\mathcal{O}_{\bf 20'}^{I}$ is
	a superconformal primary scalar operator of dimension two
	in energy-momentum tensor multiplets
	transforming as $\mathbf{20'}$ representation in $SU(4)_R$.
	The $\mathcal{N}=4$ superconformal symmetry
	allows us to describe the four--point function
	in terms of the $\mathcal{N}=4$ superconformal block \cite{Dolan:2001tt,Eden:2000bk,Arutyunov:2001mh,Eden:2001ec}.
	It is conjectured  in \cite{Beem:2013qxa} that
	the bound  on the anomalous dimension $\gamma(j)$ has a global maximum at one (or both) of the duality invariant points $\tau =\tau_S=i$ or $\tau =\tau_{TS}=e^{i \pi/3}$. 

	To compare our interpolating functions with results from $\mathcal{N}=4$ superconformal bootstrap, we have to investigate where the interpolating function takes its maximal value as a function of $\tau$. We can actually expect  that the extremal value of the interpolating function is given at either of the duality invariant points in the $\tau$ direction since the building blocks of the interpolating functions 
	have extremas at precisely these duality invariant fixed points.\footnote{Global maxima of $E_s (\tau )$ is given by $\tau =\tau_{TS}$.}  
	From figure \ref{sfinj}  we observe that the interpolating function has a peak at $g = 1$, which is one of the duality invariant points $\tau_s$. We also show in figure \ref{tsfinj} that for $\theta=\pi$ the interpolating function has maximum at $g=\frac{2}{\sqrt{3}}$, which  corresponds to the other duality invariant point $\tau=\tau_{TS}$.
	
\begin{figure}[tb]
		\centering

		\includegraphics[width=.61\linewidth]{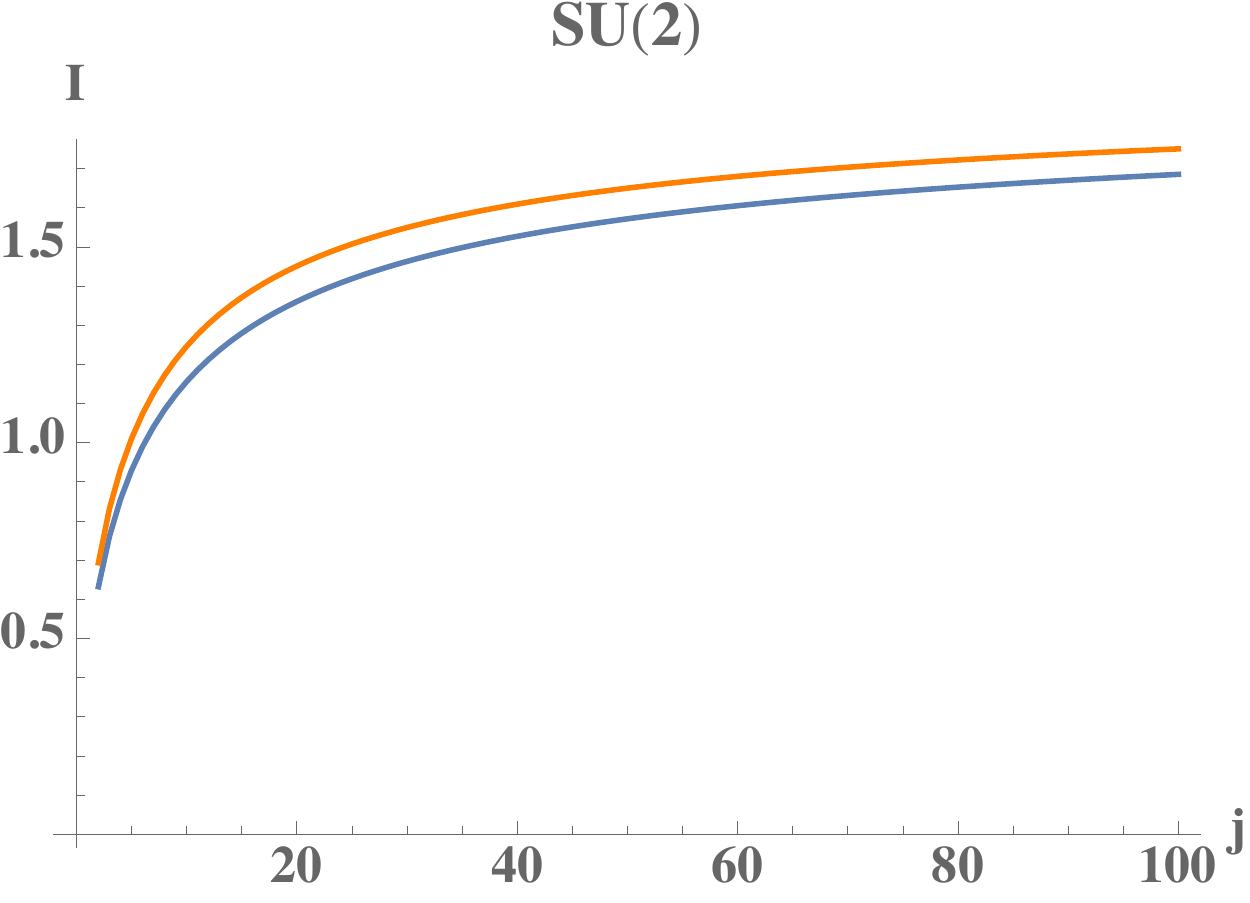}
		\caption{Maximal value of the interpolating function at the duality invariant points. The orange curve corresponds to the value at $\tau=\tau_{TS}$ and the blue curve corresponds to the value at $\tau=\tau_S$. }\label{upperbnds2}
\end{figure}

	  To predict an upper bound on the anomalous dimension it is important to find the global maxima of the interpolating function at either of this two duality invariant points.  In figure \ref{upperbnds2} we have plotted the interpolating function at these two duality invariant fixed points  for some arbitrary values of $j$ at a fixed $N=2$ and observe that the value of the interpolating function at $\tau=\tau_{TS}$ always stays larger than the value at $\tau=\tau_{S}$. It seems to be a generic feature for other values of $N$.
	
	With the observation that our interpolating function for the anomalous dimensions of leading--twist operators takes a maximum value at one of the duality invariant point $\tau=\tau_{TS}$, we could give a conservative prediction for the maximal value saturated by the anomalous dimension arising from the conformal bootstrap. In figure \ref{upperbnds2} we present this maximal value  of the anomalous dimension at $\tau=\tau_{TS}$ (orange) and $\tau=\tau_S$ (blue) for arbitrary $j$ (up to $j$=100). It should be noted that we just present a very crude approximation here based only on the maximal value of our interpolating function, which can be compared with rigorous results from superconformal bootstrap data when they become available.   

%%%%%%%%%%%%%%%%%%%%%%%%%%
%%%%%%%%%%%%%%%%%%%%%%%%%%

%%%%%%%%%%%%%%%%%%%%%%%%%%%%%%%%%%%%%%%%%%%%%%%%%
%%%%%%%%%%%%%%%%%%%%%%%%%%%%%%%%%%%%%%%%%%%%%%%%%
\section{Discussion}
\label{sec:discussion}

In this paper, we focus on one of the simplest classes of observables in ${\cal N}=4$ SYM, namely, the anomalous dimension of twist--two operators. We study their non-perturbative completion via interpolating function method, paying special attention to the dependence on the general spin parameter. 

These observables in large $N$ theory have been extensively studied in various aspects and are relatively well understood thanks to the AdS/CFT correspondence and integrability.
Based on the results both in the weak coupling and strong coupling regimes, we construct simple interpolating functions with generic spin and coupling dependence.
%In the planar large $N$ limit, there is no mixing between operators and the physics at strong coupling can be studied concretely.  
The interpolating function as a function of spin allows us to encode the cusp anomalous dimension as the large spin limit. 
Here a particular interesting aspect is the intriguing `transition' between small and large spin at strong coupling which is discussed in some details in section \ref{transition}.
When spin is small, the transition  can be described within supergravity where massive string excitations effectively decouple. In the large spin limit, the stringy effects play an important role. This shows that there is non--trivial rich physics even in the planar limit.
	
In the case of finite $N$, which is more close to the realistic QCD, the physical picture becomes significantly more complicated, mainly due to the operator mixing effect. The study in this case is more on the qualitative side. We apply the constraint from modular invariance, for which we use Eisenstein series as building blocks in our interpolating function.  We first consider the cusp anomalous dimension and construct the interpolating function by taking into account the four--loop non--planar result and the instanton contribution for the first time. Here we also solve a few technical challenges of the construction, such as correctly reproducing the strong coupling expansion and encoding the instanton contribution. We also provide a prediction to five--loop non--planar result in \eqref{eq:fiveloopNPprediction} based on the interpolating function.

We then focus on twist--two anomalous dimension with a finite spin parameter. Since the data at finite $N$ is very limited and the operator mixing may in principle be rather complicated, we have to make some assumptions to simplify the picture. Concretely, we consider a spin dependent  modular invariant function such that at weak coupling it is given by the single--trace twist--two operators $\textrm{tr}(\phi D^J \phi)$, while at strong coupling, it is dominated by the double trace twist--four operators $\textrm{tr}(\phi^2)D^J \textrm{tr}(\phi^2)$.
We make this approximation by assuming that the mixing with other operators such as higher twists (traces) are sub dominant.

\begin{figure}[t]
		\centering
		\includegraphics[width=9cm]{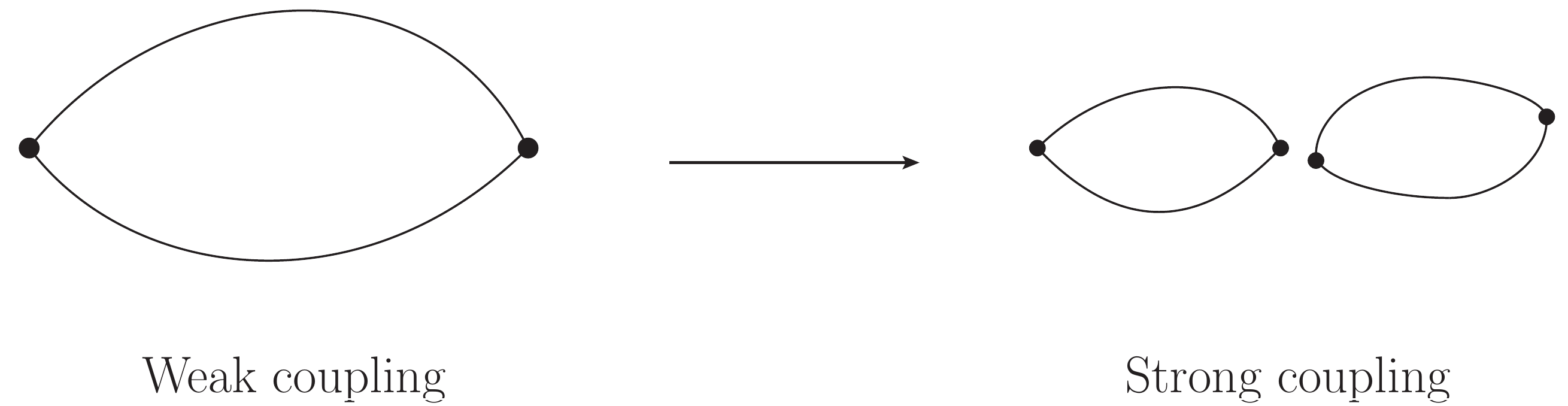}
		\caption{From the flux tube picture, at strong coupling, the energy is large enough to generate pairs of charged particles, so that a single trace operator at weak coupling splits into multiple smaller single traces. }
		\label{fig:trace_split_2} %% label for entire figure
\end{figure}

Let us comment on the relation between cusp anomalous dimension and finite spin operators. In planar limit, the large spin scaling behaviour as $\Gamma_{\rm cusp}\log j$ is true in both weak and strong coupling regimes. However, in the finite $N$ theory at strong coupling, one may expect such $\log j$ scaling is broken. In particular, in our modular invariant function the strong coupling expansion corresponds to double trace operators, which indeed have no $\log(j)$ scaling in the large spin limit. Physically, according to the flux tube picture discussed in \cite{Alday:2007mf}, the energy at strong coupling is large enough to generate pairs of color--charged particles, so that a single trace operator (at weak coupling) splits into multi--traces (at strong coupling), as illustrated in figure~\ref{fig:trace_split_2}.

As we have a connection to QCD in mind, we might hope the discussion in ${\cal N}=4$ SYM will provide certain qualitative picture of the physics of generic gauge theories.
Indeed, one may make an analogy with QCD: in the perturbative UV regime, fundamental degrees of freedom are partons (gluons or quarks) carrying color charges and a color singlet is given by a single trace operator; while in the IR non--perturbative regime, the fundamental degrees of freedom are themselves color singlets (i.e. single traces), such as pions or baryons in the effective Chiral Lagrangian theory for QCD. This is consistent with a single trace to multi-trace transition from weak to strong coupling that we discuss in ${\cal N}=4$ SYM.	
%%%%%%%%%%%%%%%%%%%%%%%%%%%%%%%%%%%%%%%%%%%%%%%%%%%%%%%%%%%%%%%
	%%%%%%%%%%%%%%%%%%%%%%%%%%%%%%%%%%%%%%%%%%%%%%%%%%%%%%%%%%%%%%%

	We have also analysed the phenomena of level--crossing between the leading twist--two and the sub--leading twist--four operators. A rigorous one loop computation of the anomalous dimension of  finite spin subleading twist operators might give us more insight into the phenomena of level--crossing.\\
	Finally, we mention that similar studies as of this paper may be applied to more general observables, such as operators with $\theta$--angle dependence or the OPE coefficients, as well as in other theories such as ABJM. 	We hope to address these questions in the future. 

\section*{Acknowledgments}

The work of AB, ST and GY is supported by the Chinese Academy of Sciences (CAS) Hundred-Talent Program, by the Key Research Program of Frontier Sciences of CAS, and by Project 11747601 supported by National Natural Science Foundation of China.
The work of AC is supported by a FWF grant with the number P 28552. AC would like to thank HRI Allahabad, JINR Dubna, Kings College London and ICTP Trieste for their hospitality during various stages of this work. AB would like to thank IIT Kanpur, ULB Brussels and CERN Theoretical Physics for kind hospitality during various stages of this work.

	\newpage
%%%%%%%%%%%%%%%%%%%%%%%%%%%%%%%%%%%%%%%%%%%%%%%%%
%%%%%%%%%%%%%%%%%%%%%%%%%%%%%%%%%%%%%%%%%%%%%%%%%
\appendix

%%%%%%%%%%%%%%%%%%%%%%%%%%%%%%%%%%%%%%%%%%%%%%%%%
%%%%%%%%%%%%%%%%%%%%%%%%%%%%%%%%%%%%%%%%%%%%%%%%%
\section{Construction of $G_{8/7}$}
\label{padfing}
In this appendix, we give the explicit construction of the finite spin interpolating Pad\'e approximant as discussed in section (\ref{padeconst}). The function in question has a form
	\be\label{padfin}
	G_{8/7} = \frac{\sum_{n=1}^8 c_n \tilde x^n}{1+ \sum_{n=1}^7 b_n \tilde x^n };~~~~\tilde x^2 = \tilde g.
	\ee
	The coefficients of the approximant are then given by, 
	\begin{tiny}
	\bea
	c_{1,2,3} &=&0,\\ \nonumber
	c_4 &=& w_0,\\ \nonumber
	c_5 &=& \tfrac{2 \sqrt{\pi } w_0 \left(-4096 \pi ^6 d_0^5 d_1 w_0+w_1 \left(-256 \pi ^4 d_0^3 \left(d_0^2 d_3-2 d_0 d_1 d_2+d_1^3\right)-w_0^2 \left(d_1^2-d_0 d_2\right) \left(d_1 d_3-d_2^2\right)\right)+16 \pi ^2 d_0 w_0^3 \left(d_0 d_1 d_3+d_0 d_2^2-2 d_1^2 d_2\right)\right)}{65536 \pi ^8 d_0^7+256 \pi ^4 d_0^2 w_0^2 \left(d_0^2 d_3-5 d_0 d_1 d_2+4 d_1^3\right)+d_2 w_0^4 \left(d_2^2-d_1 d_3\right)},\\  \nonumber
	c_6 &=& \tfrac{4 \pi  w_0^2 \left(-256 \pi ^4 d_0^3 w_0 \left(d_0 d_2-2 d_1^2\right)+16 \pi ^2 d_0 w_1 \left(d_0^2 d_1 d_3+d_0^2 d_2^2-4 d_0 d_1^2 d_2+2 d_1^4\right)+d_2 w_0^3 (d_1 d_2-d_0 d_3)\right)}{65536 \pi ^8 d_0^7+256 \pi ^4 d_0^2 w_0^2 \left(d_0^2 d_3-5 d_0 d_1 d_2+4 d_1^3\right)+d_2 w_0^4 \left(d_2^2-d_1 d_3\right)}, \\ \nonumber
	c_7 &=& \tfrac{8 \pi ^{3/2} d_0 w_0 \left(4096 \pi ^6 d_0^5 w_0+w_1 \left(256 \pi ^4 d_0^3 \left(d_1^2-d_0 d_2\right)+d_1 w_0^2 \left(d_2^2-d_1 d_3\right)\right)+16 \pi ^2 d_0 w_0^3 (d_0 d_3-2 d_1 d_2)\right)}{65536 \pi ^8 d_0^7+256 \pi ^4 d_0^2 w_0^2 \left(d_0^2 d_3-5 d_0 d_1 d_2+4 d_1^3\right)+d_2 w_0^4 \left(d_2^2-d_1 d_3\right)}, \\ \nonumber
	c_8 &=& \tfrac{16 \pi ^2 d_0 \left(32 \pi ^2 d_0 w_1 \left(128 \pi ^4 d_0^5+d_1 w_0^2 \left(d_1^2-d_0 d_2\right)\right)-512 \pi ^4 d_0^3 d_1 w_0^3+d_2^2 w_0^5\right)}{65536 \pi ^8 d_0^7+256 \pi ^4 d_0^2 w_0^2 \left(d_0^2 d_3-5 d_0 d_1 d_2+4 d_1^3\right)+d_2 w_0^4 \left(d_2^2-d_1 d_3\right)},\\ \nonumber
	b_1 &=& \tfrac{32 \pi ^{5/2} d_0 w_0 \left(w_0^2 \left(d_0 d_1 d_3+d_0 d_2^2-2 d_1^2 d_2\right)-256 \pi ^4 d_0^4 d_1\right)-2 \sqrt{\pi } w_1 \left(256 \pi ^4 d_0^3 \left(d_0^2 d_3-2 d_0 d_1 d_2+d_1^3\right)+w_0^2 \left(d_1^2-d_0 d_2\right) \left(d_1 d_3-d_2^2\right)\right)}{65536 \pi ^8 d_0^7+256 \pi ^4 d_0^2 w_0^2 \left(d_0^2 d_3-5 d_0 d_1 d_2+4 d_1^3\right)+d_2 w_0^4 \left(d_2^2-d_1 d_3\right)},\\ \nonumber
	b_2 &=&\tfrac{4 \pi  w_0 \left(-256 \pi ^4 d_0^3 w_0 \left(d_0 d_2-2 d_1^2\right)+16 \pi ^2 d_0 w_1 \left(d_0^2 d_1 d_3+d_0^2 d_2^2-4 d_0 d_1^2 d_2+2 d_1^4\right)+d_2 w_0^3 (d_1 d_2-d_0 d_3)\right)}{65536 \pi ^8 d_0^7+256 \pi ^4 d_0^2 w_0^2 \left(d_0^2 d_3-5 d_0 d_1 d_2+4 d_1^3\right)+d_2 w_0^4 \left(d_2^2-d_1 d_3\right)}  ,\\ \nonumber
	b_3 &=&\tfrac{8 \pi ^{3/2} d_0 \left(4096 \pi ^6 d_0^5 w_0+w_1 \left(256 \pi ^4 d_0^3 \left(d_1^2-d_0 d_2\right)+d_1 w_0^2 \left(d_2^2-d_1 d_3\right)\right)+16 \pi ^2 d_0 w_0^3 (d_0 d_3-2 d_1 d_2)\right)}{65536 \pi ^8 d_0^7+256 \pi ^4 d_0^2 w_0^2 \left(d_0^2 d_3-5 d_0 d_1 d_2+4 d_1^3\right)+d_2 w_0^4 \left(d_2^2-d_1 d_3\right)}  ,\\ \nonumber
	b_4 &=&\tfrac{\frac{16 \pi ^2 d_0 \left(32 \pi ^2 d_0 w_1 \left(128 \pi ^4 d_0^5+d_1 w_0^2 \left(d_1^2-d_0 d_2\right)\right)-512 \pi ^4 d_0^3 d_1 w_0^3+d_2^2 w_0^5\right)}{65536 \pi ^8 d_0^7+256 \pi ^4 d_0^2 w_0^2 \left(d_0^2 d_3-5 d_0 d_1 d_2+4 d_1^3\right)+d_2 w_0^4 \left(d_2^2-d_1 d_3\right)}-w_1}{w_0}  ,\\ \nonumber
	b_5 &=& \tfrac{2 \sqrt{\pi } \left(16 \pi ^2 d_0 w_1 \left(-256 \pi ^4 d_0^4 d_1+d_0 w_0^2 \left(d_1 d_3+d_2^2\right)-2 d_1^2 d_2 w_0^2\right)-256 \pi ^4 d_0^2 w_0^3 \left(d_0 d_2-4 d_1^2\right)-d_2 d_3 w_0^5\right)}{65536 \pi ^8 d_0^7+256 \pi ^4 d_0^2 w_0^2 \left(d_0^2 d_3-5 d_0 d_1 d_2+4 d_1^3\right)+d_2 w_0^4 \left(d_2^2-d_1 d_3\right)} ,\\ \nonumber
	b_6 &=& \tfrac{2 \sqrt{\pi } \left(16 \pi ^2 d_0 w_1 \left(-256 \pi ^4 d_0^4 d_1+d_0 w_0^2 \left(d_1 d_3+d_2^2\right)-2 d_1^2 d_2 w_0^2\right)-256 \pi ^4 d_0^2 w_0^3 \left(d_0 d_2-4 d_1^2\right)-d_2 d_3 w_0^5\right)}{65536 \pi ^8 d_0^7+256 \pi ^4 d_0^2 w_0^2 \left(d_0^2 d_3-5 d_0 d_1 d_2+4 d_1^3\right)+d_2 w_0^4 \left(d_2^2-d_1 d_3\right)} ,\\ \nonumber
	b_7 &=& \tfrac{8 \pi ^{3/2} \left(32 \pi ^2 d_0 w_1 \left(128 \pi ^4 d_0^5+d_1 w_0^2 \left(d_1^2-d_0 d_2\right)\right)-512 \pi ^4 d_0^3 d_1 w_0^3+d_2^2 w_0^5\right)}{65536 \pi ^8 d_0^7+256 \pi ^4 d_0^2 w_0^2 \left(d_0^2 d_3-5 d_0 d_1 d_2+4 d_1^3\right)+d_2 w_0^4 \left(d_2^2-d_1 d_3\right)} .\\ \nonumber
	\eea
		\end{tiny}
Here for simplicity, we have used the weak coupling expansion of the equation as
	\be
	D_w = w_0 \tilde g^2 + w_1  \tilde g^4 + w_2 \tilde g^6 + \mathcal{O}(\tilde g^8),
	\ee
	and the strong coupling data reads
	\be
	D_s =  d_0 (4\pi\tilde g)^\frac{1}{2} + d_1 (4\pi\tilde g)^{-\frac{1}{2}}  + d_2  (4\pi\tilde g)^{-\frac{3}{2}} + d_3  (4\pi\tilde g)^{-\frac{5}{2}} + \mathcal{O}(\tilde g^{-\frac{7}{2}}).
	\ee

%%%%%%%%%%%%%%%%%%%%%%%%%%%%%%%%%%%%%%%%%%%%%%%%%
%%%%%%%%%%%%%%%%%%%%%%%%%%%%%%%%%%%%%%%%%%%%%%%%%
\section{Construction of FPP's for finite spin}

We constructed FPP's to improve upon our discussion of Pad\'e approximants in large $N$ case in section \ref{fppconst}. Here we will give the explicit FPP functions for finite spin case, which read
	\begin{tiny}
	\be
	{F_{\rm small}}(j,x)=\tfrac{{w_0} x^4}{\sqrt[5]{\frac{5 {w_0}^5 x^{11} \left(3 {d_1}^2-{d_0} {d_2}\right)}{512 \pi ^{9/2} {d_0}^7}-\frac{5 {d_1} {w_0}^5 x^{13}}{128 \pi ^{7/2} {d_0}^6}+\frac{{w_0}^5 x^{15}}{32 \pi ^{5/2} {d_0}^5}-\frac{5 {w_0}^5 x^9 \left({d_0}^2 {d_3}-6 {d_0} {d_1} {d_2}+7 {d_1}^3\right)}{2048 \pi ^{11/2} {d_0}^8}+\frac{5 x^8 \left(3 {w_1}^2-{w_0} {w_2}\right)}{{w_0}^2}-\frac{5 {w_1} x^4}{{w_0}}+1}} .
	\ee
	\end{tiny}
	For Cusp, the function reads
	\be
	\Gamma_{\rm cusp}^{(F_{6,4})}(\tilde{g})=\tfrac{4\tilde{g}^2}{\sqrt[11]{2048 \tilde{g}^{11}+3727.82 \tilde{g}^{10}+3831.83 \tilde{g}^9+2946.55 \tilde{g}^8+1900.08 \tilde{g}^7+2421.26 \tilde{g}^6+452.411 \tilde{g}^4+36.1885 \tilde{g}^2+1}} \,.
	\ee

%%%%%%%%%%%%%%%%%%%%%%%%%%%%%%%%%%%%%%%%%%%%%%%%%
%%%%%%%%%%%%%%%%%%%%%%%%%%%%%%%%%%%%%%%%%%%%%%%%%
\section{Construction of S--duality invariant interpolation for cusp anomalous dimensions}
\label{resultsdcp}

%%%%%%%%%%%%%%%%%%%%%%%%%%%%%%%%%%%%%%%%%
%%%%%%%%%%%%%%%%%%%%%%%%%%%%%%%%%%%%%%%%%
\subsection*{With only weak coupling and non--planar corrections}

The cusp interpolating function with just the weak coupling data and $s=10$ is as follows:
	\begin{equation}\label{withst}
	I_1 = I_4^{(10,0,1)}=F_4^{(10,1 )} (\tau ) 
	= \Biggl[ \frac{\sum_{k=1}^2 c_k E_{10+k} (\tau )}{\sum_{k=1}^3 d_k E_{10+k} (\tau )} \Biggr] ,
	\end{equation}
	where the coefficients $c_k$ and $d_k$ reads
	\bea
	c_1 &=& \frac{657931 \left(756 N^2 \zeta (3)^2+10 \pi ^6 N^2+189   \frac{\Gamma^{np}}{8}\right)}{3230059140 \pi ^2}, \\ \nonumber
	c_2 &=& \frac{119743442 \pi  N}{1181820455},  \\ \nonumber
	d_1 &=& \frac{657931 \left(18900 N^2 \zeta (3)^2+124 \pi ^6 N^2+4725  \frac{\Gamma^{np}}{8} \right)}{969017742000}, \\ \nonumber
	d_2 &=& \frac{8553103 \left(3780 N^2 \zeta (3)^2+71 \pi ^6 N^2+945 \frac{\Gamma^{np}}{8}\right)}{21272768190 \pi ^3 N},  \\ \nonumber
	d_3 &=& 1 \,,
	\eea
	where $N$ is the order of the $SU(N)$ group and $\Gamma^{np}$ denotes the  non--planar correction \eqref{nonplan}. It can be put to zero if we don't want to include the correction.
	As a check of the construction, we can predict the data for the 5th--loop $\mathcal{O}\left(\lambda^5 \right)$ order,
\begin{equation}
\label{eq:fiveloopNPpredictionC3}
\Gamma^{(5)}_{{\rm cusp,w}}=\left (9765.01+\frac{57569.7}{ N^2}\right) \times \left(\frac{1}{4 \pi}\right)^{10} .
\end{equation} 
The $\gamma^{(5)}_{{\rm cusp,w}}$ predicted from our construction can be checked against the result from the BES equation\eqref{higherloop} \footnote{Multiplied by $\left(\frac{1}{4 \pi}\right)^{10}$ as we are working with a $\lambda$--expansion.} and our result is within $7.8\% $ error bar. 
In figure 15, we plot the above interpolating function (using weak coupling data only) for cusp anomalous dimension at $N=2$. Effect of adding non-planar data is also shown in the figure. 
	
	The strong coupling expansion is given by taking large $\tilde g$ expansion of \ref{withst}
	\begin{equation}
2.22105\, +\frac{24005}{(4\pi\tilde g)^4}-\frac{227.99}{(4\pi\tilde g)^2 }.
	\end{equation}
	\begin{figure}[tb]
		\centering
	\centering
			\includegraphics[width=.7\linewidth]{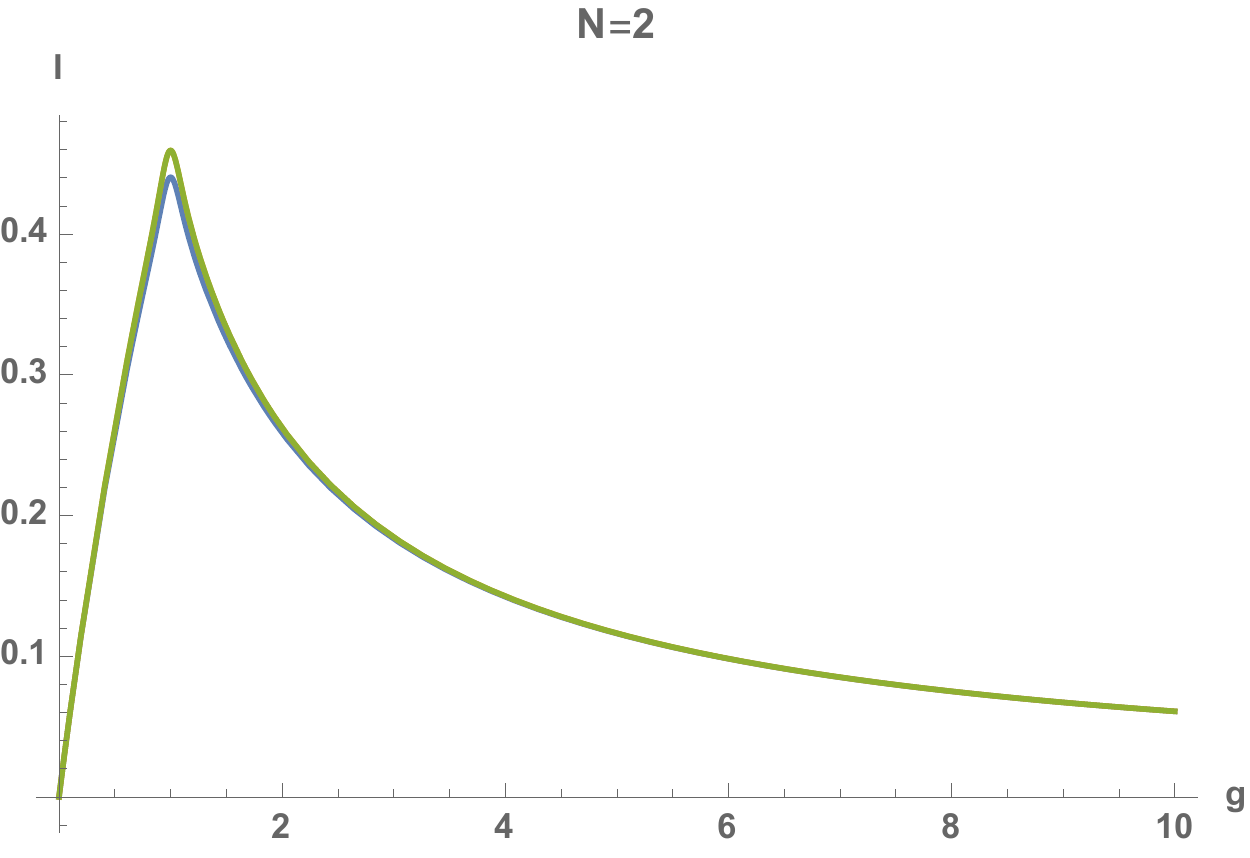}
			% \captionof{figure}{A figure}
			\label{fig:test2}
			\caption{Interpolating function with non--planar correction (green) and without non--planar correction (blue) with S--duality constraint on the weak coupling data only. }
			\end{figure}
This doesn't reproduce the correct expansion in the strong coupling limit \eqref{cuspst}. Hence in the main text we restrict to the construction of the interpolation function which is constrained both by the weak coupling and the strong coupling expansions.	
			
%%%%%%%%%%%%%%%%%%%%%%%%%%%%%%%%%%%%%%%%%%%
%%%%%%%%%%%%%%%%%%%%%%%%%%%%%%%%%%%%%%%%%%%
\subsection*{With weak coupling, strong coupling and non--planar corrections}
\label{apc.1}

	The infinite spin (cusp) interpolating function for $s=10$ and both weak and strong coupling data taken into account has two components:
	\begin{equation}
	I_5^{(10,-1,\frac{1}{2})}=F_5^{(10,\frac{1}{2} )} (\tau ) 
	= \Biggl[ \frac{ c_1 E_{10}(\tau ) + c_2 E_{12} (\tau )}
	{\sum_{k=1}^4 d_k E_{10+k} (\tau )} \Biggr]^{\frac{1}{2}} ,
	\end{equation}
	where the coefficients $c_k$ and $d_k$ reads
	\bea
	c_1 &=& 0.0116503 N^4, \\ \nonumber
	c_2 &=& \frac{47498922058 \pi ^2 N^2}{4626827081325},  \\ \nonumber
	d_1 &=& 0.131325 N^3+0.00201572 N \frac{\Gamma^{np}}{8}, \\ \nonumber
	d_2 &=& 0.0190294 N^2,  \\ \nonumber
	d_3 &=& \frac{3392780147 \pi ^3 N}{200912389470}, \\ \nonumber 
	d_4 &=& 1 \,,
	\eea
	and 
	\begin{equation}
	I_5^{(10,0,\frac{1}{2})}=F_5^{(10,\frac{1}{2} )} (\tau ) 
	= \Biggl[ \frac{ c_1 E_{11}(\tau ) + c_2 E_{12} (\tau )}
	{\sum_{k=1}^4 d_k E_{10+k} (\tau )} \Biggr]^{\frac{1}{2}} ,
	\end{equation}
	with
	\bea
	c_1 &=& 0.0086231 N^3, \\ \nonumber
	c_2 &=& \frac{47498922058 \pi ^2 N^2}{4626827081325},  \\ \nonumber
	d_1 &=& 0.0629537 N^3+0.00201572 N \frac{\Gamma^{np}}{8}, \\ \nonumber
	d_2 &=& -0.0513927 N^2,  \\ \nonumber
	d_3 &=& 0.608705 N, \\ \nonumber 
	d_4 &=& 1, \\ \nonumber
	w_1 &=&  (1-w_2) = 1.89422 \,.
	\eea

%%%%%%%%%%%%%%%%%%%%%%%%%%%%%%%%%%%%%%%%%%%%%%%%%
%%%%%%%%%%%%%%%%%%%%%%%%%%%%%%%%%%%%%%%%%%%%%%%%%

%%%%%%%%%%%%%%%%%%%%%%%%%%%%%%%%%%%%%%%%%%%%%%%%%
%%%%%%%%%%%%%%%%%%%%%%%%%%%%%%%%%%%%%%%%%%%%%%%%%
\section{Adding instanton corrections to cusp anomalous dimension}
\label{inst}
	
Here we discuss systematic inclusion of the  instanton corrections to cusp anomalous dimension. Note that the Eisenstein \eqref{eisendef} series used in the construction of the interpolating functions has in--built within itself an infinite series of non--perturbative terms taking the form $ e^{-\frac{2 \pi n}{g}} ($where $n \in \mathbb{Z}$) from the weak coupling expansion of the modified Bessel ($K$) terms. From section \ref{data1},  eq. \eqref{inss2} we see that we have to include such a correction at $ \mathcal{O}\left(g^{4} e^{-\frac{2 \pi}{g}}\right)$. 
	
	To begin with, from the construction of interpolating function mentioned in section 
	(\ref{fprsdual:})  and the weak coupling expansion (\ref{cuspdef1}), we notice that if an interpolating function with parameters $(\alpha,s)$ generates first weak coupling term at $g^{\beta}$, the we get the first instanton term having a form  $\mathcal{O}\left(g^{\min (p,q)+\beta+s} e^{-\frac{2 \pi}{g}}\right)$. As $min(p,q)$ is at least $1$, the choice for $s$ is very limited and thereby we lose the possible infinite class of modular invariant interpolating functions parametrized by $s$. Even if we construct a valid interpolating function say with  $ \alpha =\frac{1}{3}, \, m=4$ and $s=2$ where $p=1$ and $q=4$, the structure of the interpolating function is as
	\begin{equation}
	\left ( c_{1} X_{1} g + \cdots \right) + \left(c_{2} X_{1} g^{4} + \cdots \right) e^{-\frac{2 \pi}{g}}+ \left(\cdots \right) e^{-\frac{4 \pi}{g}} + \cdots \,,
	\end{equation}
	where $ c_{1} \, \&\, c_{2}$ are some numerical constants and $X_{1}$ is an unsolved coefficient always shared by  $\mathcal{O}(g)$ and $ \mathcal{O}\left(g^{4} e^{-\frac{2 \pi}{g}}\right)$ terms. Note that from weak coupling data $ X_1 \sim N$ but from the instanton data $X_1 \sim N^{-\frac{3}{2}}$ and hence it is impossible to consolidate both trends in a single interpolating function. As an alternative, we could have two interpolating functions $I_1 \, \& \, I_2$ with $I_1$ carrying a large $s$ such that the power of $g$ multiplying $\mathcal{O}(e^{-\frac{ 2\pi}{g}})$ is large and $I_2$ a minimal interpolating function used just to reproduce the $\mathcal{O}\left(g^{4} e^{-\frac{2 \pi}{g}}\right)$ term.
	
	To construct $I_2$ we only use the $\mathcal{O}\left(g^4 e^{-\frac{2 \pi}{g}}\right)$ data with $ \alpha =1, \, m=2 \, \& \, s=2$. The structure of the interpolating function turns out as follows,
	\begin{equation}
	\label{interinst}
	I_2=\frac{c_1}{N^\frac{3}{2}} g + \frac{c_2}{N^\frac{3}{2}} g^6 + \cdots +\left(\frac{c_3}{N^\frac{3}{2}} g^4 + \frac{c_4}{N^\frac{3}{2}} g^5 + \cdots \right) e^{-\frac{2 \pi}{g}} + \cdots
	\end{equation} 
	where $c's$ are some constants. Firstly, note that we have generated a set of extra weak coupling terms, but they are largely suppressed at reasonably large $N$ compared to the weak coupling data (\ref{cuspdef1}). Secondly, at $\mathcal{O}\left(e^{-\frac{2 \pi}{g}}\right) $ we have a non zero coefficient for $\mathcal{O} \left(\frac{g^5}{N^\frac{3}{2}}\right) $ which probably should have been at  $\mathcal{O}\left(\frac{g^5}{N^\frac{1}{2}}\right) $. Though we have no data to match at this order but if we had we could easily incorporate it either by creating another interpolating function with $s=3$ starting at  $\mathcal{O}\left(\frac{g^5}{N^\frac{1}{2}} \,  e^{-\frac{2 \pi}{g}} \right) $ or by allowing more unsolved coefficients in $I_2$. 
		Now, to remove the extra weak coupling terms from eq. (\ref{interinst}) up to some order in $g$ we could consider it as input weak coupling data for yet another interpolating function ($I_3$) with a large value of $s$ and finally subtract it from $I_2$. \\
	As an concrete example of construction and comparison with previous constructions, lets take $I_1$ with $ \alpha = 1, \, m=4 \, \& \, s=15$, $I_2$ as above and $I_3$ with $ \alpha =1, \, m=14 \, \& \, s=10$. Instanton corrections for both $I_1 \, \& \, I_3 $ contribute at $\mathcal{O}\left(g^{18} e^{-\frac{2 \pi}{g}}\right)$. Moreover, $I_3$ cancels the extra weak coupling terms from $I_2$ up to $\mathcal{O}\left(g^{28}\right)$. As instanton is non--perturbative, we could expect that including instanton data would be insignificant and this is clearly reflected in Table \ref{inst_comp} where the differences between the interpolating function with instanton correction ($I_1 +(I_2-I_3)$) and without the correction ($I_1$) are analysed for the  critical points of the difference function  at $N=2$, $N=20$ and $N=200$. It seems instanton correction is far less significant than the strong coupling corrections or the non--planar corrections to the modular invariant interpolating functions. 
	%and hence we will not pursue it any further.
	\begin{table}[tb]
	
		\begin{subtable}{.331\linewidth}
			\centering
			\caption{$N=2$}
			\begin{tabular}{|c|c|}
				\hline
				$g$ & $I_2-I_3$ \\
				\hline
				$0.09654$ & $3.328 \times 10^{-35}$\\
				\hline
				$0.6074$ & $-2.911 \times 10^{-9}$ \\
				\hline
				$1.000$ & $ 6.124 \times 10^{-6}$ \\
				\hline
				$ 1.646$ & $-2.911 \times 10^{-9}$	\\
				\hline
				$ 7.051 $ & $ 2.376 \times 10^{-7}$ \\
				\hline
				$ 30.60$ & $ -5.672 \times 10^{-7}$\\ \hline	
			\end{tabular}
		\end{subtable}%
		\begin{subtable}{.331\linewidth}
			\centering
			\caption{$ N=20$}
			\begin{tabular}{|c|c|}
				\hline
				$g$ & $I_2-I_3$ \\
				\hline
				$0.09654$ & $1.626 \times 10^{-36}$\\
				\hline
				$0.6074$ & $-1.422 \times 10^{-10}$ \\
				\hline
				$1.000$ & $ 2.992 \times 10^{-7}$ \\
				\hline
				$ 1.646$ & $-1.422 \times 10^{-10}$	\\
				\hline
				$ 7.051 $ & $ 1.161 \times 10^{-8}$ \\
				\hline
				$ 30.60$ & $ -2.772 \times 10^{-8}$\\ \hline	
			\end{tabular}
		\end{subtable} 
		\begin{subtable}{.331\linewidth}
			\centering
			\caption{$ N=200$}
			\begin{tabular}{|c|c|}
				\hline
				$g$ & $I_2-I_3$ \\
				\hline
				$0.09654$ & $5.295 \times 10^{-38}$\\
				\hline
				$0.6074$ & $-4.631 \times 10^{-12}$ \\
				\hline
				$1.000$ & $ 9.742 \times 10^{-9}$ \\
				\hline
				$ 1.646$ & $-4.630 \times 10^{-12}$	\\
				\hline
				$ 7.051 $ & $ 3.780 \times 10^{-10}$ \\
				\hline
				$ 30.60$ & $ -9.024 \times 10^{-10}$\\ \hline	
			\end{tabular}
		\end{subtable} 
		\caption{The difference between including the instanton correction and not including it at some critical values of $g$. We would like to highlight that even at the worst case of $ N=2 \, \& \, g=1 $ the difference is $\mathcal{O}\left(10^{-6}\right)$. }		\label{inst_comp}
	\end{table}

	The planar interpolating functions (for $\Gamma^{np} =0$) and only the weak coupling data have been used in the construction of the instanton corrected modular invariant interpolating functions. The explicit forms of the functions are as follows 
	\begin{equation}
	I_1 = I_4^{(15,0,1)}=F_4^{(15,1 )} (\tau ) 
	= \Biggl[ \frac{\sum_{k=1}^2 c_k E_{15+k} (\tau )}{\sum_{k=1}^3 d_k E_{15+k} (\tau )} \Biggr] ,
	\end{equation}
	with 
	\bea
	c_1 &=& \frac{26315271553053477373 N^2 \left(378 \zeta (3)^2+5 \pi
		^6\right)}{64596336407956238642757 \pi ^2}, \\ \nonumber
	c_2 &=& \frac{26315271553053477373 \pi  N}{259721319947216543325}, \\ \nonumber
	d_1 &=& \frac{26315271553053477373 N^2 \left(4725 \zeta (3)^2+31 \pi
		^6\right)}{9689450461193435796413550}, \\ \nonumber
	d_2 &=& \frac{26315271553053477373 N \left(3780 \zeta (3)^2+71 \pi
		^6\right)}{65449772626698568917900 \pi ^3}, \\ \nonumber
	d_3 &=& 1
	\eea
	and
	\begin{equation}
	I_2 = I_2^{(2,0,1)}=F_2^{(2,1 )} (\tau ) 
	= \Biggl[ \frac{ c_1 E_{3}(\tau ) }
	{\sum_{k=1}^2 d_k E_{2+k} (\tau )} \Biggr]^{1} ,
	\end{equation}
	with
	\bea
	c_1 &=&\frac{\pi  2^{7-2 N} (2 N-2)!}{70875 N^2 (N-2)! (N-1)!}, \\ \nonumber
	d_1 &=& 0, d_2 = 1\nonumber
	\eea
	and
	\begin{equation}
	I_3 = I_{14}^{(10,0,1)}=F_{14}^{(10,1)} (\tau ) 
	= \frac{2^{3-2 N} (2 N-2)!}{N^2 (N-2)! (N-1)!} \Biggl[ \frac{\sum_{k=1}^7 c_k E_{10+k} (\tau )}{\sum_{k=1}^8 d_k E_{10+k} (\tau )} \Biggr] ,
	\end{equation}
	with
		\bea
	c_1 &=& c_3 =c_4=c_5=c_6=0, \\ \nonumber
	c_2 &=& \frac{105261086212213909492 \pi ^6 \zeta (5)}{121611979746830106693534375}, \\ \nonumber
	c_7 &=& \frac{842088689697711275936 \pi }{3681549710251794501631875}, \\ \nonumber
	d_1 &=& \frac{26315271553053477373 \pi ^7 \zeta (7)}{81287608563277749032400}, \\ \nonumber
	d_2 &=& d_3=d_4=d_5=d_6=d_7=0 ,d_8 =1
 \\ \nonumber
		\eea
		\newpage
\begin{sideways}
	\parbox{9.25in}{	
\section{Construction of S--duality invariant interpolation for finite spin anomalous dimensions}
\label{finsd}
The finite spin interpolating function mentioned in for $s=10, \, m=4 $ and $ \alpha =1 $ reads, 
	\begin{equation}
	I_4^{(10,0,1)}=F_4^{(10,1 )} (\tau ) 
	= \Biggl[ \frac{\sum_{k=1}^2 c_k E_{10+k} (\tau )}{\sum_{k=1}^3 d_k E_{10+k} (\tau )} \Biggr] ,
	\end{equation}
	}
	%\textcolor{red}{I don't know how to write it in a good format! Only way seems sideways format.}
\end{sideways}	
	\begin{sideways}
		\parbox{9.25in}{
			where the coefficients $c_k \, \& \, d_k$ reads, 
			
			\begin{tiny}
				
				\bea
				c_1 &=& \tfrac{\splitfrac{(657931 (2 + 
						j) N^2 \pi^2 (-(S_3 + \bar{S}_{-3} - 
						2 \bar{S}_{-2,1}^2 + 
						4 S_1^3 (S_3 + 3 \bar{S}_{-3} - 2 \bar{S}_{-2,1}) + 
						2 S_1^2 (3 S_4 + 8 \bar{S}_{-4} - \bar{S}_{-2}^2 - 
						12 \bar{S}_{-3,1} - \bar{S}_{-2,2} + 
						16 \bar{S}_{-2, 1}) - 
						2 S_1}{ (-S_5 - 3 \bar{S}_{-5} + 
						2 \bar{S}_{-3} \bar{S}_{-2} + 6 \bar{S}_{-4, 1}+ 
						6 \bar{S}_{-3, 2} + 
						S_2 (S_3 + \bar{S}_{-3}- 2 \bar{S}_{-2, 1})- 
						4 \bar{S}_{-2} + \bar{S}_{-2, 1} + 6 \bar{S}_{-2, 3} - 
						12 \bar{S}_{-3, 1, 1} - 12 \bar{S}_{-2, 1, 2} - 
						12 \bar{S}_{-2, 2, 1} + 
						24 \bar{S}_{-2, 1, 1, 1})))}}{(128176950 (-2 S_1^2 + 
					2 (2 + j) S_1 (S_2 + \bar{S}_{-2}) + (2 + j) (S_3 + 
					\bar{S}_{-3} - 2 \bar{S}_{-2, 1})))} ,\\ \nonumber
				%&\stackrel{\text{j=20, N=130}}{=}& 6025.18, \\ \nonumber
				c_2 &=& \tfrac{(119743442 N \pi S_1)}{1181820455}, \\ \nonumber
			%	&\stackrel{\text{j=20, N=130}}{=}& 148.875, \\ \nonumber
				d_1 &=& \tfrac{\splitfrac{(657931 N^2 \pi^2 (-(S_3 + \bar{S}_{-3} - 
						2 \bar{S}_{-2, 1}^2 + 
						4 S_1^3 (S_3 + 3 \bar{S}_{-3} - 2 \bar{S}_{-2, 1}) + 
						2 S_1^2 (3 S_4 + 8 \bar{S}_{-4} - \bar{S}_{-2}^2 - 
						12 \bar{S}{-3,1} - 10 \bar{S}_{-2, 2} + 
						16 \bar{S}_{-2, 1, 1} - }{
						2 S_1 (-S_5 - 3 \bar{S}_{-5} + 
						2 \bar{S}_{-3} \bar{S}_{-2} + 6 \bar{S}_{-4, 1} +
						6 \bar{S}_{-3, 2} + 
						S_2 (S_3 + \bar{S}_{-3} - 2 \bar{S}_{-2, 1}) - 
						4 \bar{S}_{-2} \bar{S}_{-2, 1} + 6 \bar{S}_{-2, 3} - 
						12 \bar{S}_{-3, 1, 1} - 12 \bar{S}_{-2, 1, 2} - 
						12 \bar{S}_{-2, 2, 1} + 
						24 \bar{S}_{-2, 1, 1, 1})))}}{(128176950 (-2 S_1^2 + 
					2 (2 + j) S_1 (S_2 + 
					\bar{S}_{-2}) + (2 + j) (S_3 + 
					\bar{S}_{-3} - 2 \bar{S}_{-2, 1})))}, \\ \nonumber
				%&\stackrel{\text{j=20, N=130}}{=}& 273.872 ,\\ \nonumber
				d_2 &=& \tfrac{\splitfrac{(119743442 N \pi (-2 S_1^2 (S_2 - (2 + j) S_3 - 6 \bar{S}_{-3} - 
						3 j \bar{S}_{-3} + \bar{S}_{-2} + 4 \bar{S}_{-2, 1} + 
						2 j \bar{S}_{-2, 1}) + 
						S_1 (2 (2 + j) S_2^2 - S_3 + 6 S_4 + 
						3 j S_4 + 16 \bar{S}_{-4} + 8 j \bar{S}_{-4} - 
						\bar{S}_{-3} + 4 (2 + j) S_2 \bar{S}_{-2} + 
						2 \bar{S}_{-2}^2 + j \bar{S}_{-2}^2 - 24 \bar{S}_{-3, 1} - 
						12 j \bar{S}_{-3, 1} +}{ 2 \bar{S}_{-2, 1} - 
						20 \bar{S}_{-2, 2} - 10 j \bar{S}_{-2, 2} + 
						32 \bar{S}_{-2, 1, 1} + 16 j \bar{S}_{-2, 1, 1}) + (2 + 
						j) (S_5 + 3 \bar{S}_{-5} + 2 S_3 \bar{S}_{-2} - 
						6 \bar{S}_{-4, 1} - 6 \bar{S}_{-3, 2} + 
						S_2 (S_3 + \bar{S}_{-3} - 2 \bar{S}_{-2, 1}) - 
						6 \bar{S}_{-2, 3} + 12 \bar{S}_{-3, 1, 1} + 
						12 \bar{S}_{-2, 1, 2} + 12 \bar{S}_{-2, 2, 1} - 
						24 \bar{S}_{-2, 1, 1, 1})))}}{(1181820455 (-2 S_1^2 + 
					2 (2 + j) S_1 (S_2 + \bar{S}_{-2}) + (2 + j) (S_3 + 
					\bar{S}_{-3} - 2 \bar{S}_{-2, 1})))}, \\ \nonumber
				%&\stackrel{\text{j=20, N=130}}{=}& 82.8495,\\ \nonumber
				d_3 &=& 1
				\eea
		\end{tiny} The Eisenstein series ($E_{s+k}$) has an infinite sum over modified Bessel ($K$)  which has been truncated to include the first twenty sums.}
	\end{sideways}

%%%%%%%%%%%%%%%%%%%%%%%%%%%%%%%%%%%%%%%%%%%%%%%%%
%%%%%%%%%%%%%%%%%%%%%%%%%%%%%%%%%%%%%%%%%%%%%%%%%

\bibliographystyle{JHEP}
	
%\bibliography{ReferencesCAD}

\begin{thebibliography}{10}
		
		\bibitem{Maldacena:1997re}
		J.~M. Maldacena, \emph{{The Large N limit of superconformal field theories and
				supergravity}}, \href{https://doi.org/10.1023/A:1026654312961}{\emph{Int. J.
				Theor. Phys.} {\bfseries 38} (1999) 1113}
		[\href{https://arxiv.org/abs/hep-th/9711200}{{\ttfamily hep-th/9711200}}].
		
		\bibitem{Gubser:1998bc}
		S.~S. Gubser, I.~R. Klebanov and A.~M. Polyakov, \emph{{Gauge theory
				correlators from noncritical string theory}},
		\href{https://doi.org/10.1016/S0370-2693(98)00377-3}{\emph{Phys. Lett.}
			{\bfseries B428} (1998) 105}
		[\href{https://arxiv.org/abs/hep-th/9802109}{{\ttfamily hep-th/9802109}}].
		
		\bibitem{Witten:1998qj}
		E.~Witten, \emph{{Anti-de Sitter space and holography}},
		\href{https://doi.org/10.4310/ATMP.1998.v2.n2.a2}{\emph{Adv. Theor. Math.
				Phys.} {\bfseries 2} (1998) 253}
		[\href{https://arxiv.org/abs/hep-th/9802150}{{\ttfamily hep-th/9802150}}].
		
		\bibitem{tHooft:1973alw}
		G.~'t~Hooft, \emph{{A Planar Diagram Theory for Strong Interactions}},
		\href{https://doi.org/10.1016/0550-3213(74)90154-0}{\emph{Nucl. Phys.}
			{\bfseries B72} (1974) 461}.
		
		\bibitem{Beisert:2010jr}
		N.~Beisert et~al., \emph{{Review of AdS/CFT Integrability: An Overview}},
		\href{https://doi.org/10.1007/s11005-011-0529-2}{\emph{Lett. Math. Phys.}
			{\bfseries 99} (2012) 3} [\href{https://arxiv.org/abs/1012.3982}{{\ttfamily
				1012.3982}}].
		
		\bibitem{Korchemsky:1988si}
		G.~P. Korchemsky, \emph{{Asymptotics of the Altarelli-Parisi-Lipatov Evolution
				Kernels of Parton Distributions}},
		\href{https://doi.org/10.1142/S0217732389001453}{\emph{Mod. Phys. Lett.}
			{\bfseries A4} (1989) 1257}.
		
		\bibitem{Polyakov:1980ca}
		A.~M. Polyakov, \emph{{Gauge Fields as Rings of Glue}},
		\href{https://doi.org/10.1016/0550-3213(80)90507-6}{\emph{Nucl. Phys.}
			{\bfseries B164} (1980) 171}.
		
		\bibitem{Korchemsky:1985xj}
		G.~P. Korchemsky and A.~V. Radyushkin, \emph{{Loop Space Formalism and
				Renormalization Group for the Infrared Asymptotics of {QCD}}},
		\href{https://doi.org/10.1016/0370-2693(86)91439-5}{\emph{Phys. Lett.}
			{\bfseries B171} (1986) 459}.
		
		\bibitem{Kotikov:2002ab}
		A.~V. Kotikov and L.~N. Lipatov, \emph{{DGLAP and BFKL equations in the $N=4$
				supersymmetric gauge theory}},
		\href{https://doi.org/10.1016/S0550-3213(03)00264-5,
			10.1016/j.nuclphysb.2004.02.032}{\emph{Nucl. Phys.} {\bfseries B661} (2003)
			19} [\href{https://arxiv.org/abs/hep-ph/0208220}{{\ttfamily
				hep-ph/0208220}}].
		
		\bibitem{Kotikov:2004er}
		A.~Kotikov, L.~Lipatov, A.~Onishchenko and V.~Velizhanin, \emph{{Three loop
				universal anomalous dimension of the Wilson operators in N=4 SUSY Yang-Mills
				model}}, \href{https://doi.org/10.1016/j.physletb.2004.05.078,
			10.1016/j.physletb.2004.05.078}{\emph{Phys.Lett.} {\bfseries B595} (2004)
			521} [\href{https://arxiv.org/abs/hep-th/0404092}{{\ttfamily
				hep-th/0404092}}].
		
		\bibitem{Moch:2004pa}
		S.~Moch, J.~A.~M. Vermaseren and A.~Vogt, \emph{{The Three loop splitting
				functions in QCD: The Nonsinglet case}},
		\href{https://doi.org/10.1016/j.nuclphysb.2004.03.030}{\emph{Nucl. Phys.}
			{\bfseries B688} (2004) 101}
		[\href{https://arxiv.org/abs/hep-ph/0403192}{{\ttfamily hep-ph/0403192}}].
		
		\bibitem{Brandhuber:2012vm}
		A.~Brandhuber, G.~Travaglini and G.~Yang, \emph{{Analytic two-loop form factors
				in N=4 SYM}}, \href{https://doi.org/10.1007/JHEP05(2012)082}{\emph{JHEP}
			{\bfseries 05} (2012) 082} [\href{https://arxiv.org/abs/1201.4170}{{\ttfamily
				1201.4170}}].
		
		\bibitem{Brandhuber:2017bkg}
		A.~Brandhuber, M.~Kostacinska, B.~Penante and G.~Travaglini, \emph{{Higgs
				amplitudes from $\mathcal{N}=4$ super Yang-Mills theory}},
		\href{https://doi.org/10.1103/PhysRevLett.119.161601}{\emph{Phys. Rev. Lett.}
			{\bfseries 119} (2017) 161601}
		[\href{https://arxiv.org/abs/1707.09897}{{\ttfamily 1707.09897}}].
		
		\bibitem{Jin:2018fak}
		Q.~Jin and G.~Yang, \emph{{Analytic Two-Loop Higgs Amplitudes in Effective
				Field Theory and the Maximal Transcendentality Principle}},
		\href{https://doi.org/10.1103/PhysRevLett.121.101603}{\emph{Phys. Rev. Lett.}
			{\bfseries 121} (2018) 101603}
		[\href{https://arxiv.org/abs/1804.04653}{{\ttfamily 1804.04653}}].
		
		\bibitem{Li:2014afw}
		Y.~Li, A.~von Manteuffel, R.~M. Schabinger and H.~X. Zhu, \emph{{Soft-virtual
				corrections to Higgs production at N$^3$LO}},
		\href{https://doi.org/10.1103/PhysRevD.91.036008}{\emph{Phys. Rev.}
			{\bfseries D91} (2015) 036008}
		[\href{https://arxiv.org/abs/1412.2771}{{\ttfamily 1412.2771}}].
		
		\bibitem{Li:2016ctv}
		Y.~Li and H.~X. Zhu, \emph{{Bootstrapping Rapidity Anomalous Dimensions for
				Transverse-Momentum Resummation}},
		\href{https://doi.org/10.1103/PhysRevLett.118.022004}{\emph{Phys. Rev. Lett.}
			{\bfseries 118} (2017) 022004}
		[\href{https://arxiv.org/abs/1604.01404}{{\ttfamily 1604.01404}}].
		
		\bibitem{Sen:2013oza}
		A.~Sen, \emph{{S-duality Improved Superstring Perturbation Theory}},
		\href{https://doi.org/10.1007/JHEP11(2013)029}{\emph{JHEP} {\bfseries 11}
			(2013) 029} [\href{https://arxiv.org/abs/1304.0458}{{\ttfamily 1304.0458}}].
		
		\bibitem{Beem:2013hha}
		C.~Beem, L.~Rastelli, A.~Sen and B.~C. van Rees, \emph{{Resummation and
				S-duality in N=4 SYM}},
		\href{https://doi.org/10.1007/JHEP04(2014)122}{\emph{JHEP} {\bfseries 04}
			(2014) 122} [\href{https://arxiv.org/abs/1306.3228}{{\ttfamily 1306.3228}}].
		
		\bibitem{Alday:2013bha}
		L.~F. Alday and A.~Bissi, \emph{{Modular interpolating functions for N=4 SYM}},
		\href{https://doi.org/10.1007/JHEP07(2014)007}{\emph{JHEP} {\bfseries 07}
			(2014) 007} [\href{https://arxiv.org/abs/1311.3215}{{\ttfamily 1311.3215}}].
		
		\bibitem{Honda:2014bza}
		M.~Honda, \emph{{On Perturbation theory improved by Strong coupling
				expansion}}, \href{https://doi.org/10.1007/JHEP12(2014)019}{\emph{JHEP}
			{\bfseries 12} (2014) 019} [\href{https://arxiv.org/abs/1408.2960}{{\ttfamily
				1408.2960}}].
		
		\bibitem{Honda:2015ewa}
		M.~Honda and D.~P. Jatkar, \emph{{Interpolating function and Stokes
				Phenomena}},
		\href{https://doi.org/10.1016/j.nuclphysb.2015.09.024}{\emph{Nucl. Phys.}
			{\bfseries B900} (2015) 533}
		[\href{https://arxiv.org/abs/1504.02276}{{\ttfamily 1504.02276}}].
		
		\bibitem{Chowdhury:2016hny}
		A.~Chowdhury, M.~Honda and S.~Thakur, \emph{{S-duality invariant perturbation
				theory improved by holography}},
		\href{https://doi.org/10.1007/JHEP04(2017)137}{\emph{JHEP} {\bfseries 04}
			(2017) 137} [\href{https://arxiv.org/abs/1607.01716}{{\ttfamily
				1607.01716}}].
		
		\bibitem{Osborn:1979tq}
		H.~Osborn, \emph{{Topological Charges for N=4 Supersymmetric Gauge Theories and
				Monopoles of Spin 1}},
		\href{https://doi.org/10.1016/0370-2693(79)91118-3}{\emph{Phys. Lett.}
			{\bfseries 83B} (1979) 321}.
		
		\bibitem{Korchemsky:2015cyx}
		G.~P. Korchemsky, \emph{{On level crossing in conformal field theories}},
		\href{https://doi.org/10.1007/JHEP03(2016)212}{\emph{JHEP} {\bfseries 03}
			(2016) 212} [\href{https://arxiv.org/abs/1512.05362}{{\ttfamily
				1512.05362}}].
		
		\bibitem{Kleinert:2001ax}
		H.~Kleinert and V.~Schulte-Frohlinde, \emph{{Critical properties of
				phi**4-theories}}. 2001.
		
		\bibitem{Pius:2013tla}
		R.~Pius and A.~Sen, \emph{{S-duality improved perturbation theory in
				compactified type I/heterotic string theory}},
		\href{https://doi.org/10.1007/JHEP06(2014)068}{\emph{JHEP} {\bfseries 06}
			(2014) 068} [\href{https://arxiv.org/abs/1310.4593}{{\ttfamily 1310.4593}}].
		
		\bibitem{Klevang}
		O.~Klevang, \emph{{Automorphic Forms in String Theory}}, MSc Thesis, 
		Chalmers University of Technology, Goteborg, Sweden (2010).
		
		\bibitem{Goncalves:2014ffa}
		V.~Goncalves, \emph{{Four point function of $\mathcal{N}=4$ stress-tensor
				multiplet at strong coupling}},
		\href{https://doi.org/10.1007/JHEP04(2015)150}{\emph{JHEP} {\bfseries 04}
			(2015) 150} [\href{https://arxiv.org/abs/1411.1675}{{\ttfamily 1411.1675}}].
		
		\bibitem{Alday:2018pdi}
		L.~F. Alday, A.~Bissi and E.~Perlmutter, \emph{{Genus-One String Amplitudes
				from Conformal Field Theory}},
		\href{https://arxiv.org/abs/1809.10670}{{\ttfamily 1809.10670}}.
		
		\bibitem{Bern:2005iz}
		Z.~Bern, L.~J. Dixon and V.~A. Smirnov, \emph{{Iteration of planar amplitudes
				in maximally supersymmetric Yang-Mills theory at three loops and beyond}},
		\href{https://doi.org/10.1103/PhysRevD.72.085001}{\emph{Phys. Rev.}
			{\bfseries D72} (2005) 085001}
		[\href{https://arxiv.org/abs/hep-th/0505205}{{\ttfamily hep-th/0505205}}].
		
		\bibitem{Gubser:2002tv}
		S.~S. Gubser, I.~R. Klebanov and A.~M. Polyakov, \emph{{A Semiclassical limit
				of the gauge / string correspondence}},
		\href{https://doi.org/10.1016/S0550-3213(02)00373-5}{\emph{Nucl. Phys.}
			{\bfseries B636} (2002) 99}
		[\href{https://arxiv.org/abs/hep-th/0204051}{{\ttfamily hep-th/0204051}}].
		
		\bibitem{Kruczenski:2002fb}
		M.~Kruczenski, \emph{{A Note on twist two operators in N=4 SYM and Wilson loops
				in Minkowski signature}},
		\href{https://doi.org/10.1088/1126-6708/2002/12/024}{\emph{JHEP} {\bfseries
				12} (2002) 024} [\href{https://arxiv.org/abs/hep-th/0210115}{{\ttfamily
				hep-th/0210115}}].
		
		\bibitem{Beisert:2006ez}
		N.~Beisert, B.~Eden and M.~Staudacher, \emph{{Transcendentality and Crossing}},
		\href{https://doi.org/10.1088/1742-5468/2007/01/P01021}{\emph{J. Stat. Mech.}
			{\bfseries 0701} (2007) P01021}
		[\href{https://arxiv.org/abs/hep-th/0610251}{{\ttfamily hep-th/0610251}}].
		
		\bibitem{Bern:2006ew}
		Z.~Bern, M.~Czakon, L.~J. Dixon, D.~A. Kosower and V.~A. Smirnov, \emph{{The
				Four-Loop Planar Amplitude and Cusp Anomalous Dimension in Maximally
				Supersymmetric Yang-Mills Theory}},
		\href{https://doi.org/10.1103/PhysRevD.75.085010}{\emph{Phys. Rev.}
			{\bfseries D75} (2007) 085010}
		[\href{https://arxiv.org/abs/hep-th/0610248}{{\ttfamily hep-th/0610248}}].
		
		\bibitem{Cachazo:2006az}
		F.~Cachazo, M.~Spradlin and A.~Volovich, \emph{{Four-loop cusp anomalous
				dimension from obstructions}},
		\href{https://doi.org/10.1103/PhysRevD.75.105011}{\emph{Phys. Rev.}
			{\bfseries D75} (2007) 105011}
		[\href{https://arxiv.org/abs/hep-th/0612309}{{\ttfamily hep-th/0612309}}].
		
		\bibitem{Henn:2013wfa}
		J.~M. Henn and T.~Huber, \emph{{The four-loop cusp anomalous dimension in
				$\mathcal{N} =$ 4 super Yang-Mills and analytic integration techniques for
				Wilson line integrals}},
		\href{https://doi.org/10.1007/JHEP09(2013)147}{\emph{JHEP} {\bfseries 09}
			(2013) 147} [\href{https://arxiv.org/abs/1304.6418}{{\ttfamily 1304.6418}}].
		
		\bibitem{Roiban:2007dq}
		R.~Roiban and A.~A. Tseytlin, \emph{{Strong-coupling expansion of cusp anomaly
				from quantum superstring}},
		\href{https://doi.org/10.1088/1126-6708/2007/11/016}{\emph{JHEP} {\bfseries
				11} (2007) 016} [\href{https://arxiv.org/abs/0709.0681}{{\ttfamily
				0709.0681}}].
		
		\bibitem{Roiban:2007ju}
		R.~Roiban and A.~A. Tseytlin, \emph{{Spinning superstrings at two loops:
				Strong-coupling corrections to dimensions of large-twist SYM operators}},
		\href{https://doi.org/10.1103/PhysRevD.77.066006}{\emph{Phys. Rev.}
			{\bfseries D77} (2008) 066006}
		[\href{https://arxiv.org/abs/0712.2479}{{\ttfamily 0712.2479}}].
		
		\bibitem{Boels:2017skl}
		R.~H. Boels, T.~Huber and G.~Yang, \emph{{Four-Loop Nonplanar Cusp Anomalous
				Dimension in N=4 Supersymmetric Yang-Mills Theory}},
		\href{https://doi.org/10.1103/PhysRevLett.119.201601}{\emph{Phys. Rev. Lett.}
			{\bfseries 119} (2017) 201601}
		[\href{https://arxiv.org/abs/1705.03444}{{\ttfamily 1705.03444}}].
		
		\bibitem{Boels:2017ftb}
		R.~H. Boels, T.~Huber and G.~Yang, \emph{{The Sudakov form factor at four loops
				in maximal super Yang-Mills theory}},
		\href{https://doi.org/10.1007/JHEP01(2018)153}{\emph{JHEP} {\bfseries 01}
			(2018) 153} [\href{https://arxiv.org/abs/1711.08449}{{\ttfamily
				1711.08449}}].
		
		\bibitem{Basso:2007wd}
		B.~Basso, G.~P. Korchemsky and J.~Kotanski, \emph{{Cusp anomalous dimension in
				maximally supersymmetric Yang-Mills theory at strong coupling}},
		\href{https://doi.org/10.1103/PhysRevLett.100.091601}{\emph{Phys. Rev. Lett.}
			{\bfseries 100} (2008) 091601}
		[\href{https://arxiv.org/abs/0708.3933}{{\ttfamily 0708.3933}}].
		
		\bibitem{Korchemsky:2017ttd}
		G.~P. Korchemsky, \emph{{Instanton effects in correlation functions on the
				light-cone}}, \href{https://doi.org/10.1007/JHEP12(2017)093}{\emph{JHEP}
			{\bfseries 12} (2017) 093}
		[\href{https://arxiv.org/abs/1704.00448}{{\ttfamily 1704.00448}}].
		
		\bibitem{Alday:2016tll}
		L.~F. Alday and G.~P. Korchemsky, \emph{{Revisiting instanton corrections to
				the Konishi multiplet}},
		\href{https://doi.org/10.1007/JHEP12(2016)005}{\emph{JHEP} {\bfseries 12}
			(2016) 005} [\href{https://arxiv.org/abs/1605.06346}{{\ttfamily
				1605.06346}}].
		
		\bibitem{Dorey:1998xe}
		N.~Dorey, V.~V. Khoze, M.~P. Mattis and S.~Vandoren, \emph{{Yang-Mills
				instantons in the large N limit and the AdS / CFT correspondence}},
		\href{https://doi.org/10.1016/S0370-2693(98)01233-7}{\emph{Phys. Lett.}
			{\bfseries B442} (1998) 145}
		[\href{https://arxiv.org/abs/hep-th/9808157}{{\ttfamily hep-th/9808157}}].
		
		\bibitem{Alday:2010vh}
		L.~F. Alday, J.~Maldacena, A.~Sever and P.~Vieira, \emph{{Y-system for
				Scattering Amplitudes}},
		\href{https://doi.org/10.1088/1751-8113/43/48/485401}{\emph{J. Phys.}
			{\bfseries A43} (2010) 485401}
		[\href{https://arxiv.org/abs/1002.2459}{{\ttfamily 1002.2459}}].
		
		\bibitem{Gromov:2013pga}
		N.~Gromov, V.~Kazakov, S.~Leurent and D.~Volin, \emph{{Quantum Spectral Curve
				for Planar $\mathcal{N} = 4$ Super-Yang-Mills Theory}},
		\href{https://doi.org/10.1103/PhysRevLett.112.011602}{\emph{Phys. Rev. Lett.}
			{\bfseries 112} (2014) 011602}
		[\href{https://arxiv.org/abs/1305.1939}{{\ttfamily 1305.1939}}].
		
		\bibitem{Gromov:2014bva}
		N.~Gromov, F.~Levkovich-Maslyuk, G.~Sizov and S.~Valatka, \emph{{Quantum
				spectral curve at work: from small spin to strong coupling in $ \mathcal{N} $
				= 4 SYM}}, \href{https://doi.org/10.1007/JHEP07(2014)156}{\emph{JHEP}
			{\bfseries 07} (2014) 156} [\href{https://arxiv.org/abs/1402.0871}{{\ttfamily
				1402.0871}}].
		
		\bibitem{Gromov:2014caa}
		N.~Gromov, V.~Kazakov, S.~Leurent and D.~Volin, \emph{{Quantum spectral curve
				for arbitrary state/operator in AdS$_{5}$/CFT$_{4}$}},
		\href{https://doi.org/10.1007/JHEP09(2015)187}{\emph{JHEP} {\bfseries 09}
			(2015) 187} [\href{https://arxiv.org/abs/1405.4857}{{\ttfamily 1405.4857}}].
		
		\bibitem{Gromov:2017blm}
		N.~Gromov, \emph{{Introduction to the Spectrum of $N=4$ SYM and the Quantum
				Spectral Curve}},  \href{https://arxiv.org/abs/1708.03648}{{\ttfamily
				1708.03648}}.
		
		\bibitem{Kotikov:2001sc}
		A.~V. Kotikov and L.~N. Lipatov, \emph{{DGLAP and BFKL evolution equations in
				the N=4 supersymmetric gauge theory}},  in \emph{{35th Annual Winter School
				on Nuclear and Particle Physics Repino, Russia, February 19-25, 2001}}, 2001,
		\href{https://arxiv.org/abs/hep-ph/0112346}{{\ttfamily hep-ph/0112346}}.
		
		\bibitem{Kotikov:2007cy}
		A.~V. Kotikov, L.~N. Lipatov, A.~Rej, M.~Staudacher and V.~N. Velizhanin,
		\emph{{Dressing and wrapping}},
		\href{https://doi.org/10.1088/1742-5468/2007/10/P10003}{\emph{J. Stat. Mech.}
			{\bfseries 0710} (2007) P10003}
		[\href{https://arxiv.org/abs/0704.3586}{{\ttfamily 0704.3586}}].
		
		\bibitem{Kotikov:2000pm}
		A.~V. Kotikov and L.~N. Lipatov, \emph{{NLO corrections to the BFKL equation in
				QCD and in supersymmetric gauge theories}},
		\href{https://doi.org/10.1016/S0550-3213(00)00329-1}{\emph{Nucl. Phys.}
			{\bfseries B582} (2000) 19}
		[\href{https://arxiv.org/abs/hep-ph/0004008}{{\ttfamily hep-ph/0004008}}].
		
		\bibitem{Kotikov:2003fb}
		A.~V. Kotikov, L.~N. Lipatov and V.~N. Velizhanin, \emph{{Anomalous dimensions
				of Wilson operators in N=4 SYM theory}},
		\href{https://doi.org/10.1016/S0370-2693(03)00184-9}{\emph{Phys. Lett.}
			{\bfseries B557} (2003) 114}
		[\href{https://arxiv.org/abs/hep-ph/0301021}{{\ttfamily hep-ph/0301021}}].
		
		\bibitem{Lukowski:2009ce}
		T.~Lukowski, A.~Rej and V.~N. Velizhanin, \emph{{Five-Loop Anomalous Dimension
				of Twist-Two Operators}},
		\href{https://doi.org/10.1016/j.nuclphysb.2010.01.008}{\emph{Nucl. Phys.}
			{\bfseries B831} (2010) 105}
		[\href{https://arxiv.org/abs/0912.1624}{{\ttfamily 0912.1624}}].
		
		\bibitem{Velizhanin:2010cm}
		V.~N. Velizhanin, \emph{{Six-Loop Anomalous Dimension of Twist-Three Operators
				in N=4 SYM}}, \href{https://doi.org/10.1007/JHEP11(2010)129}{\emph{JHEP}
			{\bfseries 11} (2010) 129} [\href{https://arxiv.org/abs/1003.4717}{{\ttfamily
				1003.4717}}].
		
		\bibitem{Marboe:2014sya}
		C.~Marboe, V.~Velizhanin and D.~Volin, \emph{{Six-loop anomalous dimension of
				twist-two operators in planar $ \mathcal{N}=4 $ SYM theory}},
		\href{https://doi.org/10.1007/JHEP07(2015)084}{\emph{JHEP} {\bfseries 07}
			(2015) 084} [\href{https://arxiv.org/abs/1412.4762}{{\ttfamily 1412.4762}}].
		
		\bibitem{Marboe:2016igj}
		C.~Marboe and V.~Velizhanin, \emph{{Twist-2 at seven loops in planar $
				\mathcal{N} $ = 4 SYM theory: full result and analytic properties}},
		\href{https://doi.org/10.1007/JHEP11(2016)013}{\emph{JHEP} {\bfseries 11}
			(2016) 013} [\href{https://arxiv.org/abs/1607.06047}{{\ttfamily
				1607.06047}}].
		
		\bibitem{Velizhanin:2009gv}
		V.~N. Velizhanin, \emph{{The Non-planar contribution to the four-loop universal
				anomalous dimension in N=4 Supersymmetric Yang-Mills theory}},
		\href{https://doi.org/10.1134/S0021364009120017}{\emph{JETP Lett.} {\bfseries
				89} (2009) 593} [\href{https://arxiv.org/abs/0902.4646}{{\ttfamily
				0902.4646}}].
		
		\bibitem{Velizhanin:2010ey}
		V.~N. Velizhanin, \emph{{The Non-planar contribution to the four-loop anomalous
				dimension of twist-2 operators: First moments in N=4 SYM and non-singlet
				QCD}}, \href{https://doi.org/10.1016/j.nuclphysb.2011.01.004}{\emph{Nucl.
				Phys.} {\bfseries B846} (2011) 137}
		[\href{https://arxiv.org/abs/1008.2752}{{\ttfamily 1008.2752}}].
		
		\bibitem{Velizhanin:2014zla}
		V.~N. Velizhanin, \emph{{Non-planar anomalous dimension of twist-2 operators:
				higher moments at four loops}},
		\href{https://doi.org/10.1016/j.nuclphysb.2014.06.021}{\emph{Nucl. Phys.}
			{\bfseries B885} (2014) 772}
		[\href{https://arxiv.org/abs/1404.7107}{{\ttfamily 1404.7107}}].
		
		\bibitem{Basso:2011rs}
		B.~Basso, \emph{{An exact slope for AdS/CFT}},
		\href{https://arxiv.org/abs/1109.3154}{{\ttfamily 1109.3154}}.
		
		\bibitem{Gromov:2011de}
		N.~Gromov, D.~Serban, I.~Shenderovich and D.~Volin, \emph{{Quantum folded
				string and integrability: From finite size effects to Konishi dimension}},
		\href{https://doi.org/10.1007/JHEP08(2011)046}{\emph{JHEP} {\bfseries 08}
			(2011) 046} [\href{https://arxiv.org/abs/1102.1040}{{\ttfamily 1102.1040}}].
		
		\bibitem{Roiban:2011fe}
		R.~Roiban and A.~A. Tseytlin, \emph{{Semiclassical string computation of
				strong-coupling corrections to dimensions of operators in Konishi
				multiplet}},
		\href{https://doi.org/10.1016/j.nuclphysb.2011.02.016}{\emph{Nucl. Phys.}
			{\bfseries B848} (2011) 251}
		[\href{https://arxiv.org/abs/1102.1209}{{\ttfamily 1102.1209}}].
		
		\bibitem{Vallilo:2011fj}
		B.~C. Vallilo and L.~Mazzucato, \emph{{The Konishi multiplet at strong
				coupling}}, \href{https://doi.org/10.1007/JHEP12(2011)029}{\emph{JHEP}
			{\bfseries 12} (2011) 029} [\href{https://arxiv.org/abs/1102.1219}{{\ttfamily
				1102.1219}}].
		
		\bibitem{Dolan:2001tt}
		F.~A. Dolan and H.~Osborn, \emph{{Superconformal symmetry, correlation
				functions and the operator product expansion}},
		\href{https://doi.org/10.1016/S0550-3213(02)00096-2}{\emph{Nucl. Phys.}
			{\bfseries B629} (2002) 3}
		[\href{https://arxiv.org/abs/hep-th/0112251}{{\ttfamily hep-th/0112251}}].
		
		\bibitem{DHoker:1999mic}
		E.~D'Hoker, S.~D. Mathur, A.~Matusis and L.~Rastelli, \emph{{The Operator
				product expansion of N=4 SYM and the 4 point functions of supergravity}},
		\href{https://doi.org/10.1016/S0550-3213(00)00523-X}{\emph{Nucl. Phys.}
			{\bfseries B589} (2000) 38}
		[\href{https://arxiv.org/abs/hep-th/9911222}{{\ttfamily hep-th/9911222}}].
		
		\bibitem{Arutyunov:2000ku}
		G.~Arutyunov, S.~Frolov and A.~C. Petkou, \emph{{Operator product expansion of
				the lowest weight CPOs in $\mathcal N=4$ SYM$_4$ at strong coupling}},
		\href{https://doi.org/10.1016/S0550-3213(01)00266-8,
			10.1016/S0550-3213(00)00439-9}{\emph{Nucl. Phys.} {\bfseries B586} (2000)
			547} [\href{https://arxiv.org/abs/hep-th/0005182}{{\ttfamily
				hep-th/0005182}}].
		
		\bibitem{Alday:2017xua}
		L.~F. Alday and A.~Bissi, \emph{{Loop Corrections to Supergravity on $AdS_5
				\times S^5$}},
		\href{https://doi.org/10.1103/PhysRevLett.119.171601}{\emph{Phys. Rev. Lett.}
			{\bfseries 119} (2017) 171601}
		[\href{https://arxiv.org/abs/1706.02388}{{\ttfamily 1706.02388}}].
		
		\bibitem{Aprile:2017bgs}
		F.~Aprile, J.~M. Drummond, P.~Heslop and H.~Paul, \emph{{Quantum Gravity from
				Conformal Field Theory}},
		\href{https://doi.org/10.1007/JHEP01(2018)035}{\emph{JHEP} {\bfseries 01}
			(2018) 035} [\href{https://arxiv.org/abs/1706.02822}{{\ttfamily
				1706.02822}}].
		
		\bibitem{Banks:2013nga}
		T.~Banks and T.~J. Torres, \emph{{Two Point Pade Approximants and Duality}},
		\href{https://arxiv.org/abs/1307.3689}{{\ttfamily 1307.3689}}.
		
		\bibitem{Gromov:2011bz}
		N.~Gromov and S.~Valatka, \emph{{Deeper Look into Short Strings}},
		\href{https://doi.org/10.1007/JHEP03(2012)058}{\emph{JHEP} {\bfseries 03}
			(2012) 058} [\href{https://arxiv.org/abs/1109.6305}{{\ttfamily 1109.6305}}].
		
		\bibitem{Tirziu:2008fk}
		A.~Tirziu and A.~A. Tseytlin, \emph{{Quantum corrections to energy of short
				spinning string in AdS(5)}},
		\href{https://doi.org/10.1103/PhysRevD.78.066002}{\emph{Phys. Rev.}
			{\bfseries D78} (2008) 066002}
		[\href{https://arxiv.org/abs/0806.4758}{{\ttfamily 0806.4758}}].
		
		\bibitem{Roiban:2009aa}
		R.~Roiban and A.~A. Tseytlin, \emph{{Quantum strings in AdS(5) x S**5:
				Strong-coupling corrections to dimension of Konishi operator}},
		\href{https://doi.org/10.1088/1126-6708/2009/11/013}{\emph{JHEP} {\bfseries
				11} (2009) 013} [\href{https://arxiv.org/abs/0906.4294}{{\ttfamily
				0906.4294}}].
		
		\bibitem{Floratos:2013cia}
		E.~Floratos, G.~Georgiou and G.~Linardopoulos, \emph{{Large-Spin Expansions of
				GKP Strings}}, \href{https://doi.org/10.1007/JHEP03(2014)018}{\emph{JHEP}
			{\bfseries 03} (2014) 018} [\href{https://arxiv.org/abs/1311.5800}{{\ttfamily
				1311.5800}}].
		
		\bibitem{Drukker:2005kx}
		N.~Drukker and B.~Fiol, \emph{{All-genus calculation of Wilson loops using
				D-branes}}, \href{https://doi.org/10.1088/1126-6708/2005/02/010}{\emph{JHEP}
			{\bfseries 02} (2005) 010}
		[\href{https://arxiv.org/abs/hep-th/0501109}{{\ttfamily hep-th/0501109}}].
		
		\bibitem{Chen:2006iu}
		B.~Chen and W.~He, \emph{{On 1/2-BPS Wilson-'t Hooft loops}},
		\href{https://doi.org/10.1103/PhysRevD.74.126008}{\emph{Phys. Rev.}
			{\bfseries D74} (2006) 126008}
		[\href{https://arxiv.org/abs/hep-th/0607024}{{\ttfamily hep-th/0607024}}].
		
		\bibitem{Boels:2012ew}
		R.~H. Boels, B.~A. Kniehl, O.~V. Tarasov and G.~Yang, \emph{{Color-kinematic
				Duality for Form Factors}},
		\href{https://doi.org/10.1007/JHEP02(2013)063}{\emph{JHEP} {\bfseries 02}
			(2013) 063} [\href{https://arxiv.org/abs/1211.7028}{{\ttfamily 1211.7028}}].
		
		\bibitem{Alday:2016jeo}
		L.~F. Alday and G.~P. Korchemsky, \emph{{Instanton corrections to twist-two
				operators}}, \href{https://doi.org/10.1007/JHEP06(2017)008}{\emph{JHEP}
			{\bfseries 06} (2017) 008}
		[\href{https://arxiv.org/abs/1609.08164}{{\ttfamily 1609.08164}}].
		
		\bibitem{Beisert:2003tq}
		N.~Beisert, C.~Kristjansen and M.~Staudacher, \emph{{The Dilatation operator of
				conformal N=4 superYang-Mills theory}},
		\href{https://doi.org/10.1016/S0550-3213(03)00406-1}{\emph{Nucl. Phys.}
			{\bfseries B664} (2003) 131}
		[\href{https://arxiv.org/abs/hep-th/0303060}{{\ttfamily hep-th/0303060}}].
		
		\bibitem{Arutyunov:2002rs}
		G.~Arutyunov, S.~Penati, A.~C. Petkou, A.~Santambrogio and E.~Sokatchev,
		\emph{{Nonprotected operators in N=4 SYM and multiparticle states of AdS(5)
				SUGRA}}, \href{https://doi.org/10.1016/S0550-3213(02)00679-X}{\emph{Nucl.
				Phys.} {\bfseries B643} (2002) 49}
		[\href{https://arxiv.org/abs/hep-th/0206020}{{\ttfamily hep-th/0206020}}].
		
		\bibitem{Beem:2013qxa}
		C.~Beem, L.~Rastelli and B.~C. van Rees, \emph{{The $\mathcal N=4$
				Superconformal Bootstrap}},
		\href{https://doi.org/10.1103/PhysRevLett.111.071601}{\emph{Phys. Rev. Lett.}
			{\bfseries 111} (2013) 071601}
		[\href{https://arxiv.org/abs/1304.1803}{{\ttfamily 1304.1803}}].
		
		\bibitem{Alday:2013opa}
		L.~F. Alday and A.~Bissi, \emph{{The superconformal bootstrap for structure
				constants}}, \href{https://doi.org/10.1007/JHEP09(2014)144}{\emph{JHEP}
			{\bfseries 09} (2014) 144} [\href{https://arxiv.org/abs/1310.3757}{{\ttfamily
				1310.3757}}].
		
		\bibitem{Alday:2014qfa}
		L.~F. Alday and A.~Bissi, \emph{{Generalized bootstrap equations for $
				\mathcal{N}=4 $ SCFT}},
		\href{https://doi.org/10.1007/JHEP02(2015)101}{\emph{JHEP} {\bfseries 02}
			(2015) 101} [\href{https://arxiv.org/abs/1404.5864}{{\ttfamily 1404.5864}}].
		
		\bibitem{Alday:2014tsa}
		L.~F. Alday, A.~Bissi and T.~Lukowski, \emph{{Lessons from crossing symmetry at
				large N}}, \href{https://doi.org/10.1007/JHEP06(2015)074}{\emph{JHEP}
			{\bfseries 06} (2015) 074} [\href{https://arxiv.org/abs/1410.4717}{{\ttfamily
				1410.4717}}].
		
		\bibitem{Eden:2000bk}
		B.~Eden, A.~C. Petkou, C.~Schubert and E.~Sokatchev, \emph{{Partial
				nonrenormalization of the stress tensor four point function in N=4 SYM and
				AdS / CFT}}, \href{https://doi.org/10.1016/S0550-3213(01)00151-1}{\emph{Nucl.
				Phys.} {\bfseries B607} (2001) 191}
		[\href{https://arxiv.org/abs/hep-th/0009106}{{\ttfamily hep-th/0009106}}].
		
		\bibitem{Arutyunov:2001mh}
		G.~Arutyunov, B.~Eden, A.~C. Petkou and E.~Sokatchev, \emph{{Exceptional
				nonrenormalization properties and OPE analysis of chiral four point functions
				in N=4 SYM(4)}},
		\href{https://doi.org/10.1016/S0550-3213(01)00569-7}{\emph{Nucl. Phys.}
			{\bfseries B620} (2002) 380}
		[\href{https://arxiv.org/abs/hep-th/0103230}{{\ttfamily hep-th/0103230}}].
		
		\bibitem{Eden:2001ec}
		B.~Eden and E.~Sokatchev, \emph{{On the OPE of 1/2 BPS short operators in N=4
				SCFT(4)}}, \href{https://doi.org/10.1016/S0550-3213(01)00492-8}{\emph{Nucl.
				Phys.} {\bfseries B618} (2001) 259}
		[\href{https://arxiv.org/abs/hep-th/0106249}{{\ttfamily hep-th/0106249}}].
		
		\bibitem{Alday:2007mf}
		L.~F. Alday and J.~M. Maldacena, \emph{{Comments on operators with large
				spin}}, \href{https://doi.org/10.1088/1126-6708/2007/11/019}{\emph{JHEP}
			{\bfseries 11} (2007) 019} [\href{https://arxiv.org/abs/0708.0672}{{\ttfamily
				0708.0672}}].
		
	\end{thebibliography}

%%%%%%%%%%%%%%%%%%%%%%%%%%%%%%%%%%%%%%%%%%%%%%%%%
%%%%%%%%%%%%%%%%%%%%%%%%%%%%%%%%%%%%%%%%%%%%%%%%%
\end{document}